\documentclass[11pt,a4paper]{article}%
\usepackage{array}
\usepackage{float}
\usepackage{amsmath}
\usepackage{amsfonts}
\usepackage{amssymb}
\usepackage{latexsym}
\usepackage{epsfig}
\usepackage{graphicx}
\usepackage{sw20elba}
\usepackage{mathrsfs}
\usepackage{fancyhdr}
\usepackage{sectsty}
\usepackage[longnamesfirst]{natbib}
\usepackage[justification=normal,labelsep=colon,font=it,labelfont=sc]%
{caption}
\usepackage{theorem}
\usepackage{url}
\usepackage{rotating}
\usepackage[truetex]{color}%
\newtheorem{theorem}{Theorem}

{
\theorembodyfont{\upshape}
\theoremheaderfont{\scshape}

}

\newtheorem{corollary}{Corollary}

\newtheorem{definition}{Definition}

{\theorembodyfont{\upshape}
\newtheorem{remark}{Remark}[section]
}

\newtheorem{lemma}{{\sc Lemma}}

\newenvironment{proof}[1][Proof]{\bigskip \noindent \textbf{#1:} }{\  \rule{0.5em}{0.5em}}

\theoremheaderfont{\bfseries}
\theorembodyfont{\upshape}
\sectionfont{\centering \sc}
\subsectionfont{\centering \sc}
\subsubsectionfont{\centering \sc}
\allsectionsfont{\centering \sc}
\allowdisplaybreaks

\bibpunct[:~]{(}{)}{;}{a}{,}{;}
\renewcommand{\cite}{\citet}

\graphicspath{ {./graphics/} }

\begin{document}

\begin{center}
\bigskip

{\Large INFERENCE UNDER RANDOM\ LIMIT\ BOOTSTRAP MEASURES{\LARGE \textsc{*}}}

\mbox{}

{\normalsize \ \vspace{-0.15cm}}

\textsc{Giuseppe Cavaliere}$^{a,b}$\textsc{\ and Iliyan Georgiev}$^{a,c}$

\textsc{\vspace{-0.15in}}

{\normalsize \vspace{0.45cm}}

{\small November 30, 2019}%

%TCIMACRO{\TeXButton{Acknowledgement}{\renewcommand{\thefootnote}{}
%\footnote{
%\hspace{-7.2mm}
%$^{*}%
%$ We are grateful to Ulrich Müller (handling Co-Editor) and three anonymous referees for the
%many constructive comments
%and suggestions on an earlier version of the paper. We have also benefited from discussions and feedback from
%Matias Cattaneo, Graham Elliott, Michael Jansson, Søren Johansen, Ye Lu,
%Adam McCloskey, Marcelo Moreira, Rasmus Søndergaard Pedersen, Anders Rahbek, Mervyn Silvapulle, Michael Wolf,
%seminar participants at Boston U, Cambridge, ETH, Exeter, FGV Sao Paulo, Geneve, Hitotsubashi, Kyoto, LSE, Melbourne,
%Oxford, PUC Rio, Queen's, QUT, SMU, Tilburg, UCL, UPF, Verona, Vienna,
%as well as participants to the $7^{th}$ Italian Congress of
%Econometrics and Empirical Economics (ICEEE), the $14^{th}%
%$ SETA meeting and the
%NBER/NSF 2018 Time Series Conference. Financial support from the
%University of Bologna, ALMA IDEA 2017 grants, and from the Danish Council for Independent Research
%(DSF Grant 7015-00028), is gratefully acknowledged. Address
%correspondence to: Giuseppe Cavaliere, Department of Economics, University of Bologna,
%Piazza Scaravilli 2, 40126 Bologna, Italy; email: giuseppe.cavaliere@unibo.it.
%\\
%$^{a}$ Department of Economics, University of Bologna, Italy.
%\\
%$^{b}$ Department of Economics, Exeter Business School, UK.
%\\
%$^{c}%
%$ Nova School of Business and Economics, Universidade Nova de Lisboa, Portugal.}
%\addtocounter{footnote}{-2}
%\renewcommand{\thefootnote}{\arabic{footnote}}}}%
%BeginExpansion
\renewcommand{\thefootnote}{}
\footnote{
\hspace{-7.2mm}
$^{*}%
$ We thank Matias Cattaneo, Graham Elliott, Michael Jansson, Søren Johansen, Ye Lu,
Adam McCloskey, Marcelo Moreira, Ulrich Müller, Rasmus Søndergaard Pedersen, Anders Rahbek,
Mervyn Silvapulle, Michael Wolf, for comments and suggestions. We also thank
seminar participants at Boston U, Cambridge, ETH, Exeter, FGV Sao Paulo, Geneve, Hitotsubashi, Kyoto, LSE, Melbourne,
Oxford, PUC Rio, Queen's, QUT, SMU, Tilburg, UCL, UPF, Verona, Vienna,
as well as participants to the $7^{th}$ Italian Congress of
Econometrics and Empirical Economics (ICEEE), the $14^{th}%
$ SETA meeting and the
NBER/NSF 2018 Time Series Conference. Financial support from the
University of Bologna, ALMA IDEA 2017 grants, and from the Danish Council for Independent Research
(DSF Grant 7015-00028), is gratefully acknowledged. Address
correspondence to: Giuseppe Cavaliere, Department of Economics, University of Bologna,
Piazza Scaravilli 2, 40126 Bologna, Italy; email: giuseppe.cavaliere@unibo.it.
\\
$^{a}$ Department of Economics, University of Bologna, Italy.
\\
$^{b}$ Department of Economics, Exeter Business School, UK.
\\
$^{c}%
$ Nova School of Business and Economics, Universidade Nova de Lisboa, Portugal.}
\addtocounter{footnote}{-2}
\renewcommand{\thefootnote}{\arabic{footnote}}%
%EndExpansion

\end{center}%

%TCIMACRO{\TeXButton{Begin intro}{\par\begingroup\leftskip=1 cm
%\rightskip=1 cm
%\small}}%
%BeginExpansion
\par\begingroup\leftskip=1 cm
\rightskip=1 cm
\small
%EndExpansion

\begin{center}
\textsc{Abstract\vspace{-0.15cm}}
\end{center}

Asymptotic bootstrap validity is usually understood as consistency of the
distribution of a bootstrap statistic, conditional on the data, for the
unconditional limit distribution of a statistic of interest. From this
perspective, randomness of the limit bootstrap measure is regarded as a
failure of the bootstrap. We show that such limiting randomness does not
necessarily invalidate bootstrap inference if validity is understood as
control over the frequency of correct inferences in large samples. We first
establish sufficient conditions for asymptotic bootstrap validity in cases
where the unconditional limit distribution of a statistic can be obtained by
averaging a (random) limiting bootstrap distribution. Further, we provide
results ensuring the asymptotic validity of the bootstrap as a tool for
conditional inference, the leading case being that where a bootstrap
distribution estimates consistently a conditional (and thus, random) limit
distribution of a statistic. We apply our framework to several inference
problems in econometrics, including linear models with possibly non-stationary
regressors, functional CUSUM statistics, conditional Kolmogorov-Smirnov
specification tests, the `parameter on the boundary' problem and tests for
constancy of parameters in dynamic econometric models. \bigskip


%TCIMACRO{\TeXButton{End intro}{\par\endgroup\normalsize}}%
%BeginExpansion
\par\endgroup\normalsize
%EndExpansion

\setcounter{footnote}{0}%

%TCIMACRO{\TeXButton{numbering of equations}{\numberwithin{equation}{section}%
%}}%
%BeginExpansion
\numberwithin{equation}{section}%
%EndExpansion
%

%TCIMACRO{\TeXButton{numbering of theorems, corollaries and lemmas}%
%{\numberwithin{theorem}{section}
%\numberwithin{corollary}{section}
%\numberwithin{lemma}{section}}}%
%BeginExpansion
\numberwithin{theorem}{section}
\numberwithin{corollary}{section}
\numberwithin{lemma}{section}%
%EndExpansion
\newpage

\section{Introduction}

\textsc{Consider a data sample }$D_{n}$ of size $n$ and a statistic $\tau
_{n}:=\tau_{n}\left(  D_{n}\right)  $, say a test statistic or a parameter
estimator, possibly normalized. Interest is in a distributional approximation
of $\tau_{n}$. Let a bootstrap procedure generate a bootstrap analogue
$\tau_{n}^{\ast}$ of $\tau_{n}$; i.e., computed on a bootstrap sample. Assume
that $\tau_{n}$ converges in distribution to a non-degenerate random variable
[rv], say $\tau$. In classic bootstrap inference, asymptotic bootstrap
validity is usually understood and established as convergence in probability
(or almost surely) of the cumulative distribution function {[}cdf{]} of the
bootstrap statistic $\tau_{n}^{\ast}$ conditional on the data $D_{n}$, say
$F_{n}^{\ast}$, to the unconditional cdf of $\tau$, say $F$. This convergence,
along with continuity of $F$, implies by Polya's theorem that $\sup
_{x\in\mathbb{R}}|F_{n}^{\ast}\left(  x\right)  -F\left(  x\right)
|\rightarrow0$, in probability (or almost surely).

In many applications, however, the bootstrap statistic $\tau_{n}^{\ast}$ may
possess, conditionally on the data, a \emph{random }limit distribution. Cases
of random bootstrap limit distributions appear in various areas of
econometrics and statistics; for instance, they are documented for
infinite-variance processes (Athreya, 1987; Knight, 1989; Aue, Berkes and
Horv\'{a}th, 2008; Cavaliere, Georgiev and Taylor, 2016), time series with
unit roots (Basawa, Mallik, McCormick, Reeves and Taylor, 1991; Cavaliere,
Nielsen and Rahbek, 2015), parameters on the boundary of the parameter space
(Andrews, 2000), subsample inference based on fixed-\emph{b} asymptotics (Shao
and Politis, 2013), cube-root consistent estimators (Sen, Banerjee and
Woodroofe, 2010; Cattaneo, Jansson and Nagasawa, 2017), Hodges-LeCam
superefficient estimators (Beran, 1997). In most of these cases, the
occurrence of a random limit distribution for the bootstrap statistic
$\tau_{n}^{\ast}$ given the data --{} in contrast to a non-random limit of the
unconditional distribution of the original statistic $\tau_{n}$ --{} is taken
as evidence of failure of the bootstrap.

In this paper we show that randomness in the limiting distribution of a
bootstrap statistic need not invalidate bootstrap inference. On the contrary,
although the bootstrap no longer estimates the limiting unconditional
distribution of the statistic of interest, it may\textbf{\ }still deliver
hypothesis tests (or confidence intervals) with the desired null rejection
probability (or coverage probability) when the sample size diverges. This
happens because asymptotic control over the frequency of wrong inferences can
be guaranteed by the asymptotic distributional uniformity of the bootstrap
\emph{p}-values, which in its turn can occur without the convergence in
probability (or almost surely) of the bootstrap cdf $F_{n}^{\ast}$ of
$\tau_{n}^{\ast}$ to the asymptotic cdf $F$ of $\tau$.

Therefore, instead of assessing bootstrap validity in terms of the convergence
of $F_{n}^{\ast}$ to $F$, in cases where the limit of the bootstrap
distribution is random we study bootstrap validity in terms of the property of
asymptotic distributional uniformity of bootstrap \emph{p}-values.
Specifically, let $p_{n}^{\ast}$ denote the bootstrap \emph{p}-value, usually
defined as $p_{n}^{\ast}:=F_{n}^{\ast}\left(  \tau_{n}\right)  $. We define
`bootstrap validity' or `unconditional bootstrap validity' the fact that
\begin{equation}
P(p_{n}^{\ast}\leq q)\rightarrow q\label{eq unconditional validity}%
\end{equation}
for $q\in\left(  0,1\right)  $. The focus on this property is not new in the
literature on bootstrap and simulation-based inference (see, e.g., Hansen,
1996, and Lockhart, 2012, among others).

Our first set of results provides sufficient conditions for bootstrap validity
in the sense of (\ref{eq unconditional validity}) in situations where the
bootstrap distribution is random in the limit. Classic results for bootstrap
validity when the limit bootstrap measure is not random can be obtained as
special cases. The main requirement in our results is that the unconditional
limit distribution of $\tau_{n}$ should be an average of the random limit
distribution of $\tau_{n}^{\ast}$ given the data.

It is often the case that bootstrap\ validity can be addressed through the
lens of a conditioning argument. In this regard, our second set of results
concerns the possibility that, for a sequence of random elements $X_{n}$, it
holds that the bootstrap \emph{p}-value is uniformly distributed in large
sample \emph{conditionally }on $X_{n}$:
\begin{equation}
P(p_{n}^{\ast}\leq q|X_{n})\overset{p}{\rightarrow}%
q\label{eq conditional validity}%
\end{equation}
for $q\in\left(  0,1\right)  $. This property, that we define `bootstrap
validity conditional on $X_{n}$', implies unconditional validity in the sense
of (\ref{eq unconditional validity}). Moreover, conditional bootstrap validity
given $X_{n}$ implies that the bootstrap replicates asymptotically the
property of conditional tests and confidence intervals to have, conditionally
on $X_{n}$, constant null rejection probability and coverage probability,
respectively (for further roles of conditioning in inference, like the
relevance of the drawn inferences and information recovery, see Reid, 1995,
and the references therein). The leading case where we show
(\ref{eq conditional validity}) to hold --\thinspace under regularity
conditions that will be discussed in the paper --\thinspace is that where the
(random) limit of the conditional distribution of $\tau_{n}$ given $X_{n}$
matches the (random) limit distribution of the bootstrap statistic. A property
like (\ref{eq conditional validity}) was initially established by LePage and
Podgorski (1996) for permutation tests in location models. Their approach has
not been developed further in the bootstrap literature, in particular because
it requires probabilistic tools that are not widely popular in this field.

When dealing with random limit distributions, the usual convergence concept
employed to establish bootstrap validity, i.e. weak convergence in
probability, can only be employed in some very special cases. Instead, our
formal discussion makes extensive use of the probabilistic concept of weak
convergence of random measures; see e.g. Kallenberg (2017, Ch.4). To our
knowledge, in the bootstrap context this concept has so far been mostly used
to obtain negative results of lack of validity for specific bootstrap
procedures (as e.g. in Knight, 1989, and Basawa et al., 1991), rather than
positive validity results, as we do here. As an ingredient of our analysis, we
also present some novel results on the weak convergence of conditional expectations.

To illustrate the practical relevance of our results, we initially present
them by using a simple linear model with either stationary or non-stationary
regressors, and later we analyze four well-known cases in the econometric
literature where the bootstrap features a random limit distribution. The first
is a standard CUSUM-type test of the i.i.d. property for a random sequence
with infinite variance. This is a case where the limit distribution of the
CUSUM statistic depends on unknown nuisance parameters (e.g., the tail index)
and bootstrap or permutation tests fail to estimate this distribution
consistently. We argue that a simple bootstrap based on permutations, albeit
having a random limit distribution and hence being invalid in the usual sense,
provides \emph{exact }conditional inference and hence is also unconditionally
valid in the sense of (\ref{eq unconditional validity}). The second
application considers a Kolmogorov-Smirnov-type test for correct specification
of the conditional distribution of a response variable given a vector of
covariates. Andrews (1997) considers a parametric bootstrap implementation
where the covariates are kept fixed across bootstrap samples. While in the
independent case the limit of the bootstrap distribution is non-random, this
is not the case in general. Using our theory we discuss conditions for
validity of the bootstrap within this framework. The third application is the
implementation of the bootstrap in `parameter on the boundary' problems
(Andrews, 1999,2000). Taking hypothesis testing in a predictive regression
framework as an illustration, we show that the kind of randomness in the limit
distribution of the bootstrap statistics of interest depends, when the true
parameter lies on the boundary of the parameter space, on how well the mutual
position of the boundary, the set identified by the null hypothesis and the
true parameter value, is approximated in the bootstrap world. Although the
standard bootstrap may fail to be valid in the sense of
(\ref{eq unconditional validity}), we provide conditions for the validity of
alternative bootstrap schemes.\textbf{\ }The fourth application includes an
analysis of the much applied bootstrap `$\sup F$' tests of parameter constancy
in regression models where the design matrix could be random but be
conditioned upon; see Hall (1992, p.170). In the resampling process forming
the bootstrap sample, it appears natural to take the design matrix as fixed
across the bootstrap repetitions. Under a set of assumptions proposed by
Hansen (2000), we argue that the fixed-regressor bootstrap `$\sup F$'
statistic has a random limit distribution, thus invalidating previous claims
in the literature that the bootstrap is consistent for the unconditional limit
distribution of the original `$\sup F$' test statistic. We then provide
conditions under which the fixed-regressor bootstrap is unconditionally valid
and, additionally, valid conditionally on the chosen set of regressors.

\subsection*{Structure of the paper}

The paper is organized as follows. In Section \ref{sec example} we outline the
main concepts and ideas using a simple linear regression model. Our main
theoretical results are presented in Section \ref{sec g}. Section
\ref{Section on Applications} contains the four applications of the theory,
whereas Section \ref{sec conclusion} concludes. The paper has two Appendices.
In Appendix \ref{sec itere} we collect some results on weak convergence in
distribution which are useful to prove our main theorems and develop the
applications. Appendix \ref{sec all proofs} contains the proofs of all theory
results given in the paper. The proofs of the results from Appendix
\ref{sec itere} and some additional material are collected in the accompanying
supplement, Cavaliere and Georgiev (2019).

\subsection*{Notation and definitions}

We use the following notation throughout. The spaces of c\`{a}dl\`{a}g
functions $[0,1]\rightarrow\mathbb{R}^{n}$, $[0,1]\rightarrow\mathbb{R}%
^{m\times n}$ and $\mathbb{R}\rightarrow\mathbb{R}$ (all equipped with the
respective Skorokhod $J_{1}$-topologies; see Kallenberg, 1997, Appendix
A2),\textbf{\ }are denoted by $\mathscr{D}{}_{n}$, $\mathscr{D}{}_{m\times n}$
and $\mathscr{D}({\mathbb{R}})$, respectively; for the first one, when $n=1$
the subscript is suppressed. Integrals are over $[0,1]$ unless otherwise
stated, $\Phi$ is the standard Gaussian cdf, $U(0,1)$ is the uniform
distribution on $[0,1]$ and $\mathbb{I}_{\{\cdot\}}$ is the indicator
function. If $F$ is a (random) cdf, $F^{-1}$ stands for the right-continuous
generalized inverse, i.e., $F^{-1}(u):=\sup\{v\in\mathbb{R}:F\left(  v\right)
\leq u\}$, $u\in\mathbb{R}$. Unless differently specified, limits are for
$n\rightarrow\infty$.

Polish (i.e., complete and separable metric) spaces are always equipped with
their Borel $\sigma$-algebras. Throughout, we assume that all random elements
are Polish-space valued and that well-defined conditional distributions exist.
For random elements of a Polish space, the existence of regular conditional
distributions is guaranteed and we assume without loss of generality that
conditional probabilities are regular (Kallenberg, 1997, Theorem 5.3).
Equality of conditional distributions is understood in the almost sure [a.s.]
sense and, for random cdf's as random elements of $\mathscr{D}({\mathbb{R}})$,
equalities are up to indistinguishability.

Let $\mathcal{C}_{b}(\mathcal{S})$ be the set of all continuous and bounded
real-valued functions on a metric space $\mathcal{S}$. For random elements
$Z,Z_{n}$ ($n\in\mathbb{N}$) of a metric space $\mathcal{S}_{Z}$, we employ
the usual notation $Z_{n}\overset{w}{\rightarrow}Z$ for the property that the
distribution of $Z_{n}$ weakly converges to the distribution of $Z$, defined
by the convergence $E\left\{  g\left(  Z_{n}\right)  \right\}  {\rightarrow
}E\left\{  g\left(  Z\right) \right\} $ for all $g\in\mathcal{C}%
_{b}(\mathcal{S}_{Z})$. For random elements $(Z,X)$, $(Z_{n},X_{n})$ of the
metric spaces $\mathcal{S}_{Z}\times\mathcal{S}$ and $\mathcal{S}_{Z}%
\times\mathcal{S}_{n} $ ($n\in\mathbb{N}$), and defined on a common
probability space, we denote by $Z_{n}|X_{n} \overset{w}{\rightarrow}_{p}Z|X$
(resp. $Z_{n}|X_{n}\overset{w}{\rightarrow}_{a.s.}Z|X$) the fact that
$E\left\{  g\left(  Z_{n}\right)  |X_{n}\right\} {\rightarrow}E\left\{
g\left(  Z\right)  |X\right\}  $ in probability (resp. a.s.) for all
$g\in\mathcal{C}_{b}(\mathcal{S}_{Z})$. In the special case where $E\left\{
g\left(  Z_{n}\right)  |X_{n}\right\}  \overset{w}{\rightarrow}E\left\{
g\left(  Z\right) \right\} $ in probability (resp. a.s.) for all
$g\in\mathcal{C}_{b}(\mathcal{S}_{Z})$, we write $Z_{n}|X_{n} \overset
{w}{\rightarrow}_{p}Z$ (resp. $Z_{n}|X_{n}\overset{w}{\rightarrow}_{a.s.}Z$).
In such a case the weak limit (in probability or a.s.) of the random
conditional distribution $Z_{n}|X_{n}$ is the non-random distribution of $Z$,
thus reducing our definition to the one of weak convergence in probability
(resp. a.s.) usually employed in the bootstrap literature.

In order to deal with random limit measures, we need a further convergence
concept. For $(Z,X)$,$\,(Z_{n},X_{n})$ ($n\in\mathbb{N}$) defined on possibly
different probability spaces, we denote by $Z_{n}|X_{n}\overset{w}%
{\rightarrow}_{w}Z|X$ the fact that $E\{g(Z_{n})|X_{n}\}\overset
{w}{\rightarrow}E\{g\left(  Z\right)  |X\}$ for all $g\in\mathcal{C}%
_{b}(\mathcal{S}_{Z})$ and label it `weak convergence in distribution'. It
coincides with the probabilistic concept of weak convergence of random
measures (here, of the random conditional distributions $Z_{n}|X_{n}$; see
Kallenberg, 2017, Ch.4). Whenever $Z_{n}$ and $Z$ are rv's and the conditional
distribution of $Z$ given $X$ is \emph{diffuse} (non-atomic), this is
equivalent to the weak convergence $P\left(  Z_{n}\leq\cdot|X_{n}\right)
\overset{w}{\rightarrow}P\left(  Z\leq\cdot|X\right)  $ of the random cdf's as
random elements of $\mathscr{D}(\mathbb{R})$ (see Kallenberg, 2017, Theorem
4.20). Finally, on probability spaces where both the data $D_{n}$ and the
auxiliary variates used in the construction of the bootstrap data are defined,
we use $Z_{n}\overset{w^{\ast}}{\rightarrow}_{p}Z|X$ (resp. $\overset{w^{\ast
}}{\rightarrow}_{a.s}$, $\overset{w^{\ast}}{\rightarrow}_{w}$) interchangeably
with $Z_{n}|D_{n}\overset{w}{\rightarrow}_{p}Z|X$ (resp. $\overset
{w}{\rightarrow}_{a.s}$, $\overset{w}{\rightarrow}_{w}$), and write $P^{\ast
}(\cdot)$ for $P(\cdot|D_{n})$.

\section{A linear regression example}

\label{sec example}

In this section we provide an overview of the main results established in the
sections below, and the concepts employed, by using a simple linear regression
model. Further applications will be given in Section
\ref{Section on Applications}. We observe that even for this basic model
bootstrap statistics may have a random limit distribution. Then, we show that
convergence of the bootstrap statistic to a random limit may imply
(asymptotic) bootstrap validity in the unconditional sense of eq.
(\ref{eq unconditional validity}). Finally, we illustrate the possibility that
bootstrap inference may have a conditional interpretation.

\subsection{Model, bootstrap and random limit bootstrap measures}

\label{sec example model and BS}

Assume that the data are given by $D_{n}:=\{y_{t},x_{t}\}_{t=1}^{n}$ and
consider the linear model
\begin{equation}
y_{t}=\beta x_{t}+\varepsilon_{t}\text{\hspace{1cm}(}t=1,2,...,n\text{)}%
\label{eq:lm}%
\end{equation}
where $x_{t},y_{t}$ are scalar rv's and $\varepsilon_{t}$ are unobservable
zero-mean errors with $\omega_{\varepsilon}:=\operatorname*{Var}%
(\varepsilon_{t})\in(0,\infty)$, $t=1,...,n$. Assume that $M_{n}%
:=\sum\nolimits_{t=1}^{n}x_{t}^{2}>0$ a.s. for all $n$; further assumptions
will be introduced gradually. Interest is in inference on $\beta$ based on
$T_{n}:=\hat{\beta}-\beta$, with $\hat{\beta}$ the OLS estimator of $\beta$;
for instance, a confidence interval or a test of a null hypothesis of the form
$\mathsf{H}_{0}:\beta=0$.

The classic (parametric) fixed-design bootstrap, see e.g. Hall (1992), entails
generating a bootstrap sample $\{y_{t}^{\ast},x_{t}\}_{t=1}^{n}$ as
\begin{equation}
y_{t}^{\ast}=\hat{\beta}x_{t}+\hat{\omega}_{\varepsilon}^{1/2}\varepsilon
_{t}^{\ast}\text{\hspace{1cm}(}t=1,2,...,n\text{)}\label{eq:blm}%
\end{equation}
where $\{\varepsilon_{t}^{\ast}\}_{t=1}^{n}$ are i.i.d. $N\left(  0,1\right)
$, independent of the original data, and $\hat{\omega}_{\varepsilon}$ is an
estimator of $\omega_{\varepsilon}$, e.g., the residual variance $n^{-1}%
\sum_{t=1}^{n}(y_{t}-\hat{\beta}x_{t})^{2}$. The OLS estimator of $\beta$ from
the bootstrap sample is denoted by $\hat{\beta}^{\ast}$ and, conditionally on
the original data, $T_{n}^{\ast}:=\hat{\beta}^{\ast}-\hat{\beta}\sim N\left(
0,\hat{\omega}_{\varepsilon}M_{n}^{-1}\right)  $. As is standard, the
distribution of $T_{n}$ is approximated by the distribution of $T_{n}^{\ast}$
conditional on the data. With $F_{n}^{\ast}$ denoting the cdf of $T_{n}^{\ast
}$ under $P^{\ast}$ (the probability measure induced by the bootstrap; i.e.,
conditional on the original data), the bootstrap \emph{p}-value is given by
$p_{n}^{\ast}:=F_{n}^{\ast}\left(  T_{n}\right)  $.

\begin{remark}
\label{Remark on exact conditional inference}A special case where the ensuing
bootstrap inference is exact in finite samples, such that $p_{n}^{\ast}$ is
uniformly distributed for finite $n$, obtains when the original $\varepsilon
_{t}$'s are $N(0,\omega_{\varepsilon})$, independent of $X_{n}:=\{x_{t}%
\}_{t=1}^{n}$, and $\omega_{\varepsilon}$ is known to the econometrician
(hence $\hat{\omega}_{\varepsilon}=\omega_{\varepsilon} $). Then the
conditional distribution of $T_{n}^{\ast}$ given the data $D_{n}$ equals the
distribution of the original statistic $T_{n}$ \emph{conditional }%
on\emph{\ }the regressor $X_{n}$ (equivalently, on the ancillary statistic
$M_{n}$): $T_{n}^{\ast}|D_{n}\overset{d}{=}T_{n}|X_{n}\sim N\left(
0,\omega_{\varepsilon}M_{n}^{-1}\right)  |M_{n}$. Put differently,
\[
F_{n}^{\ast}\left(  u\right)  :=P(T_{n}^{\ast}\leq u|D_{n})=P(T_{n}\leq
u|X_{n})=\Phi(\omega_{\varepsilon}^{-1/2}M_{n}^{1/2}u)\text{, }u\in
\mathbb{R}\text{.}%
\]
Then, as $\omega_{\varepsilon}^{-1/2}M_{n}^{1/2}T_{n}|M_{n}\sim N(0,1)$, it is
straightforward that in this special case bootstrap inference is exact:
$p_{n}^{\ast}=F_{n}^{\ast}\left(  T_{n}\right)  =\Phi(\omega_{\varepsilon
}^{-1/2}M_{n}^{1/2}T_{n})\overset{d}{=}\Phi(N\left(  0,1\right)  )\sim
U\left(  0,1\right)  $, and that this result also holds conditionally on
$M_{n}$: $p_{n}^{\ast}|M_{n}\sim U(0,1)$.$\hfill\square$
\end{remark}

Although bootstrap inference is not exact in general, it may be still be
asymptotically valid. To show this, we distinguish between the cases of a
stationary and a non-stationary regressor $x_{t}$. It is the second case that
anticipates the main results of the paper. We assume $\hat{\omega
}_{\varepsilon}\overset{p}{\rightarrow}\omega_{\varepsilon}$ throughout.

\subsubsection{Classic bootstrap validity when the regressor is stationary}

Suppose initially that $\{x_{t}\}_{t\in\mathbb{N}}$ is weakly stationary and
$n^{-1}M_{n}\overset{p}{\rightarrow}M:=Ex_{1}^{2}>0$. Define $\tau
_{n}:=n^{1/2}(\hat{\beta}-\beta)$ and $\tau_{n}^{\ast}:=n^{1/2}(\hat
{\beta^{\ast}}-\hat{\beta})$; the bootstrap \emph{p}-values based on
$(\tau_{n},\tau_{n}^{\ast})$ and $(T_{n},T_{n}^{\ast})$ are identical. The
distribution of the bootstrap statistic $\tau_{n}^{\ast}$ conditional on the
original data $D_{n}$ satisfies
\begin{equation}
P^{\ast}(\tau_{n}^{\ast}\leq u)=\Phi(n^{-1/2}\hat{\omega}_{\varepsilon}%
^{-1/2}M_{n}^{1/2}u)\overset{p}{\rightarrow}\Phi(\omega_{\varepsilon}%
^{-1/2}M^{1/2}u),\text{ }u\in\mathbb{R}.\label{eq lin model asy bootstrap cdf}%
\end{equation}
Hence, $\tau_{n}^{\ast}\overset{w^{\ast}}{\rightarrow}_{p}\tau\sim
N(0,\omega_{\varepsilon}M^{-1})$ and the limit distribution is non-random.

If the initial assumptions are strengthened such that a central limit theorem
[CLT]\ holds for $\{x_{t}\varepsilon_{t}\}_{t\in\mathbb{N}}$; that is,
$n^{-1/2}\sum_{t=1}^{n}x_{t}\varepsilon_{t}\overset{w}{\rightarrow}N\left(
0,\omega_{\varepsilon}M\right)  $, then it also holds that $\tau_{n}%
\overset{w}{\rightarrow}\tau\sim N(0,\omega_{\varepsilon}M^{-1})$. Hence, the
bootstrap distribution of $\tau_{n}^{\ast}$ consistently estimates the
unconditional limit distribution of $\tau_{n}$ in the usual\textbf{\ }sense
that $\sup_{u\in\mathbb{R}}|P^{\ast}\left(  \tau_{n}^{\ast}\leq u\right)
-P\left(  \tau\leq u\right)  |$$\overset{p}{\rightarrow}0$, by Polya's
theorem. As the limit cdf is continuous, the \emph{p}-value $p_{n}^{\ast}$
associated with $(\tau_{n},\tau_{n}^{\ast}) $ is asymptotically uniformly
distributed; i.e., (\ref{eq unconditional validity}) holds.

\subsubsection{Random limit bootstrap measures when the regressor is
non-stationary}

\label{sec Random limit bootstrap measures when the regressor is non-stationary}%

Suppose now that $\{x_{t}\}_{t\in\mathbb{N}}$ is such that, for some constant
$\alpha$, $n^{-\alpha}M_{n}\overset{w}{\rightarrow}M$, with $M>0$ a.s. having
a non-degenerate distribution. A well-known special case is that where $x_{t}$
is a finite-variance random walk and $\alpha=2$. Redefine $\tau_{n}%
:=n^{\alpha/2}(\hat{\beta}-\beta)$ and $\tau_{n}^{\ast}:=n^{\alpha/2}%
(\hat{\beta^{\ast}}-\hat{\beta})$; bootstrap \emph{p}-values remain unchanged.
Now the bootstrap distribution of $\tau_{n}^{\ast}$, conditional on the data,
remains random in the limit. Specifically, by the continuous mapping theorem
{[}CMT{{]},}
\begin{equation}
P^{\ast}(\tau_{n}^{\ast}\leq u)=\Phi(n^{-\alpha/2}\hat{\omega}_{\varepsilon
}^{-1/2}M_{n}^{1/2}u)\overset{w}{\rightarrow}\Phi(\omega_{\varepsilon}%
^{-1/2}M^{1/2}u)\text{, }u\in\mathbb{R}%
,\label{eq limit of BS measure - example}%
\end{equation}
which is a random cdf. In terms of weak convergence in distribution, this
amounts to
\begin{equation}
\tau_{n}^{\ast}\overset{w^{\ast}}{\rightarrow}_{w}\left.  N(0,\omega
_{\varepsilon}M^{-1})\right\vert M\text{.}\label{eq blim}%
\end{equation}
As a result, with $\tau_{n}^{\ast}$ and $M$ generally defined on different
probability spaces, weak convergence in probability of $\tau_{n}^{\ast}$ does
not occur. Moreover, whatever the (unconditional) limit distribution of
$\tau_{n}$ is, provided that it exists, $P\left(  \tau_{n}\leq u\right)  $,
$u\in\mathbb{R}$, will tend to a deterministic cdf. Therefore, the bootstrap
cannot estimate consistently the limit distribution of $\tau_{n}$ and it
cannot hold that $\sup_{u\in\mathbb{R}}|P^{\ast}\left(  \tau_{n}^{\ast}\leq
u\right)  -P\left(  \tau_{{}}\leq u\right)  |$$\overset{p}{\rightarrow}0 $.
Nevertheless, bootstrap inference need not become meaningless, as it may even
be exact (see Remark \ref{Remark on exact conditional inference}). We proceed,
therefore, to identify in what sense bootstrap inference could remain meaningful.

\subsection{Bootstrap validity}

\label{sec example BS validity}

Within the framework of the linear regression model, we discuss two concepts
of bootstrap validity in the case of a random limit bootstrap measure. These
are employed to interpret the bootstrap as a tool for unconditional or
conditional inference.

\subsubsection{Unconditional bootstrap validity}

\label{sec intro to on average bs validity}\label{sec 2.2.1}

Under the assumption in Section
\ref{sec Random limit bootstrap measures when the regressor is non-stationary}%
, consider the random-walk special case, where $x_{t}:=\sum_{s=1}^{t}\eta_{s}$
with $e_{t}:=(\varepsilon_{t},\eta_{t})^{\prime}$ forming a stationary,
ergodic and conditionally homoskedastic martingale difference sequence
{[}mds{]} with p.d.\thinspace variance matrix $\Omega:=\operatorname*{diag}%
\{\omega_{\varepsilon},\omega_{\eta}\}$.\footnote{Non-diagonal $\Omega$ could
be handled by augmenting the estimated regression with $\Delta x_{t}$ (as we
do in section \ref{sec par on the boundary}), leading to no qualitative
differences from the case of diagonal $\Omega$.} Then, for $\beta\neq0$ eq.
(\ref{eq:lm}) is an instance of a cointegration regression. It holds that
$(n^{-1/2}\sum_{t=1}^{\left\lfloor n\cdot\right\rfloor }e_{t}^{\prime}%
,n^{-1}\sum_{t=1}^{n}x_{t-1}\varepsilon_{t})\overset{w}{\rightarrow
}(B_{\varepsilon},B_{\eta},\int B_{\eta}dB_{\varepsilon})$ in
$\mathscr{D}{}_{2}\times\mathbb{R}$, where $(B_{\varepsilon},B_{\eta}%
)^{\prime}$ is a bivariate Brownian motion with covariance matrix $\Omega$;
see Theorem 2.4 of Chan and Wei (1988). Moreover, $n^{-2}M_{n}\overset
{w}{\rightarrow}M:=\int B_{\eta}^{2}$ by the CMT, jointly with the convergence
to a stochastic integral above, so that the assumption in Section
\ref{sec Random limit bootstrap measures when the regressor is non-stationary}
holds with $\alpha=2$ and
\begin{equation}
\tau_{n}:=n(\hat{\beta}-\beta)\overset{w}{\rightarrow}\left(  \int B_{\eta
}^{2}\right)  ^{-1}\int B_{\eta}dB_{\varepsilon}\sim N(0,\omega_{\varepsilon
}M^{-1})\text{,}\label{eq:ucl}%
\end{equation}
the limit being (by independence of $B_{\eta}$ and $B_{\varepsilon}$) a
variance mixture of normals, with mixing variable $M^{-1}$ and cdf
$\int_{\mathbb{R}}\Phi(\omega_{\varepsilon}^{-1/2}M^{1/2}u)dP\left(  M\right)
$.

A comparison between the limit distributions of $\tau_{n}^{\ast}$ and
$\tau_{n}$, resp. in (\ref{eq blim}) and (\ref{eq:ucl}), shows that the
bootstrap mimics a component of the mixture limit distribution of $\tau_{n}$,
since the limit distribution of $\tau_{n}$ can be recovered by integrating
over $M$ the conditional limit distribution of $\tau_{n}^{\ast} $ given the
data. This turns out to be sufficient for bootstrap unconditional validity in
the sense of eq. (\ref{eq unconditional validity}). A direct argument is as
follows: the bootstrap \emph{p}-value $p_{n}^{\ast}:=P^{\ast}(\tau_{n}^{\ast
}\leq\tau_{n})$ satisfies, by the CMT,
\begin{align}
& p_{n}^{\ast}\overset{}{=}\Phi(\hat{\omega}_{\varepsilon}^{-1/2}M_{n}%
^{1/2}(\hat{\beta}-\beta))\overset{w}{\rightarrow}\Phi((\omega_{\varepsilon}%
%TCIMACRO{\tint }%
%BeginExpansion
{\textstyle\int}
%EndExpansion
B_{\eta}^{2})^{-1/2}%
%TCIMACRO{\tint }%
%BeginExpansion
{\textstyle\int}
%EndExpansion
B_{\eta}dB_{\varepsilon})\label{eq:piv}\\
& \hspace{1.5in}\overset{d}{=}\Phi(N(0,1))\overset{}{\sim}U(0,1)\text{.}%
\nonumber
\end{align}
Thus, when inference on $\beta$ is based on the distribution of $\tau
_{n}^{\ast}$ conditional on the data, the large-sample frequency of wrong
inferences can be controlled.

\subsubsection{Conditional bootstrap validity}

\label{sec intro to conditional bs validity}

In the case of unconditional bootstrap validity, it may be possible to find an
interpretation of bootstrap inference as also\textbf{\ }valid in the sense of
(\ref{eq conditional validity}), \textit{i.e.} conditionally on some $X_{n}$
defined on the probability space of the original data\textbf{\ }$D_{n}$ (for
instance, but not necessarily, the regressor $X_{n}:=\{x_{t}\}_{t=1}^{n}$).

In the linear regression case considered here, conditional bootstrap validity
with respect to the regressor $X_{n}$ can be obtained under a tightening of
our previous assumptions such that the invariance principle $n^{-1/2}%
\sum_{t=1}^{\left\lfloor n\cdot\right\rfloor }e_{t}\overset{w}{\rightarrow
}(B_{\varepsilon},B_{\eta})^{\prime}$ holds \emph{conditionally }(on $X_{n}$
for finite $n$ and on $B_{\eta}$ in the limit, in the sense of weak
convergence in distribution). A sufficient condition for the conditional
invariance principle is that, additionally to the assumptions on $e_{t}$ in
Section \ref{sec intro to on average bs validity}, $\varepsilon_{t}$ is an mds
with respect to $\mathcal{G}_{t}=\sigma(\{\varepsilon\}_{s=-\infty}^{t}%
\cup\{\eta_{s}\}_{s\in\mathbb{Z}})$, and that $n^{-1}\sum_{t=1}^{n}%
E(\varepsilon_{t}^{2}|\{\eta_{s}\}_{s\in\mathbb{Z}})\rightarrow\omega
_{\varepsilon}$ a.s. (see the proof of Theorem 2 in Rubshtein, 1996). Then, by
using Theorem 3 of Georgiev, Harvey, Leybourne and Taylor (2018), it follows
that
\[
\tau_{n}|X_{n}\overset{w}{\rightarrow}_{w}\left.  N(0,\omega_{\varepsilon
}M^{-1})\right\vert M\text{,}%
\]
which compared to (\ref{eq blim}) shows that the distribution of $\tau
_{n}^{\ast}$ conditional on the data estimates consistently the random limit
distribution of $\tau_{n}$ conditional on the regressor $X_{n}$. This fact is
stated more precisely in Remark \ref{Remark 3.5} where it is concluded that
$p_{n}^{\ast}|X_{n}\overset{w}{\rightarrow}_{p}U(0,1)$, \textit{i.e.}, the
bootstrap is valid conditionally on the regressor.

\subsubsection{A numerical illustration}

\label{sec uncond validity W/O conditional validity}

The result in Section \ref{sec intro to conditional bs validity} implies that
unconditional bootstrap validity can sometimes be established by means of a
conditioning argument; for example, by showing validity conditional on the
regressor $X_{n}$. To illustrate, in Figure 1, panels (a)
and (b), we summarize for two different data generating processes [DGPs] the
cdf's of $p_{n}^{\ast}|X_{n}$ across $M=1,000$ independent realizations of
$X_{n}$ for samples of size $n=10$ (upper panels) and $n=1,000$ (lower
panels). Specifically, the DGP used for panel (i) is based on i.i.d. shocks,
while the one for (ii) features ARCH-type shocks (details are reported in the
accompanying Supplement, Section \ref{Appendix MC}). In both cases, the
conditions of Section \ref{sec intro to conditional bs validity} are
satisfied. For both DGPs, the conditional distributions of $p_{n}^{\ast}$
given $X_{n}$ are, as expected, close to the $45%
%TCIMACRO{\U{b0}}%
%BeginExpansion
{{}^\circ}%
%EndExpansion
$ line, which corresponds to the implied asymptotic $U\left(  0,1\right)  $
distribution. Unconditional validity follows accordingly.%

\begin{figure}[tbh]
\centerline{\includegraphics[width=0.95\textwidth]{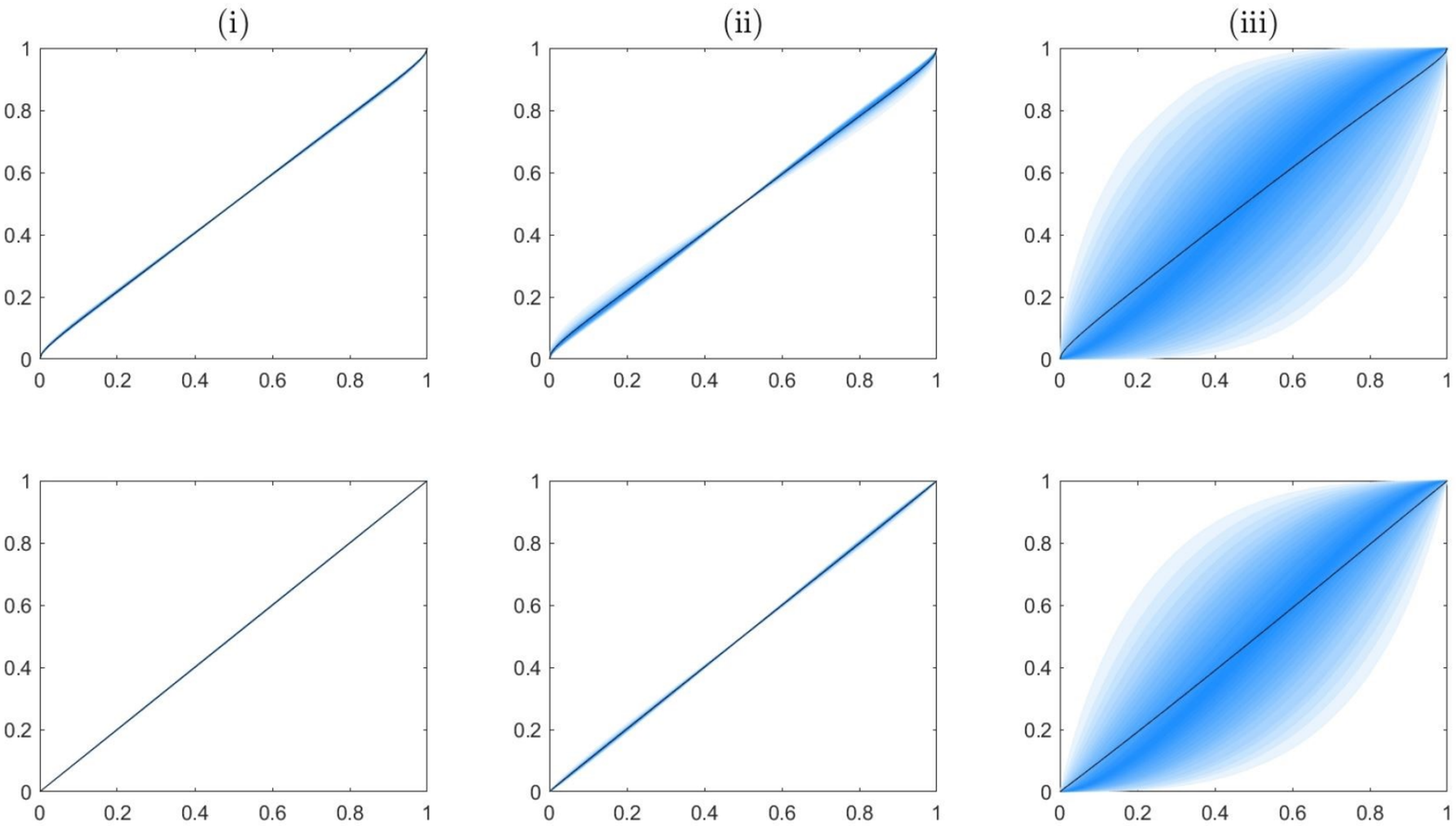}}
\caption{Fan chart of the simulated cdfs (conditional on $X_{n}$) of the
bootstrap $p$-values for the three DGPs (i)--(iii) and $n=10$ (upper panels),
$1000$ (lower panels).}
\end{figure}

Nevertheless, unconditional validity may also hold without validity
conditional on an apparently `natural' conditioning variable $X_{n}$, like the
regressor in a fixed-regressor bootstrap design. For instance, suppose that
for the DGP\ in Sections \ref{sec 2.2.1} and
\ref{sec intro to conditional bs validity} it holds that $\eta_{t}=\xi
_{t}(1+\mathbb{I}_{\{\varepsilon_{t}<0\}})$, with $\{\varepsilon_{t}\}$ and
$\{\xi_{t}\}$ two independent i.i.d. sequences of zero-mean, unit-variance
rv's. Since $\eta_{t}$ is informative about the sign of $\varepsilon_{t}$, the
$\varepsilon_{t}$'s conditionally on their own past and the regressor $X_{n}$
do not form an mds. It is shown in Appendix
\ref{sec Proofs of the results in Section <ref>sec g</ref>}, eq.
(\ref{eq:phi}), that this endogeneity fact, not replicated in the bootstrap
world, induces the original statistic $\tau_{n}$ to satisfy
\begin{equation}
\text{$\tau_{n}|X_{n}$}\overset{w}{\rightarrow}_{w}\left.  M^{-1/2}%
(\omega_{\varepsilon|\eta}^{1/2}\xi_{1}+(1-\omega_{\varepsilon|\eta})^{1/2}%
\xi_{2})\right\vert (M,\xi_{2})\text{,}\label{eq:xit}%
\end{equation}
where $\omega_{\varepsilon|\eta}:=E\{\operatorname*{Var}(\varepsilon_{s}%
|\eta_{s})\}\in(0,1)$, and $M$, $\xi_{1}$, $\xi_{2}$ are jointly independent
with $\xi_{i}\sim N(0,1)$,$\,i=1,2$. The limit in (\ref{eq:xit}) contains more
randomness (through $\xi_{2}$) than the bootstrap limit in eq. (\ref{eq blim}%
), thus resulting in a random limit for the distribution of the bootstrap
\emph{p}-value $p_{n}^{\ast}$ conditional on the regressor $X_{n}$; see panel
(iii) of Figure 1, where for this DGP\ the cdf's of
$p_{n}^{\ast}|X_{n}$ are reported for 1,000 realizations of $X_{n}$. These
cdf's display substantial dispersion around the $45^{%
%TCIMACRO{\U{b0}}%
%BeginExpansion
{{}^\circ}%
%EndExpansion
}$ line, and this feature does not vanish as $n$\ increases. However, and in
agreement with the earlier discussion, their unconditional average (plotted in
black) is very close to the $45%
%TCIMACRO{\U{b0}}%
%BeginExpansion
{{}^\circ}%
%EndExpansion
$ line, showing indeed unconditional validity of the bootstrap. This follows
because $e_{t}:=(\varepsilon_{t},\eta_{t})^{\prime}$ is a zero-mean i.i.d.
sequence with a diagonal covariance matrix and $p_{n}^{\ast}\overset
{w}{\rightarrow}U(0,1)$ as derived in Section
\ref{sec intro to on average bs validity}.

\begin{remark}
Although not valid conditionally on the regressor $X_{n}$, in the previous
example the bootstrap may be valid conditionally on a non-trivial function of
the regressor. See, in particular, Section \ref{subsec:id} and Remark
\ref{Remark 3.45} therein.$\hfill\square$
\end{remark}

\section{Main results}

\label{sec g}

We provide general conditions for bootstrap validity in cases where a
bootstrap statistic conditionally on the data possesses a random limit
distribution. Before all else, we formally distinguish between two concepts of
bootstrap validity.

\subsection{Definitions}

\label{sec 3 definitions}

The following definition employs the bootstrap \emph{p}-value as a summary
indicator of the accuracy of bootstrap inferences (see also Remark
\ref{Remark 3.4} below). The original and the bootstrap statistic are denoted
by $\tau_{n}$ and $\tau_{n}^{\ast}$, respectively.

\begin{definition}
\label{def def}Let $\tau_{n}:=\tau_{n}(D_{n})$ and $\tau_{n}^{\ast}:=\tau
_{n}^{\ast}(D_{n},W_{n}^{\ast})$, $n\in\mathbb{N}$, where $D_{n}$ denotes the
data whereas $W_{n}^{\ast}$ are auxiliary variates defined jointly with
$D_{n}$ on a possibly extended probability space. Let $p_{n}^{\ast}:=P\left(
\left.  \tau_{n}^{\ast}\leq\tau_{n}\right\vert D_{n}\right)  $ be the
bootstrap p-value.

We say that the bootstrap based on $\tau_{n}$ and $\tau_{n}^{\ast}$ is valid
unconditionally if $p_{n}^{\ast}$ is asymptotically $U(0,1)$ distributed:
\begin{equation}
P\left(  p_{n}^{\ast}\leq q\right)  \rightarrow q,\text{ }q\in
(0,1),\label{eq BS unconditionally valid}%
\end{equation}
where $P(\cdot)$ denotes probability w.r.t. the distribution of $D_{n}$.

Let further $X_{n}$ be a random element defined on the probability space of
$D_{n}$ and $W_{n}^{\ast}$. We say that the bootstrap based on $\tau_{n}$ and
$\tau_{n}^{\ast}$ is valid \textit{conditionally on} $X_{n}$ if $p_{n}^{\ast}$
is asymptotically $U(0,1)$ distributed conditionally on $X_{n}:$
\begin{equation}
P\left(  \left.  p_{n}^{\ast}\leq q\right\vert X_{n}\right)  \overset
{p}{\rightarrow}q\text{, }q\in(0,1),\label{eq BS conditionally valid}%
\end{equation}
where $P(\cdot|X_{n})$ is determined up to a.s. equivalence by the
distribution of $(D_{n},X_{n})$.
\end{definition}

\begin{remark}
\label{Remark 3.3}Bootstrap validity conditionally on some $X_{n}$ implies
unconditional validity, by the dominated convergence theorem. In applications,
therefore, the discussion of conditional validity may represent an
intermediate step to assess unconditional validity.
\end{remark}

\begin{remark}
\label{Remark 3.4 - part about right sided tests}The validity properties in
Definition \ref{def def} ensure correct asymptotic null rejection probability,
unconditionally or conditionally on some $X_{n}$, for bootstrap hypothesis
tests which reject the null when the bootstrap \emph{p}-value $p_{n}^{\ast}$
does not exceed a chosen nominal level, say $\alpha\in(0,1)$. If $P\left(
\tau_{n}^{\ast}\leq\cdot|D_{n}\right)  $ converges weakly in
$\mathscr{D}(\mathbb{R})$ to a sample-path continuous random cdf, then correct
asymptotic null rejection probability is ensured also for bootstrap tests
rejecting the null hypothesis when $\tilde{p}_{n}^{\ast}:=P\left(  \left.
\tau_{n}^{\ast}\geq\tau_{n}\right\vert D_{n}\right)  \leq\alpha$ (for
applications, see Sections \ref{sec par on the boundary} and
\ref{Section on BS tests for parameter constancy}).
\end{remark}

\begin{remark}
\label{Remark 3.4} Validity as in Definition \ref{def def} has also
implications on the properties of bootstrap (percentile) confidence sets.
Suppose, for instance, that $T_{n}$ is an estimator of a population (scalar)
parameter, whose true value is denoted by $\theta_{0}$, and assume for
simplicity that $\tau_{n}$ is of the form $\tau_{n}=\rho(n)(T_{n}-\theta_{0}%
)$, where $\rho(n)$ is a normalizing factor such that $\tau_{n}$ has a
non-degenerate limiting distribution (see Horowitz, 2001, p.3174). Its
bootstrap analog is denoted by $\tau_{n}^{\ast}$, and we assume that the
bootstrap is valid in the unconditional sense of
(\ref{eq BS unconditionally valid}). Interest is in constructing a right-sided
confidence interval for $\theta_{0}$, with (asymptotic) coverage $1-\alpha
\in\left(  0,1\right)  $, using a simple bootstrap percentile method. With
$F_{n}^{\ast}(x):=P\left(  \tau_{n}^{\ast}\leq\cdot|D_{n}\right)  $, let
$q_{n}^{\ast}\left(  1-\alpha\right)  :=\inf\{x\in\mathbb{R}:F_{n}^{\ast
}(x)\geq1-\alpha\}$ be the $(1-\alpha)$ quantile of the bootstrap distribution
$F_{n}^{\ast}$. Then, it is straightforward to show that, if $F_{n}^{\ast}$
converges weakly to a sample-path continuous random cdf, then
\[
P\left(  \tau_{n}\leq q_{n}^{\ast}\left(  1-\alpha\right)  \right)  =P\left(
p_{n}^{\ast}\leq1-\alpha\right)  +o\left(  1\right)  \rightarrow1-\alpha
\]
This implies that a confidence interval of the form $[T_{n}-\rho(n)^{-1}%
q_{n}^{\ast}\left(  1-\alpha\right)  ,+\infty)$ has (unconditional) asymptotic
coverage probability of $1-\alpha$. If the bootstrap is valid conditionally on
some $X_{n}$, as in (\ref{eq BS conditionally valid}), then the (asymptotic)
coverage is $1-\alpha$ also conditionally on this $X_{n}$.$\hfill\square$
\end{remark}

Our main results make extensive use of \emph{joint }weak convergence in
distribution. Should the related notation not be self-explanatory, we refer
the reader to Appendix A for the formal definitions.

\subsection{Unconditional bootstrap validity}

\label{sec dis}\label{sec gen on unc val}

The unconditional validity results in this section have in common the
requirement, explicit or implicit, that the unconditional limit distribution
of $\tau_{n}$ should be an average of the random limit distribution of
$\tau_{n}^{\ast}$ given the data. Applications of Theorem \ref{th2} do not
require a conditional analysis of $\tau_{n}$, in contrast to applications of
Theorem \ref{p2 copy(2)}.

\begin{theorem}
\label{th2}Let there exist a rv $\tau$ and a random element $X$, both defined
on the same probability space, such that $\left(  \tau_{n},F_{n}^{\ast
}\right)  \overset{w}{\rightarrow}\left(  \tau,F\right)  $ in
$\mathscr{\mathbb{R}}\times\mathscr{D}(\mathbb{R})$ for $F_{n}^{\ast
}(u):=P(\tau_{n}^{\ast}\leq u|D_{n})$ and $\text{\text{$F(u):=P(\tau\leq
u|X)$}, $u\in\mathbb{R}$.}$ If the (possibly)\textbf{\ }random cdf $F$ is
sample-path continuous, then the bootstrap based on $\tau_{n}$ and $\tau
_{n}^{\ast}$ is valid unconditionally.
\end{theorem}

\noindent Some remarks are in order.

\begin{remark}
\label{Remark 3.7} A trivial special case of Theorem \ref{th2} is obtained for
independent $\tau$ and $X$. In this case the bootstrap distribution of
$\tau_{n}^{\ast}$ estimates consistently the limiting unconditional
distribution of $\tau_{n}$ and the bootstrap is valid in the usual sense.
\end{remark}

\begin{remark}
\label{Remark 3.9}An important special case of Theorem \ref{th2} involves
stable convergence of the original statistic $\tau_{n}$ (see H\"{a}usler and
Luschgy, 2015, p.33, for a definition). With the notation of Theorem
\ref{th2}, let the data $D_{n}$ and the random element $X$ be defined on the
same probability space, whereas the rv $\tau$ be defined on an extension of
this probability space. Assume that $\tau_{n}\rightarrow\tau$ stably and
$F_{n}^{\ast}\overset{p}{\rightarrow}F$. Then $(\tau_{n},F_{n}^{\ast}%
)\overset{w}{\rightarrow}(\tau,F)$ by Theorem 3.7(b) of H\"{a}usler and
Luschgy (2015). For instance, in the statistical literature on integrated
volatility, a result of the form $\tau_{n}\rightarrow\tau$ stably is contained
in Theorem 3.1 of Jacod, Mykland, Podolskij and Vetter (2009) for $\tau_{n}$
defined as a $t$-type statistic for integrated volatility, whereas the
corresponding $F_{n}^{\ast}\overset{p}{\rightarrow}F$ result is established in
Theorem 3.1 of Hounyo, Gon\c{c}alves and Meddahi (2017) for a combined wild
and blocks-of-blocks bootstrap introduced in the latter paper.
\end{remark}

\begin{remark}
\label{Remark conj}More generally, if $\tau_{n}^{\ast}\overset{w^{\ast}%
}{\rightarrow}_{w}\tau|X$ and $\left(  \tau_{n}^{\ast},\tau_{n},X_{n}\right)
\overset{w}{\rightarrow}\left(  \tau^{\ast},\tau,X\right)  $ with $D_{n}%
$-measurable $X_{n}$ ($n\in\mathbb{N}$), then the joint convergence
$((\tau_{n}^{\ast}|D_{n}),\tau_{n},X_{n})\overset{w}{\rightarrow}_{w}%
((\tau^{\ast}|X),\tau,X)$ follows (see Lemma \ref{le crpr}(b) in Appendix
\ref{sec itere}). If $\tau^{\ast}|X\overset{d}{=}\tau|X$ and $F$ is
sample-path continuous, then $\left(  \tau_{n},F_{n}^{\ast}\right)
\overset{w}{\rightarrow}\left(  \tau,F\right)  $ by Lemma \ref{le kal}(b) in
Appendix \ref{sec itere}.
\end{remark}

\begin{remark}
\label{Remark redet}Alternatively, the convergence $\left(  \tau_{n}%
,F_{n}^{\ast}\right)  \overset{w}{\rightarrow}\left(  \tau,F\right)  $ could
be obtained from (a) the convergence $(\tau_{n},X_{n})\overset{w}{\rightarrow
}(\tau,X)$ for some $D_{n}$-measurable random elements $X_{n}$ of the space of
$X$, and (b) the implication (were it to hold) from the strong version
$(\tau_{n},X_{n})\overset{a.s.}{\rightarrow}(\tau,X)$ to $\tau_{n}^{\ast
}\overset{w^{\ast}}{\rightarrow}_{p}\tau|X$. The idea is to choose $X_{n}$
such that $\tau_{n}^{\ast}$ depends on the data essentially through $X_{n}$.
Applications of Theorem \ref{th2} along these lines could proceed in two
steps: (i) prove that $(\tau_{n},X_{n})\overset{w}{\rightarrow}(\tau,X)$; (ii)
consider, by extended Skorokhod coupling (Corollary 5.12 of Kallenberg, 1997),
a representation of $D_{n}$ and $(\tau,X)$ such that, with an abuse of
notation, $(\tau_{n},X_{n})\overset{a.s.}{\rightarrow}(\tau,X)$ and, on a
product extension of the Skorokhod-representation space, prove that $\tau
_{n}^{\ast}\overset{w^{\ast}}{\rightarrow}_{p}\tau|X$. The latter conditional
assertion, due to the product structure of the probability space, reduces to a
collection of unconditional assertions by fixing the outcomes in the
factor-space of the data. It then holds that $\left(  \tau_{n},F_{n}^{\ast
}\right)  \overset{p}{\rightarrow}\left(  \tau,F\right)  $ on the
Skorokhod-representation space, whereas on a general probability space
$\left(  \tau_{n},F_{n}^{\ast}\right)  \overset{w}{\rightarrow}\left(
\tau,F\right)  $. We proceed like this in the applications of Section
\ref{sec example K/S} (eq. (\ref{eq tri})) and Section
\ref{Section on BS tests for parameter constancy} (Theorem \ref{th fb} under
Assumption \textsc{$\mathcal{H}$}).$\hfill\square$
\end{remark}

Unconditional bootstrap validity could also be established by means of an
auxiliary conditional analysis of the original statistic $\tau_{n}$. In the
next theorem the conditioning sequence $X_{n}$ is chosen such that the
bootstrap statistic $\tau_{n}^{\ast}$ depends on the data $D_{n}$
approximately through $X_{n}$ (condition (\dag)). Then, the main requirement
for bootstrap validity is that the limit bootstrap distribution should be a
conditional average of the limit distribution of $\tau_{n}$ given $X_{n}$.

\begin{theorem}
\label{p2 copy(2)}With the notation of Definition \ref{def def}, let $X_{n}$
be $D_{n}$-measurable ($n\in\mathbb{N}$). Let it hold that
\begin{equation}
\left(  P\left(  \left.  \tau_{n}\leq\cdot\right\vert X_{n}\right)
,\,P\left(  \left.  \tau_{n}^{\ast}\leq\cdot\right\vert D_{n}\right)  \right)
\overset{w}{\rightarrow}\left(  F,F^{\ast}\right) \label{eq joint conv}%
\end{equation}
in $\mathscr{D}\left(  \mathbb{R}\right)  \times\mathscr{D}\left(
\mathbb{R}\right)  $, where $F$ and $F^{\ast}$ are sample-path continuous
random cdf's, and let

\medskip

\noindent($\dagger$) there exist random elements $X^{\prime},X_{n}^{\prime}$
such that $F^{\ast}$ is $X^{\prime}$-measurable, $X_{n}^{\prime}$ are $X_{n}%
$-measurable and $X_{n}^{\prime}\overset{w}{\rightarrow}X^{\prime}$ jointly
with (\ref{eq joint conv}).

\medskip

\noindent Then, if $E\{F(\cdot)|F^{\ast}\}=F^{\ast}(\cdot)$, the bootstrap
based on $\tau_{n}$ and $\tau_{n}^{\ast}$ is valid unconditionally.
\end{theorem}

\begin{remark}
\label{Remark exp}Condition ($\dagger$) of Theorem \ref{p2 copy(2)} implies
that $\tau_{n}^{\ast}$ depends on the data $D_{n}$ approximately through
$X_{n}$ alone, for under this condition $P\left(  \left.  \tau_{n}^{\ast}%
\leq\cdot\right\vert X_{n}\right)  \,$and $P\left(  \left.  \tau_{n}^{\ast
}\leq\cdot\right\vert D_{n}\right)  $ are both close to $P\left(  \left.
\tau_{n}^{\ast}\leq\cdot\right\vert X_{n}^{\prime}\right)  $. Condition
($\dagger$) is trivially satisfied in the case $F=F^{\ast}$ with the choice
$X_{n}^{\prime}=P(\tau_{n}\leq\cdot|X_{n})$. It is also satisfied if $\tau
_{n}^{\ast}=\tilde{\tau}_{n}^{\ast}+o_{p}(1)$ for some $\tilde{\tau}_{n}%
^{\ast}$ which is a measurable transformation of $X_{n}$ and $W_{n}^{\ast}$,
w.r.t. the probability measure on the space where $D_{n} $ and $W_{n}^{\ast}$
are jointly defined. In this case, $X_{n}^{\prime}=P(\tilde{\tau}_{n}^{\ast
}\leq\cdot|X_{n})$ satisfies condition ($\dagger$);\ see Appendix
\ref{sec prg}. An example of a pair $\tau_{n}^{\ast}$, $\tilde{\tau}_{n}%
^{\ast}$ is given in eq. (\ref{ur zvez}) in Section \ref{sec example K/S}%
.$\hfill\square$
\end{remark}

Convergence (\ref{eq joint conv}) in Theorem \ref{p2 copy(2)}\textbf{\ }could
be deduced from the weak convergence of the conditional distributions of
$\tau_{n}$ and $\tau_{n}^{\ast}$, as in the next corollary.

\begin{corollary}
\label{p2 part 2}Let $D_{n}$ and $X_{n}$ ($n\in\mathbb{N}$) be as in Theorem
\ref{p2 copy(2)}. Let the rv $\tau$ and the random elements $X$, $X^{\prime}$
be defined on a single probability space and
\begin{equation}
(\tau_{n}|X_{n},\tau_{n}^{\ast}|D_{n})\overset{w}{\rightarrow}_{w}(\tau
|X,\tau|X^{\prime})\label{eq coroll joint conv}%
\end{equation}
in the sense of eq. (\ref{eq:wcrm}). Let further $F\left(  u\right)
:=P(\tau\leq u|X)$ and $F^{\ast}\left(  u\right)  :=P(\tau\leq u|X^{\prime})$,
$u\in\mathbb{R}$, define sample-path continuous random cdf's. Then convergence
(\ref{eq joint conv}) holds. Moreover, the bootstrap based on $\tau_{n}$ and
$\tau_{n}^{\ast}$ is valid unconditionally provided that one of the following
extra conditions holds:

\medskip\noindent(a) $X^{\prime}=X$;

\medskip\noindent(b) $X=(X^{\prime},X^{\prime\prime})$ and $X_{n}^{\prime
}\overset{w}{\rightarrow}X^{\prime}$ jointly with (\ref{eq coroll joint conv})
for some $X_{n}$-measurable random elements $X_{n}^{\prime}$.
\end{corollary}

\begin{remark}
\label{Remark 3.6}An instance of (\ref{eq joint conv}) with $F\neq F^{\ast}$
is implied by the setup of Section
\ref{sec uncond validity W/O conditional validity}. There
(\ref{eq coroll joint conv}) holds with $\tau:=M^{-1/2}(\omega_{\varepsilon
|\eta}^{1/2}\xi_{1}+(1-\omega_{\varepsilon|\eta})^{1/2}\xi_{2})$ and
$X=(X^{\prime},X^{\prime\prime})=(M,(1-\omega_{\varepsilon|\eta})^{1/2}\xi
_{2})$. Moreover, (\ref{eq coroll joint conv}) is joint with the convergence
$X_{n}^{\prime}\overset{w}{\rightarrow}X^{\prime}$ for $X_{n}^{\prime}%
=n^{-2}M_{n}$ (see Appendix \ref{sec prg}). Hence, Corollary \ref{p2 part 2}%
(b) implies that the bootstrap is unconditionally valid, as was already
concluded in Section \ref{sec 2.2.1}.
\end{remark}

\begin{remark}
\label{Re renext}Convergence (\ref{eq coroll joint conv}) could be proved by
replacing in Remark \ref{Remark redet} the convergence $(\tau_{n}%
,X_{n})\overset{}{\rightarrow}(\tau,X)$ (weakly and a.s.) by $((\tau_{n}%
|X_{n}),X_{n}^{\prime})\rightarrow((\tau|X),X^{\prime})$ (weakly in
distribution and weakly a.s.) Other ways of proving
(\ref{eq coroll joint conv}), that could be relevant if conditional bootstrap
validity is of interest, are discussed in the next section.$\hfill\square$
\end{remark}

\subsection{Conditional bootstrap validity}

\label{subsec:id}Theorem \ref{p2} below states the asymptotic behavior of the
bootstrap \emph{p}-value conditional on an $X_{n}$ chosen to satisfy condition
($\dagger$) of Theorem \ref{p2 copy(2)}. It also characterizes the cases where
the bootstrap is valid conditionally on such an $X_{n}$. Should validity
conditional on such an $X_{n}$ fail, in Corollary \ref{c1}(b) we provide a
result for validity conditional on a transformation of it.

\begin{theorem}
\label{p2}Under the conditions of Theorem \ref{p2 copy(2)}, the bootstrap
p-value $p_{n}^{\ast}$ satisfies
\begin{equation}
P\left(  \left.  p_{n}^{\ast}\leq q\right\vert X_{n}\right)  \overset
{w}{\rightarrow}F(F^{\ast-1}(q))\label{eq:pis}%
\end{equation}
for almost all $q\in(0,1)$, and the bootstrap based on $\tau_{n}$ and
$\tau_{n}^{\ast}$ is valid conditionally on $X_{n}$ if and only if $F=F^{\ast
}$ such that
\begin{equation}
\sup_{u\in\mathbb{R}}\left\vert P\left(  \left.  \tau_{n}\leq u\right\vert
X_{n}\right)  -P\left(  \left.  \tau_{n}^{\ast}\leq u\right\vert D_{n}\right)
\right\vert \overset{p}{\rightarrow}0.\label{eq s}%
\end{equation}

\end{theorem}

\begin{remark}
Under (\ref{eq s}), the bootstrap distribution of $\tau_{n}^{\ast}$
consistently estimates the limit of the conditional distribution of $\tau_{n}$
given $X_{n}$. Although under condition ($\dagger$) the proximity of $P\left(
\left.  \tau_{n}\leq\cdot\right\vert X_{n}\right)  $\ and $P\left(  \left.
\tau_{n}^{\ast}\leq\cdot\right\vert D_{n}\right)  $ is necessary for bootstrap
validity conditional on $X_{n}$, no such proximity is necessary for
conditional validity in the general case. In fact, validity conditional on
some $X_{n}$ implies validity conditional on any measurable transformation
$X_{n}^{\prime}=\psi_{n}(X_{n})$ and an analogue of (\ref{eq s}) with
$X_{n}^{\prime}$ in place of $X_{n}$ cannot generally hold for all $\psi_{n}$,
unless $F^{\ast}$ is non-random. This is similar to what happens with
unconditional bootstrap validity which, according to Theorem \ref{th2}, may
occur even if $P\left(  \tau_{n}\leq\cdot\right)  $\ and $P\left(  \left.
\tau_{n}^{\ast}\leq\cdot\right\vert D_{n}\right)  $ are not close.
$\hfill\square$
\end{remark}

A corollary in the terms of weak convergence in distribution is given next.

\begin{corollary}
\label{c1}Let $D_{n},X_{n}$ ($n\in\mathbb{N}$), $\tau,F, F^{\ast}$ be as in
Corollary \ref{p2 part 2}. Let (\ref{eq coroll joint conv}) hold and $F,
F^{\ast}$ be sample-path continuous random cdf's. Then:

(a) If $X^{\prime}=X$, the bootstrap based on $\tau_{n}$ and $\tau_{n}^{\ast}$
\ is valid conditionally on $X_{n}$ and (\ref{eq s}) holds.

(b) If $X=(X^{\prime},X^{\prime\prime})$, $(X_{n}^{\prime},X_{n}^{\prime
\prime})\overset{w}{\rightarrow}(X^{\prime},X^{\prime\prime})$ jointly with
(\ref{eq coroll joint conv}) for some $X_{n}$-measurable random elements
$(X_{n}^{\prime},X_{n}^{\prime\prime})$, and $X_{n}^{\prime\prime}%
|X_{n}^{\prime}\overset{w}{\rightarrow}_{w}X^{\prime\prime}|X^{\prime}$, then
the bootstrap is valid conditionally on $X_{n}^{\prime} $ and (\ref{eq s})
holds with $X_{n}$ replaced by $X_{n}^{\prime}$.
\end{corollary}

Corollary \ref{c1} requires checking the joint convergence in
(\ref{eq coroll joint conv}); see Remark \ref{Re renext}. We provide further
strategies to establish this convergence, useful if interest is in conditional
bootstrap validity, in Remarks \ref{Remark remalt}--\ref{Remark joco} below.

\begin{remark}
\label{Remark 3.5}Consider the linear regression example under the extra
assumptions of Section \ref{sec intro to conditional bs validity} and
set\textbf{\ }$\tau=(\int B_{\eta}^{2})^{-1}\int B_{\eta}dB_{\varepsilon}$$, $
$X=M$. It then follows (by using Theorem 3 of Georgiev \textit{et al}., 2018)
that condition (\ref{eq coroll joint conv}) holds in the form
\begin{equation}
(\tau_{n}|X_{n},\tau_{n}^{\ast}|D_{n})\overset{w}{\rightarrow}_{w}%
(\tau|B_{\eta},\tau|B_{\eta})\overset{d}{=}(1,1)N(0,\omega_{\varepsilon}%
M^{-1})|M\text{,}\label{eq taun}%
\end{equation}
where $X_{n}:=\{x_{t}\}_{t=1}^{n}$; equivalently, (\ref{eq joint conv}) holds
with $F=F^{\ast}=\Phi(\omega_{\varepsilon}^{-1/2}M^{1/2}(\cdot))$. Hence, the
bootstrap is consistent for the limit distribution of $\tau_{n} $ conditional
on the regressor and, by Corollary \ref{c1}(a), the bootstrap is valid
conditionally on the regressor.
\end{remark}

\begin{remark}
\label{Remark remalt}The joint convergence in (\ref{eq coroll joint conv})
would follow from the separate convergence facts $\tau_{n}|X_{n}\overset
{w}{\rightarrow}_{w}\tau|X$, $\tau_{n}^{\ast}\overset{w^{\ast}}{\rightarrow
}_{w}\tau^{\ast}|X^{\prime}$ and $(\tau_{n},\tau_{n}^{\ast},\phi_{n}%
(X_{n}),\psi_{n}(D_{n}))\overset{w}{\rightarrow}\left(  \tau,\tau^{\ast
},X,X^{\prime}\right)  $ for some measurable $\phi_{n},\psi_{n}$, provided
that $\tau|X^{\prime}\overset{d}{=}\tau^{\ast}|X^{\prime}$; see Appendix
\ref{sec prg}. We use this approach in Section \ref{sec example K/S}, point
(ii). The convergence $\tau_{n}|X_{n}\overset{w}{\rightarrow}_{w}\tau|X$ is
the new ingredient compared to Remark \ref{Remark conj}.
\end{remark}

\begin{remark}
\label{Remark joco}Convergence (\ref{eq coroll joint conv}) would also follow
from $(\tau_{n},\phi_{n}(X_{n}),\psi_{n}(D_{n}))\overset{w}{\rightarrow}%
(\tau,X,$ $X^{\prime})$ and $\tau_{n}|X_{n}\overset{w}{\rightarrow}_{w}\tau|X$
together with the implication (were it to hold) from $\psi_{n}(D_{n}%
)\overset{a.s.}{\rightarrow}X^{\prime}$ to $\tau_{n}^{\ast}\overset{w^{\ast}%
}{\rightarrow}_{p}\tau^{\ast}|X^{\prime}$, with $\tau|X^{\prime}\overset{d}%
{=}\tau^{\ast}|X^{\prime}$. A possible implementation strategy is: (i) prove
that $\tau_{n}|X_{n}\overset{w}{\rightarrow}_{w}\tau|X$ and $(\tau_{n}%
,\phi_{n}(X_{n}),\psi_{n}(D_{n}))\overset{w}{\rightarrow}\left(
\tau,X,X^{\prime}\right)  $; (ii) consider a Skorokhod representation of
$D_{n}$ and $\left(  \tau,X,X^{\prime}\right)  $ such that, maintaining the
notation, $(\tau_{n},\phi_{n}(X_{n}),\psi_{n}(D_{n}))\overset{a.s.}%
{\rightarrow}\left(  \tau,X,X^{\prime}\right)  \ $and, as a result, $\tau
_{n}|X_{n}\overset{w}{\rightarrow}_{w}\tau|X$ strengthens to $\tau_{n}%
|X_{n}\overset{w}{\rightarrow}_{p}\tau|X$ (see Lemma \ref{le crpr} in Appendix
\ref{sec itere}); (iii) redefine the bootstrap variates $W_{n}^{\ast}$ on a
product extension of the Skorokhod-representation space and prove there that
$\tau_{n}^{\ast}\overset{w^{\ast}}{\rightarrow}_{p}\tau^{\ast}|X^{\prime}$.
Then (\ref{eq coroll joint conv}) holds on a general probability space. We
proceed like this in the proof of Theorem \ref{th fb} under Assumption
$\mathcal{C}$. The convergence $\tau_{n}|X_{n}\overset{w}{\rightarrow}_{w}%
\tau|X$ is the extra ingredient compared to Remark \ref{Remark redet}. Notice
also that if $\phi_{n}(X_{n})=(X_{n}^{\prime},X_{n}^{\prime\prime})$ and
$\psi_{n}(D_{n})=X_{n}^{\prime}$, then the convergence $(X_{n}^{\prime}%
,X_{n}^{\prime\prime})\overset{w}{\rightarrow}(X^{\prime},X^{\prime\prime})$
in Corollary \ref{c1}(b) would be joint with (\ref{eq coroll joint conv}).
\end{remark}

\begin{remark}
\label{Remark 3.45} In the setup of Section
\ref{sec uncond validity W/O conditional validity},
(\ref{eq coroll joint conv}) holds with $\tau$ and $X=(X^{\prime}%
,X^{\prime\prime})\ $ given in Remark \ref{Remark 3.6}. Moreover,
(\ref{eq coroll joint conv}) is joint with the convergence $(X_{n}^{\prime
},X_{n}^{\prime\prime})\overset{w}{\rightarrow}(X^{\prime},X^{\prime\prime})$
for $X_{n}^{\prime}=n^{-2}M_{n}$ and $X_{n}^{\prime\prime}=M_{n}^{-1/2}%
\sum_{t=1}^{n}x_{t}E(\varepsilon_{t}|\eta_{t})$ (see Appendix \ref{sec prg}).
By Corollary \ref{c1}(b), the bootstrap would be valid conditionally on
$M_{n}$ if it additionally holds that $X_{n}^{\prime\prime}|M_{n}\overset
{w}{\rightarrow}_{w}(1-\omega_{\varepsilon|\eta})^{1/2}\xi_{2}|M \overset
{d}{=}N(0,1-\omega_{\varepsilon|\eta})$ .$\hfill\square$
\end{remark}

\subsection{Local power of bootstrap tests}

\label{sec lop}

When the limit bootstrap measure is random, the power function of the
bootstrap test, conditionally on the data, is also random, even
asymptotically. Its unconditional power function can be investigated using the
following generalization of Theorem \ref{th2}.

\begin{theorem}
\label{th2a}Let there exist rv's $\tau,\tau^{\ast}$ and a random element $X$,
the three defined on the same probability space, such that $\left(  \tau
_{n},F_{n}^{\ast}\right)  \overset{w}{\rightarrow}\left(  \tau,F^{\ast
}\right)  $ in $\mathscr{\mathbb{R}}\times\mathscr{D}(\mathbb{R})$ for
$F_{n}^{\ast}(u):=P(\tau_{n}^{\ast}\leq u|D_{n})$ and $\text{\text{$F^{\ast
}(u):=P(\tau^{\ast}\leq u|X)$}, $u\in\mathbb{R}$.}$ If $F^{\ast}$ is
sample-path continuous, then the bootstrap p-value $p_{n}^{\ast}$ satisfies
$P(p_{n}^{\ast}\leq q)\rightarrow E\{F(F^{\ast-1}(q))\}$, $q\in(0,1)$ with
$F\left(  \cdot\right)  :=P\left(  \tau\leq\cdot|X\right)  $.
\end{theorem}

To illustrate, with $y_{t}$, $x_{t}$ and $\varepsilon_{t}$ as in Section
\ref{sec intro to conditional bs validity}, let interest be in the
large-sample behavior of the bootstrap test for the hypothesis $\mathsf{H}%
_{0}:\beta=0$ against $\mathsf{H}_{1}:\beta<0$, under the local alternative
$\beta=\beta_{n}:=b/n$ in (\ref{eq:lm}). The original test statistic is
$\tau_{n}=n\hat{\beta}$.

Without recourse to the explicit expression in
(\ref{eq lin model asy bootstrap cdf})\ for the bootstrap \emph{p}-value
$p_{n}^{\ast}$ in terms of the Gaussian cdf, which in many applications may
have no analogue, we can instead\textbf{\ }use, for $\tau_{n}=b+n(\hat{\beta
}-\beta)$ and $\tau_{n}^{\ast}$, the joint convergence\footnote{The
conditional analysis of $\tau_{n}^{\ast}$, needed to show that under local
alternatives it behaves asymptotically as under $\mathsf{H}_{0}$, is
straightforward and is omitted.}%
\[
(\tau_{n},(\tau_{n}^{\ast}|D_{n}))\overset{w}{\rightarrow}_{w}(\tau
,(\tau^{\ast}|M)),
\]
with $\tau:=b+\tau^{\ast}$ and $\tau^{\ast}:=\omega_{\varepsilon}%
^{1/2}M^{-1/2}\xi$ for $\xi\sim N\left(  0,1\right)  $ independent of $M$. By
using Lemma \ref{le kal} in Appendix \ref{sec itere}, we can conclude that the
conditions of Theorem \ref{th2a} hold with $X=M$, $F\left(  u\right)
=P\left(  \tau\leq u|M\right)  =\Phi(\omega_{\varepsilon}^{-1/2}M^{1/2}(u-b))$
and $F^{\ast}(u)=\Phi(\omega_{\varepsilon}^{-1/2}M^{1/2}u)$. The unconditional
asymptotic local power function of the one-sided, $q$-level, bootstrap test
then follows as
\begin{equation}
P(p_{n}^{\ast}\leq q)\rightarrow E\{F(F^{\ast-1}(q))\}=E\{\Phi(\Phi
^{-1}(q)-\omega_{\varepsilon}^{-1/2}M^{1/2}b)\}\text{.}\label{eq uncond loc}%
\end{equation}

Notice that this power function is distinct from the asymptotic local power of
the unconditional test based on critical values from the null asymptotic
(unconditional) distribution of $\tau_{n}$, which is that of $\omega
_{\varepsilon}^{1/2}M^{-1/2}\xi$. Hence, when the limit bootstrap measure is
random, the bootstrap test in general does not replicate, in terms of
(unconditional) power, the standard asymptotic test (in this specific case,
numerical evidence shows that for small $b$, where both local powers are
relatively low, the bootstrap test is more powerful, whereas for large
negative $b$, where the local power of both tests is high, the asymptotic
test\ is preferable).

The unconditional power function in (\ref{eq uncond loc}) can also be derived
through a conditioning argument. This can be done using the results in Section
\ref{subsec:id} by considering the joint convergence
\[
(\tau_{n}|X_{n},\tau_{n}^{\ast}|D_{n})\overset{w}{\rightarrow}_{w}%
(b,0)+\left.  (1,1)N(0,\omega_{\varepsilon}M^{-1})\right\vert M\text{,}%
\]
see (\ref{eq taun}), which implies (\ref{eq joint conv}) with $F$ and
$F^{\ast}$ as defined above. Hence, by Theorem \ref{p2}, (\ref{eq:pis}) holds
and%
\begin{equation}
P\left(  \left.  p_{n}^{\ast}\leq q\right\vert X_{n}\right)  \overset
{w}{\rightarrow}\Phi(\omega_{\varepsilon}^{-1/2}M^{1/2}(F^{\ast}{}%
^{-1}(q)-b))=\Phi(\Phi^{-1}(q)-\omega_{\varepsilon}^{-1/2}M^{1/2}%
b)\label{eq conditional local power function}%
\end{equation}
for $q\in(0,1)$. The latter expression is the (random) asymptotic local power,
conditional on\textbf{\ }$X_{n}$, of the one-sided, $q$-level test, bootstrap
test. By averaging the rhs of (\ref{eq conditional local power function}) over
$M$, the unconditional power function in (\ref{eq uncond loc}) follows.

\section{Applications}

\label{Section on Applications}

\subsection{A permutation CUSUM test under infinite variance}

\label{sec example CUSUM} Consider a standard CUSUM\ test for the null
hypothesis (say, $\mathsf{H}_{0}$) that $\{\varepsilon_{t}\}_{t=1}^{n}$ is a
sequence of i.i.d. random variables. The test statistic is of the form
\[
\tau_{n}:=\nu_{n}^{-1}\max_{t=1,...,n}\left\vert \sum\nolimits_{i=1}%
^{t}(\varepsilon_{i}-\overline{\varepsilon}_{n})\right\vert \text{, }%
\overline{\varepsilon}_{n}:=n^{-1}\sum\nolimits_{t=1}^{n}\varepsilon_{t},
\]
where $\nu_{n}$ is a permutation-invariant normalization sequence. Standard
choices are $\nu_{n}^{2}=\sum_{t=1}^{n}(\varepsilon_{t}-\overline{\varepsilon
}_{n})^{2}$ in the case where $E\varepsilon_{t}^{2}<\infty$, and $\nu_{n}%
=\max_{t=1,...,n}|\varepsilon_{t}|$ when $E\varepsilon_{t}^{2}=\infty$. If
$\varepsilon_{t}$ is in the domain of attraction of a strictly $\alpha$-stable
law with $\alpha\in(0,2)$, such that $E\varepsilon_{t}^{2}=\infty$, the
asymptotic distribution of $\tau_{n}$ depends on unknown parameters (e.g., the
characteristic exponent $\alpha$), which makes the test difficult to apply
(see also Politis, Romano and Wolf, 1999, and the references therein). To
overcome this problem, Aue \emph{et al.} (2008) consider a permutation
analogue of $\tau_{n}$, defined as
\[
\text{$\tau_{n}^{\ast}:=\nu_{n}^{-1}\max_{t=1,...,n}\left\vert \sum
\nolimits_{i=1}^{t}(\varepsilon_{\pi(i)}-\overline{\varepsilon}_{n}%
)\right\vert $}%
\]
where $\pi$ is a (uniformly distributed) random permutation of $\{1,2,...,n\}$%
, independent of the data.\footnote{The normalization of $\nu_{n}$ is only of
theoretical importance for obtaining non-degenerate limit distributions. In
practice, any bootstrap procedure comparing $\tau_{n}$ to the quantiles of
$\tau_{n}^{\ast}$ is invariant to the choice of $\nu_{n}$ and can be
implemented by setting $\nu_{n}=1$.} In terms of Definition \ref{def def}, the
data is $D_{n}:=\{\varepsilon_{t}\}_{t=1}^{n}$ and the auxiliary `bootstrap'
variate is $W_{n}^{\ast}:=\pi$. With $X_{n}:=\{\varepsilon_{(t)}\}_{t=1}^{n}$
denoting the vector of order statistics of $\{\varepsilon_{t}\}_{t=1}^{n}$,
there exists a random permutation $\varpi$ of $\{1,...,n\}$ (under
$\mathsf{H}_{0}$, uniformly distributed conditionally on $X_{n}$) for which it
holds that $\varepsilon_{t}=\varepsilon_{(\varpi(t))}$ ($t=1,...,n$), whereas
the `bootstrap' sample is $\{\varepsilon_{\pi(t)}\}_{t=1}^{n}$ . The results
in Aue et al. (2008, Corollary 2.1, Theorem 2.4) imply that, if $\mathsf{H}%
_{0}$ holds and $\varepsilon_{t}$ is in the domain of attraction of a strictly
$\alpha$-stable law with $\alpha\in(0,2)$, then $\tau_{n}$$\overset
{w}{\rightarrow}\rho_{\alpha}(S)$ and $\tau_{n}^{\ast}\overset{w^{\ast}%
}{\rightarrow}_{w}\rho_{\alpha}(S)|S$ for a certain random function
$\rho_{\alpha}$ and $S=(S_{1},S_{2})^{\prime}$, with $S_{i}=\{S_{ij}%
\}_{j=1}^{\infty}$ ($i=1,2$) being partial sums of sequences of i.i.d.
standard exponential rv's, and with $\rho_{\alpha}$ independent of
$S$.\footnote{To avoid centering terms, Aue \textit{et al}. (2008) assume
additionally that the location parameter of the limit stable law is zero when
$\alpha\in\lbrack1,2).$ Moreover, although they provide conditional
convergence results only for the finite-dimensional distributions of the CUSUM
process, these could be strengthened to conditional functional convergence as
in Proposition 1 of LePage \textit{et al}. (1997) in order to obtain the
conditional convergence of $\tau_{n}^{*}.$}

\smallskip{}

Aue et al. (2008) do not report the fact that statistical inferences are not
invalidated by the failure of the permutation procedure to estimate
consistently the distribution of $\rho_{\alpha}(S)$. In fact, the situation is
similar to that of Remark \ref{Remark on exact conditional inference}, as
$\tau_{n}|X_{n}\overset{d}{=}\tau_{n}^{\ast}|D_{n}$ under $\mathsf{H}_{0}$. As
a consequence, under $\mathsf{H}_{0}$ the permutation test implements
\emph{exact\footnote{Here by `exact' we mean that bootstrap inference
replicates the finite-sample (conditional) distribution of the test statistic
for any sample size with no error. }} finite-sample inference conditional on
$X_{n}$ and, additionally, the distribution of $\tau_{n}^{\ast}$ given the
data estimates consistently the limit of the conditional distribution
$\tau_{n}|X_{n}$, in the sense of joint weak convergence in distribution (see
eq. (\ref{eq:wcrm})):
\begin{equation}
\left(  \tau_{n}|X_{n},\tau_{n}^{\ast}|D_{n}\right)  ^{\prime}\overset
{w}{\rightarrow}_{w}(\rho_{\alpha}(S)|S,\rho_{\alpha}(S)|S)\text{
.}\label{eq:cus}%
\end{equation}

CUSUM tests can also be applied to residuals from an estimated model in order
to test for correct model specification or stability of the parameters (see
e.g., Ploberger and Kr\"{a}mer, 1992). Consider thus the case where
$\{\varepsilon_{t}\}_{t=1}^{n}$ are the disturbances in a statistical model
(e.g., the regression model of Section 2), and we observe residuals
$\hat{\varepsilon}_{t}$ obtained upon estimation of the model using a sample
$D_{n} $ not containing the unobservable $\{\varepsilon_{t}\}_{t=1}^{n}$. The
residual-based CUSUM\ statistic is $\hat{\tau}_{n}:=$$\hat{\nu}_{n}^{-1}%
\max_{t=1,...,n}|\sum_{i=1}^{t}(\hat{\varepsilon}_{i}-\overline{\hat
{\varepsilon}}_{n})|$, where $\hat{\nu}_{n}$ and $\overline{\hat{\varepsilon}%
}_{n}$ are the analogues of $\nu_{n}$ and $\bar{\varepsilon}_{n} $ computed
from $\hat{\varepsilon}_{t}$ instead of $\varepsilon_{t}$. The bootstrap
statistic could be defined as $\hat{\tau}_{n}^{\ast}:=$$\hat{\nu}_{n}^{-1}%
\max_{t=1,...,n}|\sum_{i=1}^{t}(\hat{\varepsilon}_{\pi(i)}-\overline
{\hat{\varepsilon}}_{n})|$. If $\hat{\tau}_{n}-\tau_{n}\overset{p}%
{\rightarrow}0$ and $(\hat{\tau}_{n}^{\ast}-\tau_{n}^{\ast})|D_{n}\overset
{w}{\rightarrow}_{p}0$ under $\mathsf{H}_{0}$ (e.g., due to consistent
parameter estimation), then also $(\hat{\tau}_{n}-\tau_{n})|X_{n}\overset
{w}{\rightarrow}_{p}0$, such that the (L\'{e}vy) distances between the pairs
of conditional distributions $\hat{\tau}_{n}|X_{n}$ and $\tau_{n}|X_{n}$ on
the one hand, and $\hat{\tau}_{n}^{\ast}|D_{n}\ $and $\tau_{n}^{\ast}|D_{n}$
on the other hand, converge in probability to zero. Hence, in view of
(\ref{eq:cus}), and under the conjecture that $P(\rho_{\alpha}(S)\leq\cdot|S)$
defines a sample-path continuous cdf, the residual-based permutation procedure
is consistent in the sense that
\begin{equation}
\left(  \hat{\tau}_{n}|X_{n},\hat{\tau}_{n}^{\ast}|D_{n}\right)  \overset
{w}{\rightarrow}_{w}(\rho_{\alpha}(S)|S,\rho_{\alpha}(S)|S)\label{eq:cusr}%
\end{equation}
for $X_{n}:=\{\varepsilon_{(t)}\}_{t=1}^{n}$ again. It follows that:

\smallskip{}

(i) The permutation residual-based test is valid conditionally on $X_{n}$, by
Corollary \ref{c1}(a) with condition (\ref{eq coroll joint conv}) taking the
form (\ref{eq:cusr}).

\smallskip

(ii) This test is valid unconditionally, as a results of either the validity
conditional on $X_{n}$, or by Corollary \ref{p2 part 2}.

\subsection{A parametric bootstrap goodness-of-fit test}

\label{sec example K/S}

The parametric bootstrap is a standard technique for the approximation of a
conditional distribution of goodness-of-fit test statistics (Andrews, 1997;
Lockhart, 2012). When these are discussed in the i.i.d. finite-variance
setting, the limit of the bootstrap distribution is non-random. However, if we
return to the relation (\ref{eq:lm}), there exist relevant settings where a
random limit of the normalized $M_{n}$ implies that parametrically
bootstrapped goodness-of-fit test statistics have random limit distributions.

\subsubsection{Set up and a random limit bootstrap measure}

Let the null hypothesis of interest, say $\mathsf{H}_{0}$, be that the
standardized errors $\omega_{\varepsilon}^{-1/2}\varepsilon_{t}$ in
(\ref{eq:lm}) have a certain known density $f$ with mean 0 and variance 1. For
expositional ease we assume that $\omega_{\varepsilon}=1$ and is known to the
econometrician. Then, the Kolmogorov-Smirnov statistic based on OLS residuals
$\hat{\varepsilon}_{t}$ is
\[
\tau_{n}:=n^{1/2}\sup_{u\in\mathbb{R}}\left\vert n^{-1}\sum_{t=1}%
^{n}\mathbb{I}_{\{\hat{\varepsilon}_{t}\leq u\}}-\int_{-\infty}^{u}%
f\right\vert \text{.}%
\]
A (parametric) bootstrap counterpart, $\tau_{n}^{\ast}$, of $\tau_{n}$ could
be constructed under $\mathsf{H}_{0}$ by (i) drawing $\{\varepsilon_{t}^{\ast
}\}_{t=1}^{n}$ as i.i.d. from $f$, independent of the data; (ii), regressing
them on $x_{t}$, thus obtaining an estimator $\hat{\beta}^{\ast}$ and
associated residuals $\hat{\varepsilon}_{t}^{\ast}$; and (iii) calculating
$\tau_{n}^{\ast}$ as $\tau_{n}^{\ast}:=n^{1/2}\sup_{u\in\mathbb{R}}|n^{-1}%
\sum_{t=1}^{n}\mathbb{I}_{\{\hat{\varepsilon}_{t}^{\ast}\leq u\}}%
-\int_{-\infty}^{u}f|$.

To see that the distribution of the bootstrap statistic $\tau_{n}^{\ast}$
conditional on the data $D_{n}:=\{x_{t},y_{t}\}_{t=1}^{n}$ may have a random
limit, consider the Gaussian case, $f=\Phi^{\prime}$. Under the assumptions of
Johansen and Nielsen (2016, Sec. 4.1-4.2), it holds (\textit{ibidem}) that
$\tau_{n}^{\ast}=\tilde{\tau}_{n}^{\ast}+o_{p}(1)$ under the product
probability on the product probability space where the data and $\{\varepsilon
_{t}^{\ast}\}$ are jointly defined, with
\begin{equation}
\tilde{\tau}_{n}^{\ast}:=\sup_{u\in\lbrack0,1]}\left\vert n^{-1/2}\sum
_{t=1}^{n}(\mathbb{I}_{\{\varepsilon_{t}^{\ast}\leq q(u)\}}-u)+\Phi^{\prime
}(q(u))\hat{\beta}^{\ast}n^{-1/2}\sum_{t=1}^{n}x_{t}\right\vert \text{,}%
\label{ur zvez}%
\end{equation}
where $q(u)=\Phi^{-1}(u)$ is the $u$-th quantile of $\Phi$. The expansion of
$\tau_{n}^{\ast}$ holds also conditionally on the data, i.e., $\tau_{n}^{\ast
}-\tilde{\tau}_{n}^{\ast}\overset{w^{\ast}}{\rightarrow}_{p}0$, since
convergence in probability to a constant is preserved upon such conditioning.
Hence, if $\left.  \tilde{\tau}_{n}^{\ast}\right\vert D_{n}$ converges to a
random limit, so does $\left.  \tau_{n}^{\ast}\right\vert D_{n}$ for the same
limit. Assume that $X_{n}:=n^{-\alpha/2}x_{\left\lfloor n\cdot\right\rfloor }%
$$\overset{w}{\rightarrow}X$ in $\mathscr{D}$ for some $\alpha>0$ and that
$M:=\int X^{2}>0$ a.s. (e.g., $X=B_{\eta}$ if $x_{t}=\sum_{s=1}^{t-1}\eta_{s}$
with $\{\eta_{t}\}$ introduced in Section 2.2). Then $(\text{$M_{n},\xi_{n})$%
}:=(\sum_{t=1}^{n}x_{t}^{2},\sum_{t=1}^{n}x_{t})$ satisfies
$(\text{$n^{-\alpha-1}M_{n},n^{-\alpha/2-1}\xi_{n})$}\overset{w}{\rightarrow
}(M,\xi)$, $\xi:=\int X$.\ Furthermore, if $W_{n}^{\ast}(u):=n^{-1}\sum
_{t=1}^{n}(\mathbb{I}_{\{\varepsilon_{t}^{\ast}\leq q(u)\}}-u)$,~$u\in
\lbrack0,1],$ is the bootstrap empirical process in probability scale, then
$W_{n}^{\ast}$ and $M_{n}^{1/2}\hat{\beta}^{\ast}$ are independent of the data
individually (the second one being conditionally standard Gaussian), but not
jointly independent of the data, because
\[
\operatorname*{Cov}\nolimits^{\ast}(n^{1/2}W_{n}^{\ast}(u),M_{n}^{1/2}%
\hat{\beta}^{\ast})=(n^{-\alpha-1}M_{n})^{-1/2}n^{-\alpha/2-1}\xi_{n}%
\psi(u)\overset{w}{\rightarrow}M^{-1/2}\xi\psi(u)\text{,}%
\]
$u\in\lbrack0,1]$, where $\psi(\cdot):=E^{\ast}[\varepsilon_{1}^{\ast
}\mathbb{I}_{\{\varepsilon_{1}^{\ast}\leq q(\cdot)\}}]=-\Phi^{\prime}%
(q(\cdot))$ is a trimmed mean function, with $\operatorname*{Cov}%
\nolimits^{\ast}(\cdot)$ and $E^{\ast}(\cdot)$ calculated under $P^{\ast}$. It
is shown in Appendix \ref{sec Proofs of the results in Section applications}
that, more strongly,
\begin{equation}
(n^{1/2}W_{n}^{\ast},\,n^{(\alpha+1)/2}\hat{\beta}^{\ast},\,n^{-\alpha/2-1}%
\xi_{n})\overset{w^{\ast}}{\rightarrow}_{w}\left.  (W,M^{-1/2}b,\xi
)\right\vert (M,\xi)\label{eq:jbb}%
\end{equation}
on $\mathscr{D}\times\mathbb{R}^{2},$ where $(W,b$$)$ is a pair of a standard
Brownian bridge and a standard Gaussian rv individually independent of $X$
(and thus, of $M,\xi$), but with Gaussian joint conditional (on $X$)
distributions having covariance $\operatorname*{Cov}(W(u),b|X)=M^{-1/2}\xi
\psi(u),$ $u\in\lbrack0,1]$. Combining the expansion of $\tau_{n}^{\ast}$,
(\ref{ur zvez}) and (\ref{eq:jbb}) with the extended CMT (Theorem \ref{th cmt}
in Appendix \ref{sec itere}) yields
\begin{equation}
\tau_{n}^{\ast}\overset{w^{\ast}}{\rightarrow}_{w}\{\sup_{u\in\lbrack
0,1]}|W(u)+\Phi^{\prime}(q(u))M^{-1/2}b\xi|\}\big\vert(M,\xi)\overset{d}%
{=}\tau\left\vert (M,\xi)\right.  \text{,}\label{eq:tst}%
\end{equation}
where $\tau:=\sup_{u\in\lbrack0,1]}|\tilde{W}(u)|$ for a process $\tilde{W}$
which conditionally on $X$ (and thus, on $M,\xi$), is a zero-mean Gaussian
process with $\tilde{W}(0)=\tilde{W}(1)=0$ a.s. and conditional covariance
function $K(u,v)=u(1-v)-M^{-1}\xi^{2}\psi(u)\psi(v)$ for $0\leq u\leq v\leq1$.
In summary, the limit bootstrap distribution is random because the latter
conditional covariance is random whenever $M$ or $\xi$ are such.

\subsubsection{Bootstrap validity}

We now discuss in what sense $\tau_{n}^{\ast}$ can provide a distributional
approximation of $\tau_{n}$ and whether the bootstrap can be valid in the
sense of Definition \ref{def def}.

\smallskip{}

\noindent(i)\ Under $\mathsf{H}_{0}$ that $\varepsilon_{t}\sim\text{i.i.d.}%
$$N(0,1)$, the bootstrap could be shown to be unconditionally valid using
Theorem \ref{th2}. Specifically, under $\mathsf{H}_{0}$, the assumptions and
results of Johansen and Nielsen (2016, Sec. 4.1-4.2) guarantee that $\tau_{n}$
has the expansion $\tau_{n}=\tilde{\tau}_{n}+o_{p}(1)$, with $\tilde{\tau}%
_{n}:=\sup_{u\in\lbrack0,1]}|n^{-1/2}\sum_{t=1}^{n}(\mathbb{I}_{\{\varepsilon
_{t}\leq q(u)\}}-u)+\Phi^{\prime}(q(u))(\hat{\beta}-\beta)n^{-1/2}\sum
_{t=1}^{n}x_{t}|$ defined similarly to $\tilde{\tau}_{n}^{\ast} $. Assume that
$\hat{\beta}$ is asymptotically mixed Gaussian, such that jointly with
$n^{-\alpha/2}x_{\left\lfloor n\cdot\right\rfloor }$$\overset{w}{\rightarrow
}X$ it holds that
\[
(n^{-1/2}\sum_{t=1}^{n}(\mathbb{I}_{\{\varepsilon_{t}\leq q(u)\}}%
-u),\,n^{(\alpha+1)/2}(\hat{\beta}-\beta),\,n^{-\alpha/2-1}\xi_{n})\overset
{w}{\rightarrow}(W,M^{-1/2}b,\xi)\text{ ;}%
\]
then $\tau_{n}=\tilde{\tau}_{n}+o_{p}(1)\overset{w}{\rightarrow}\tau
=\sup_{u\in\lbrack0,1]}|\tilde{W}(u)|$. Thus, the unconditional limit of
$\tau_{n}$ obtains by averaging (over $M,\xi$) the conditional limit of
$\tau_{n}^{\ast}$. This is the main prerequisite for establishing
unconditional bootstrap validity via Theorem \ref{th2}. More precisely, it is
proved in Appendix \ref{sec Proofs of the results in Section applications}
that
\begin{equation}
\left(  \tau_{n},F_{n}^{\ast}\right)  \overset{w}{\rightarrow}\left(
\tau,F\right)  \text{, }F_{n}^{\ast}(\cdot):=P^{\ast}(\tau_{n}^{\ast}\leq
\cdot)\text{, }F(\cdot):=P(\tau\leq\cdot|M,\xi).\label{eq tri}%
\end{equation}
As $F$ is sample-path continuous (e.g., by Proposition 3.2 of Linde, 1989,
applied conditionally on $M,\xi$), Theorem \ref{th2} guarantees the
unconditional validity of the bootstrap.\smallskip{}

\noindent(ii) As $\tau_{n}=\tilde{\tau}_{n}+o_{p}(1)$ under $\mathsf{H}_{0} $,
with $\tilde{\tau}_{n}$ related to $\left(  M_{n},\xi_{n}\right)  $ through
the same functional form as $\tilde{\tau}_{n}^{\ast}$, it is possible for
$\tau_{n}|X_{n}$ to have the same random limit distribution under
$\mathsf{H}_{0}$ as $\tau_{n}^{\ast}$ given the data, i.e., $\tau_{n}%
|X_{n}\overset{w}{\rightarrow}_{w}\tau\left\vert (M,\xi)\right.  $. For
instance, this occurs if $\{\varepsilon_{t}\}$ is an i.i.d. sequence
independent of $X_{n}$, by the same argument as for $\tilde{\tau}_{n}^{\ast}
$. According to Remark \ref{Remark remalt}, the convergence $\tau_{n}%
|X_{n}\overset{w}{\rightarrow}_{w}\tau|(M,\xi)$ and the convergence $(\tau
_{n},\tau_{n}^{\ast},n^{-\alpha-1}M_{n},n^{-\alpha/2-1}\xi_{n})\overset
{w}{\rightarrow}\left(  \tau,\tau^{\ast},M,\xi\right)  $ with $\tau^{\ast
}|(M,\xi)\overset{d}{=}\tau|(M,\xi)$ (shown in the proof of (\ref{eq tri}),
see Appendix \ref{sec Proofs of the results in Section applications}) are
sufficient for eq. (\ref{eq coroll joint conv}) to hold in the form%
\[
(\tau_{n}|X_{n},\tau_{n}^{\ast}|D_{n})\overset{w}{\rightarrow}_{w}(\tau
|(M,\xi),\tau|(M,\xi))\text{.}%
\]
As $F$ is sample-path continuous, the bootstrap is valid conditionally on
$X_{n}$ by Corollary \ref{c1}(a).

\subsection{Parameters on the boundary in predictive regression}

\label{sec par on the boundary}Here we consider an instance of the `parameter
on the boundary' problem in the framework of predictive regressions for
financial returns; see e.g. Phillips (2014) and the references therein. While
in this context the bootstrap is potentially useful (e.g., when there is
uncertainty about the degree of persistence of the posited predicting
variable), its application is not straightforward if some of the parameters
may lie on the boundary of the parameter space; see Andrews (2000).

We show that in the presence of parameters on the boundary, the distribution
of the bootstrap statistic may be random in the limit. Moreover, the type of
randomness induced by parameters on the boundary depends on how well the
bootstrap scheme approximates the mutual position of three objects, namely
(i)\ the boundary, (ii)\ the set identified by the null hypothesis, and
(iii)\, the true parameter value. Standard bootstrap approximations may not be
sufficiently precise, giving rise to complex conditioning in the limit
bootstrap distribution, with ensuing unconditional bootstrap validity only for
special statistics. Conversely, non-standard, or ad hoc, bootstrap schemes,
designed to provide a better match with the original geometry, may feature
limit bootstrap distributions where no randomness attributable to the possibly
boundary value of a parameter is present.

\subsubsection{General setup}

\label{sec par on the boundary setup}

Consider the predictive regression%
\begin{equation}
y_{t}=\theta_{1}+\theta_{2}x_{n,t-1}+\varepsilon_{t}\text{ (}t=1,...,n;\text{
}n=1,2,...\text{)}\label{eq:prr}%
\end{equation}
under Assumption 1 of Georgiev et al. (2018), specialized for simplicity to
unconditionally homoskedastic errors. The posited predicting variable
$x_{n,t}$ is such that, in $\mathscr{D}{}$, $x_{n,\left\lfloor n\cdot
\right\rfloor }\overset{w}{\rightarrow}X$, e.g. a Brownian motion or an
Ornstein-Uhlenbeck process, and hence features low frequency variability in
the sense of M\"{u}ller and Watson (2008). We assume that the parameter space,
say $\Theta$, is defined by an inequality constraint and that the true value
of the parameter $\theta:=(\theta_{1},\theta_{2})^{\prime}$, say $\theta
_{0}:=(\theta_{1,0},\theta_{2,0})^{\prime}$, may lie on the boundary of
$\Theta$. An important example is when $\theta$ is assumed to belong to the
set $\mathbb{R}\times\lbrack0,\infty)$, with the boundary corresponding to the
case $\theta_{2}=0$ of no predictability of $y_{t}$ by $x_{n,t-1}$ and the
interior corresponding to (sign-restricted) predictability.

More specifically, assume that $\Theta:=\{\theta\in\mathbb{R}^{2}%
:g(\theta)\geq0\}$, where $g:$ $\mathbb{R}^{2}\rightarrow\mathbb{R}$ is a real
function, continuously differentiable on some neighborhood of $\theta_{0}$ and
with gradient $\tfrac{\partial}{\partial\theta^{\prime}}g(\theta)\neq0$ on
that neighborhood, with $\dot{g}:=\tfrac{\partial}{\partial\theta^{\prime}%
}g(\theta_{0})$. The boundary of $\Theta$ is denoted by $\partial
\Theta:=\{\theta\in\mathbb{R}^{2}:g(\theta)=0\}$. The aforementioned example
$\theta_{2}\geq0$ is obtained by setting $g(\theta)=(0,1)\theta=\theta_{2}$.

Interest is in bootstrap inference on a null hypothesis $\mathsf{H}_{0}$
identifying a set of parameter values that has a non-empty intersection with
the boundary of the parameter space. In particular, we consider the following
mutual positions of the boundary, the parameter set identified by
$\mathsf{H}_{0}$ and the true value $\theta_{0}$:

\begin{description}
\item[$\mathscr{G}{}_{1}$.] $\mathsf{H}_{0}$ is the hypothesis that
$\theta_{0}$ belongs to the boundary: $\mathsf{H}_{0}:g(\theta_{0})=0$;

\item[$\mathscr{G}{}_{2}$.] $\mathsf{H}_{0}$ is a simple null hypothesis on
the boundary: $\mathsf{H}_{0}:\theta_{0}=\bar{\theta}$, $g(\bar{\theta})=0$;

\item[$\mathscr{G}{}_{3}$.] $\mathsf{H}_{0}:h(\theta_{0})=0$, where
$\{\theta\in\mathbb{R}^{2}:h\left(  \theta\right)  =0\}$ is not a subset of
the boundary $\partial\Theta$, but meets $\partial\Theta$ at a singleton set.
\end{description}

\noindent For example, let again $g(\theta)=\theta_{2}\text{.}$ Then the
hypothesis of no predictability $\mathsf{H}_{0}:\theta_{2,0}=0$ falls under
$\mathscr{G}{}_{1}$; the hypothesis $\mathsf{H}_{0}:\theta_{0}=(0,0)^{\prime}$
that $y_{t}$ is unpredictable with zero mean falls under $\mathscr{G}{}_{2}$;
the hypothesis $\mathsf{H}_{0}:(1,1)^{\prime}\theta_{0}=\theta_{1,0}%
+\theta_{2,0}=0$ falls under $\mathscr{G}{}_{3}$. In the latter case, the
intersection point of the boundary and $\mathsf{H}_{0}$ is $(0,0)^{\prime}$
which might, but need not, be the true value under $\mathsf{H}_{0}$.

Let $\hat{\theta}$ be the OLS estimator of the first two coefficients in the
equation
\begin{equation}
y_{t}=\theta_{1}+\theta_{2}x_{n,t-1}+\delta\Delta x_{n,t}+e_{t}\label{eq:pr}%
\end{equation}
subject to the constraint $\hat{\theta}\in\Theta$, i.e. $g(\hat{\theta})\geq0$
(here $\Delta x_{n,t}$ is included in order to obtain residuals asymptotically
uncorrelated with the innovations driving $x_{n,t}$). It holds that
$n^{1/2}(\hat{\theta}-\theta_{0})\overset{w}{\rightarrow}\ell(\theta_{0})$,
with $\ell(\theta_{0})$ depending on the position of $\theta_{0}$ relative to
the boundary $\partial\Theta$. Thus, $\ell(\theta_{0})=\tilde{\ell}%
:=M^{-1/2}\xi$ if $\theta_{0}\in\operatorname*{int}(\Theta):=\Theta
\setminus\partial\Theta$, where $M:=\int\tilde{X}\tilde{X}^{\prime}$,
$\tilde{X}:=(1,X)^{\prime}$, $\xi\sim N\left(  0,\sigma_{e}^{2}I_{2}\right)  $
is independent of $X$, and $\sigma_{e}>0$, whereas (see Section 12 in the
working paper version of Andrews, 1999),
\begin{equation}
n^{1/2}(\hat{\theta}-\theta_{0})\overset{w}{\rightarrow}\ell(\theta_{0}%
)=\ell:=\underset{\lambda\in\Lambda:=\{\lambda\in\mathbb{R}^{2}:\dot
{g}^{\prime}\lambda\geq0\}}{\arg\min}||\lambda-M^{-1/2}\xi||_{M}%
\label{eq asy distribution}%
\end{equation}
if $g(\theta_{0})=0$, where we use the notation $||x||_{M}:=(x^{\prime
}Mx)^{1/2}$ for $x\in\mathbb{R}^{2}$.

Consider now a bootstrap sample generated as
\begin{equation}
y_{t}^{\ast}=\hat{\theta}_{1}+\hat{\theta}_{2}x_{n,t-1}+\varepsilon_{t}^{\ast
}\text{,}\label{eq BS for PR}%
\end{equation}
where $\varepsilon_{t}^{\ast}=\hat{e}_{t}w_{t}^{\ast}$, $t=1,...n$, with
$\hat{e}_{t}$ the residuals of (\ref{eq:pr}) and $w_{t}$ i.i.d. $N(0,1)$,
independent of the original data.\footnote{The conclusions do not change if,
instead of this wild (fixed regressor) bootstrap, a standard residual-based
i.i.d. bootstrap or a parametric bootstrap is used.} Then the distribution of
$n^{1/2}(\hat{\theta}-\theta_{0})$ could be tentatively approximated by the
distribution of $n^{1/2}(\hat{\theta}^{\ast}-\hat{\theta})$ conditional on the
original data, where $\hat{\theta}^{\ast}$ is obtained by regressing
$y_{t}^{\ast}$ on $(1,x_{n,t-1})^{\prime}$ (the term $\Delta x_{n,t}$ is no
longer necessary) under the constraint $\hat{\theta}^{\ast}\in\Theta^{\ast
}=\Theta$ (as for the original estimator), i.e., $g(\hat{\theta}^{\ast})\geq
0$; see Andrews (2000).

For $\theta_{0}\in\operatorname*{int}(\Theta)$, it turns out that the
bootstrap statistic converges to a conditional version of the limit of
$n^{1/2}(\hat{\theta}-\theta_{0})$ found earlier:
\begin{equation}
n^{1/2}(\hat{\theta}^{\ast}-\hat{\theta})=n^{1/2}(\tilde{\theta}^{\ast}%
-\hat{\theta})+o_{p}(1)\overset{w^{\ast}}{\rightarrow}_{w}\tilde{\ell
}|M\text{, }\label{eq random limit depending on M only}%
\end{equation}
where $\tilde{\theta}^{\ast}$ denotes the unconstrained OLS estimator from the
bootstrap sample.

On the other hand, if $\theta_{0}\in\partial\Theta$ the bootstrap statistic
converges as follows, jointly with $n^{1/2}(\hat{\theta}-\theta_{0})$:
\begin{equation}
n^{1/2}(\hat{\theta}^{\ast}-\hat{\theta})\overset{w^{\ast}}{\rightarrow}%
_{w}\ell^{\ast}|(M,\ell)\text{, }\ell^{\ast}:=\underset{\lambda\in
\Lambda_{\ell}^{\ast}:=\{\lambda\in\mathbb{R}^{2}:\dot{g}^{\prime}\lambda
\geq-\dot{g}^{\prime}\ell\}}{\arg\min}||\lambda-M^{-1/2}\xi^{\ast}%
||_{M},\label{eq asy for standard BS with param on the boundary}%
\end{equation}
where $\xi^{\ast}\sim N(0,1)$ is independent of $(M,\ell)$; see Theorem
\ref{Lemma bootstrap with boundary} below. In contrast with the case
$\theta_{0}\in\operatorname*{int}(\Theta)$, the limit in
(\ref{eq asy for standard BS with param on the boundary}) is not a conditional
version of the limit of $n^{1/2}(\hat{\theta}-\theta_{0})$, inasmuch as
$\Lambda_{\ell}^{\ast}$ in
(\ref{eq asy for standard BS with param on the boundary}) is a \emph{random}
half-plane, rather than the original set $\Lambda$ of
(\ref{eq asy distribution}). The reason is that the standard bootstrap does
not approximate well the original mutual position of the true value and the
boundary, unless $g(\hat{\theta})=0$. Other, non-standard bootstraps may be
designed in order to provide better approximations, at least under the null
hypothesis. This is analyzed next.

\subsubsection{Unconditionally valid bootstrap schemes}

\label{sec par on the boundary main}

In order to unify the discussion of several bootstrap schemes for inference on
$\mathsf{H}_{0}$ under the three cases $\mathscr{G}{}_{1}$, $\mathscr{G}{}_{2}%
$ and $\mathscr{G}{}_{3}$, consider a bootstrap sample generated as in
(\ref{eq BS for PR}) and, more generally than before, a bootstrap OLS
estimator $\hat{\theta}^{\ast}$ constrained to belong to the (random) set%
\[
\Theta^{\ast}:=\{\theta\in\mathbb{R}^{2}:g(\theta)\geq g^{\ast}(\hat{\theta
})\}
\]
where the function $g^{\ast}:\mathbb{R}^{2}\rightarrow\mathbb{R}$ is
continuously differentiable on some neighborhood of $\theta_{0}$ and satisfies
$g^{\ast}(\theta)\leq g(\theta)$ for $\theta\in\Theta$. The standard bootstrap
considered in Section \ref{sec par on the boundary setup} obtains by setting
$g^{\ast}=0$ (such that $\Theta^{\ast}=\Theta$, the original parameter space).
Alternatively, setting $g^{\ast}=g$ restricts the bootstrap true value
$\hat{\theta}$ to lie on the boundary of the bootstrap parameter space
$\Theta^{\ast}$ (as, in this case, $\Theta^{\ast}=\{\theta\in\mathbb{R}%
^{2}:g(\theta)\geq g(\hat{\theta})\}$); see Cavaliere, Nielsen and Rahbek
(2017) for an application of this `restricted' bootstrap to the location
model. Finally, setting $g^{\ast}=g-|g|^{1+\kappa}$ for some $\kappa>0$
introduces a correction, in the spirit of an alternative to the standard
bootstrap mentioned in Andrews (2000,p.403, Method two) and Fang and Santos
(2019, Example 2.1), where the bootstrap true value either shrinks to the
boundary of the bootstrap parameter space at a proper rate or remains bounded
away from this boundary, according to whether $\theta_{0}$ belongs to the
original boundary $\partial\Theta$ or not.\footnote{Instead of setting
$g^{\ast}=g-|g|^{1+\kappa}$, one could alternatively set $g^{\ast
}:=g-n^{-\kappa}|g|$ for $\kappa\in(0,\tfrac{1}{2})$. The results would be
unchanged.
\par
{}}

In general, the limit distribution of the resulting bootstrap estimator is
random, with randomness depending on both the stochastic regressor and the
position of $\theta_{0}$ relative to the boundary. This distribution is given
in the following theorem, where $\dot{g}^{\ast}:=\frac{\partial}%
{\partial\theta^{\prime}} g^{\ast}(\theta_{0})$.

\begin{theorem}
\label{Lemma bootstrap with boundary}Under the assumptions and the notation
introduced above, let a null hypothesis $\mathsf{H_{0}}$ as in
$\mathscr{G}{}_{1}$--$\,\mathscr{G}{}_{3}$ hold. Let also $\xi^{\ast}%
|(M,\ell(\theta_{0}))\sim N(0,1)$. Then
\begin{equation}
(n^{1/2}(\hat{\theta}-\theta_{0}),(n^{1/2}(\hat{\theta}^{\ast}-\hat{\theta
})|D_{n}))\overset{w}{\rightarrow}_{w}\left(  \ell(\theta_{0}),(\ell^{\ast
}(\theta_{0})|(M,\ell(\theta_{0})))\right)  \text{, }%
\label{eq asy distrib for general bootstraps}%
\end{equation}
where in the case $g^{\ast}(\theta_{0})<g(\theta_{0})$,
\begin{equation}
\ell^{\ast}(\theta_{0})=\tilde{\ell}^{\ast}:=M^{-1/2}\xi^{\ast}\text{ with
}\tilde{\ell}^{\ast}|(M,\ell(\theta_{0}))\overset{d}{=}\tilde{\ell}|M\text{,
}\label{eq semno}%
\end{equation}
whereas in the case $g^{\ast}(\theta_{0})=g(\theta_{0})$,
\begin{equation}
\ell^{\ast}(\theta_{0})=\ell^{\ast}:=\underset{\lambda\in\Lambda_{\ell}^{\ast
}}{\arg\min}||\lambda-M^{-1/2}\xi^{\ast}||_{M}\text{, }\Lambda_{\ell}^{\ast
}:=\{\lambda\in\mathbb{R}^{2}:\dot{g}^{\prime}\lambda\geq(\dot{g}^{\ast}%
-\dot{g})^{\prime}\ell(\theta_{0})\}\text{. }\label{eq amans}%
\end{equation}

\end{theorem}

\noindent The following conclusions could be drawn.

\medskip

\noindent(i) Consider first configurations $\mathscr{G}{}_{1}$ and
$\mathscr{G}{}_{2}$ under $\mathsf{H}_{0}$, such that $g(\theta_{0})=0$.
Consider the magnitude order, in probability, of the distance between the
bootstrap `true' value $\hat{\theta}$ and the bootstrap boundary
$\partial\Theta^{\ast}$ as a precision measure for a bootstrap approximation
to the geometry of $\mathscr{G}_{1}$ and $\mathscr{G}_{2}$. As seen above, the
standard bootstrap (corresponding to $g^{\ast}=0$) approximates the geometry
up to an exact magnitude order of $n^{-1/2}$, resulting in a situation where
the belonging of $\theta_{0}$ to the boundary contributes to the randomness of
limit bootstrap distribution given by
(\ref{eq asy for standard BS with param on the boundary}) and (\ref{eq amans})
via conditioning on the rv $\ell(\theta_{0})=\ell$. Conversely, bootstrap
schemes employing $g^{\ast}(\theta_{0})=g(\theta_{0})$ and $\dot{g}^{\ast
}=\dot{g}$, such that the bootstrap boundary is tangent to the original
boundary at $\theta_{0}$, give rise to approximations of order $o_{p}%
(n^{-1/2})$ and all the randomness in the bootstrap limit is due to the
properties of the stochastic regressor (via the rv $M$, as now $\ell^{\ast
}|(M,\ell)\overset{d}{=}\ell|M$; see (\ref{eq asy distribution}) and
(\ref{eq amans})). Moreover, for such schemes the bootstrap mimics a
conditional version of the asymptotic distribution of the original estimator:
$n^{1/2}(\hat{\theta}^{\ast}-\hat{\theta})\overset{w^{\ast}}{\rightarrow}%
_{w}\ell|M$. Examples are the `restricted' bootstrap based on $g^{\ast}=g$,
which replicates the geometry of the original data under $\mathsf{H}_{0}$ by
putting $\hat{\theta}$ on the bootstrap boundary, and the choices $g^{\ast
}=g-|g|^{1+\kappa}$ for some $\kappa>0$.

\medskip

\noindent(ii) Consider now the case in $\mathscr{G}{}_{3}$, such that
$g(\theta_{0})=0$ need not, but may hold under $\mathsf{H}_{0}$. Among the
bootstraps considered in (i), the standard one would fail to mimic a
conditional version of the original distribution if $g(\theta_{0})=0$, while
the `restricted' one would fail if $g(\theta_{0})>0$. As an alternative,
consider the bootstrap based on $g^{\ast}=g-|g|^{1+\kappa}$ for some
$\kappa>0$, see above. If $\theta_{0}\in\partial\Theta$, then this choice puts
the bootstrap true value $\hat{\theta}$ at an (asymptotically negligible)
distance of $o_{p}(n^{-1/2})$ from the bootstrap boundary, whereas if
$\theta_{0}\in\operatorname*{int}(\Theta)$, then $\hat{\theta}$ is bounded
away from the bootstrap boundary, in probability. This guarantees bootstrap
unconditional validity, see (iii)\ below.

\medskip

\noindent(iii) In general, bootstrap unconditional validity can be evaluated
through the following corollary of Theorem \ref{Lemma bootstrap with boundary}.

\begin{corollary}
\label{corollary bootstrap with boundary}Under the assumptions of Theorem
\ref{Lemma bootstrap with boundary}, a necessary and sufficient condition for
the convergence
\begin{equation}
\left(  n^{1/2}(\hat{\theta}-\theta_{0}),(n^{1/2}(\hat{\theta}^{\ast}%
-\hat{\theta})|D_{n})\right)  \overset{w}{\rightarrow}_{w}\left(  \ell
(\theta_{0}),(\ell(\theta_{0})|M)\right)  \text{ }\label{eq bound ave}%
\end{equation}
is that: (i) under $\mathscr{G}{}_{1}$ and $\mathscr{G}{}_{2}$, $g(\theta
_{0})=g^{\ast}(\theta_{0})$ and $\dot{g}=\dot{g}^{\ast}$; (ii) under
$\mathscr{G}{}_{3}$, either $g(\theta_{0})=g^{\ast}(\theta_{0})$ and $\dot
{g}=\dot{g}^{\ast}$, or $g(\theta_{0})>\max\{0,g^{\ast}(\theta_{0})\}$.

Moreover, under (\ref{eq bound ave}) the bootstrap is unconditionally valid
for any pair of statistics $\tau=\phi(n^{1/2}(\hat{\theta}-\theta_{0}%
))+o_{p}(1)$ and $\tau^{\ast}=\phi(n^{1/2}(\hat{\theta}^{\ast}-\hat{\theta
}))+o_{p}(1)$, where $\phi$ is a continuous real function such that the cdf of
$\phi(\ell(\theta_{0}))|M$ is continuous.
\end{corollary}

\noindent The class of functions $g^{\ast}=g-|g|^{1+\kappa}$ for $\kappa>0 $
satisfies both conditions (i) and (ii) of the previous corollary; hence, the
ensuing bootstrap inference is unconditionally valid under all of
$\mathscr{G}{}_{1}$-$\mathscr{G}{}_{3}$. In contrast, the standard bootstrap
violates condition (i) and, in general, is asymptotically invalid if
$g(\theta_{0})=0$. An exception is when the discrepancy between the original
and the bootstrap geometry is offset by the use of a test statistic that takes
into account the geometric position of the null hypothesis in the original
parameter space. The next section focuses on this setup.

\subsubsection{Unconditional validity of one-sided standard bootstrap tests}

Under case $\mathscr{G}{}_{1}$, consider testing $\mathsf{H}_{0}:g(\theta
_{0})=0$ against the alternative $\mathsf{H}_{1}:g(\theta_{0})>0$ using the
standard bootstrap (i.e., with $g^{\ast}=0$). For a test statistic of the form
$\tau_{n}:=n^{1/2}g(\hat{\theta})$,\footnote{What follows easily generalizes
to statistics of the form $\tau_{n}:=\phi(n^{1/2}g(\hat{\theta}))$ with
$\phi(\cdot)$ strictly increasing and normalized by $\phi(0)=0$.} its
bootstrap counterpart is given by $\tau_{n}^{\ast}:=n^{1/2}(g(\hat{\theta
}^{\ast})-g(\hat{\theta}))$ and the (one-sided) bootstrap test rejects for
\emph{large} values of the bootstrap \emph{p}-value $p_{n}^{\ast}:=P^{\ast
}(\tau_{n}^{\ast}\leq\tau_{n})$; equivalently, for small values of $\tilde
{p}_{n}^{\ast}:=1-p_{n}^{\ast}$ (see Remark
\ref{Remark 3.4 - part about right sided tests}). As for $\hat{\theta}^{\ast}%
$, also $\tau_{n}^{\ast}$ is affected in the limit by extra randomness due to
$\theta_{0}$ being on the boundary. From
(\ref{eq asy distrib for general bootstraps}), which reduces to
(\ref{eq asy distribution}) and
(\ref{eq asy for standard BS with param on the boundary}), it follows by the
Delta method that%
\[
(\tau_{n},(\tau_{n}^{\ast}|D_{n}))\overset{w}{\rightarrow}_{w}\left(  \dot
{g}^{\prime}\ell\text{,}\left(  \dot{g}^{\prime}\ell^{\ast}|(M,\ell)\right)
\right)  =(\dot{g}^{\prime}\ell,(\max\{-\dot{g}^{\prime}\ell,\dot{g}^{\prime
}\tilde{\ell}^{\ast}\}|(M,\ell)))\text{.}%
\]
For $\tau_{n}^{\ast}$, however, the randomness induced by conditioning on
$\ell$ affects the sample paths of the associated random cdf on the negative
half-line alone (because $\dot{g}^{\prime}\ell\geq0$), and is thus irrelevant
for bootstrap tests with nominal levels in $(0,\frac{1}{2})$. Put differently,
the bootstrap \emph{p}-values $\tilde{p}_{n}^{\ast}$ are (asymptotically)
uniformly distributed below $\frac{1}{2}$. This follows rigorously from the
next generalization of Theorem \ref{th2} (the proof being analogous), where
conditions for unconditional bootstrap validity restricted to a subset of
nominal testing levels are formulated.

\bigskip

\noindent\textsc{Theorem \ref{th2}}\textbf{$^{\ast}$}. \textit{Let there exist
a rv $\tau$ and a random element $X$, both defined on the same probability
space, such that the support of $\tau_{n}$ is contained in a closed interval
$\mathbb{T}$ (finite or infinite), and $(\tau_{n},F_{n}^{\ast})\overset
{w}{\rightarrow}(\tau,F)$ in $\mathbb{R}\times D(\mathbb{T})$ for $F_{n}%
^{\ast}(u):=P(\tau_{n}^{\ast}\leq u|D_{n})$ and $F(u):=P(\tau\leq u|X)$,
$u\in\mathbb{T}$. If the (possibly random) cdf $F$ is sample-path continuous
on $\mathbb{T}$, then the bootstrap }$p$\textit{-value }$p_{n}^{\ast
}:=\mathit{F_{n}^{\ast}(\tau_{n})}$\textit{\ satisfies
\[
P(p_{n}^{\ast}\leq q)\rightarrow q
\]
for $q$ such that $q\in F(\mathbb{T})$ a.s.}\bigskip

\noindent By Theorem \ref{th2}\textbf{$^{\ast}$} with $\mathbb{T}=[0,\infty)$
(which corresponds to the support of $\tau_{n}$ and $\tau:=\dot{g}^{\prime
}\ell$), it follows that the standard bootstrap applied to the one-sided
statistic $\tau_{n}$ is unconditionally valid for nominal levels in
$(0,\frac{1}{2})$.

\subsection{Bootstrap tests of parameter constancy}

\label{Section on BS tests for parameter constancy}

\subsubsection{General set up}

Here we apply the results of Section \ref{sec g} to the classic problem of
parameter constancy testing in regression models (Chow, 1960; Quandt, 1960;
Nyblom, 1989; Andrews, 1993; Andrews and Ploberger, 1994). Specifically, we
deal with bootstrap implementations when the moments of the regressors may be
unstable over time; see Hansen (2000) and Zhang and Wu (2012), inter alia.

Consider a linear regression model for $y_{nt}\in\mathbb{R}$ given $x_{nt}%
\in\mathbb{R}^{m}$, in triangular array notation:
\begin{equation}
y_{nt}=\beta_{t}^{\prime}x_{nt}+\varepsilon_{nt}\text{ \hspace{1cm}%
(}t=1,2,...,n\text{).}\label{eq reg model}%
\end{equation}
The null hypothesis of parameter constancy is $\mathsf{H}_{0}:\beta_{t}%
=\beta_{1}\,(t=2,...,n)$, which is tested here against the alternative
$\mathsf{H}_{1}:\beta_{t}=\beta_{1}+\theta\mathbb{I}_{\{t\geq n^{\star}\}} $
($t=2,...,n$), where $n^{\star}:=\lfloor r^{\star}n\rfloor$ and $\theta\neq0$
respectively denote the timing and the magnitude of the possible
break,\footnote{We suppress the possible dependence of $\beta_{t}=\beta_{nt}$
on $n$ with no risk of ambiguities.} both assumed unknown to the
econometrician. The so-called break fraction $r^{\star}$ belongs to a known
closed interval $[\underline{r},\overline{r}]\ $in $(0,1)$. In order to test
$\mathsf{H}_{0}$ against $\mathsf{H}_{1}$, it is customary to consider the
`$\sup F$' (or `$\sup$ Wald') test (Quandt, 1960; Andrews, 1993), based on the
statistic $\mathscr{F}{}_{n}:=\max_{r\in\lbrack\underline{r},\overline{r}%
]}F_{\left\lfloor nr\right\rfloor },$ where $F_{\left\lfloor nr\right\rfloor
}$ is the usual $F$ statistic for testing the auxiliary null hypothesis that
$\theta=0$ in the regression
\[
y_{nt}=\beta^{\prime}x_{nt}+\theta^{\prime}x_{nt}\mathbb{I}_{\{t\geq
\left\lfloor rn\right\rfloor \}}+\varepsilon_{nt}\text{.}%
\]

We make the following assumption, allowing for non-stationarity in the
regressors (see also Hansen, 2000, Assumptions 1 and 2).

\bigskip{}

\noindent\textsc{Assumption $\mathcal{H}$}. The following conditions on
$\{x_{nt},\varepsilon_{nt}\}$ hold:

\begin{description}
\item (i) \emph{(mda)}$\ \varepsilon_{nt}$\emph{\ is a martingale difference
array with respect to the current value of }$x_{nt}$\emph{\ and the lagged
values of }$\left(  x_{nt},\varepsilon_{nt}\right)  $\emph{;}

\item (ii) \emph{(wlln) }$\varepsilon_{nt}^{2}$\emph{\ satisfies the law of
large numbers }$n^{-1}\sum_{t=1}^{\left\lfloor nr\right\rfloor }%
\varepsilon_{nt}^{2}\overset{p}{\rightarrow}r(E\varepsilon_{nt}^{2}%
)=r\sigma^{2}>0$\emph{$,$ for all }$r\in(0,1]$\emph{;}

\item (iii) \emph{(non-stationarity) in }$\mathscr{D}{}_{m\times m}%
\times\mathscr{D}{}_{m\times m}\times\mathscr{D}{}_{m}$\emph{:}
\[
\left(  \tfrac{1}{n}\sum_{t=1}^{\left\lfloor n\cdot\right\rfloor }x_{nt}%
x_{nt}^{\prime},\tfrac{1}{n\sigma^{2}}\sum_{t=1}^{\left\lfloor n\cdot
\right\rfloor }x_{nt}x_{nt}^{\prime}\varepsilon_{nt}^{2},\tfrac{1}%
{n^{1/2}\sigma}\sum_{t=1}^{\left\lfloor n\cdot\right\rfloor }x_{nt}%
\varepsilon_{nt}\right)  \overset{w}{\rightarrow}(M,V,N),
\]
\emph{where }$M$\emph{\ and }$V$\emph{\ are a.s. continuous and (except at 0)
strictly positive-definite valued processes, whereas }$N$\emph{, conditionally
on }$\{V,M\}$\emph{,\ is a zero-mean Gaussian process with covariance kernel
}$E\{N\left(  r_{1}\right)  N\left(  r_{2}\right)  ^{\prime}\}=V\left(
r_{1}\right)  $\emph{\ }$(0\leq r_{1}\leq r_{2}\leq1)$\emph{.}
\end{description}

\begin{remark}
\label{Remark 4.1}A special case of Assumption $\mathcal{H}$ is obtained when
the regressors satisfy the weak convergence $x_{n\left\lfloor n\cdot
\right\rfloor }\overset{w}{\rightarrow}U\left( \cdot\right) $ in
$\mathscr{D}{}_{m}$, such that $M\left( \cdot\right) =\int_{0}^{\cdot
}UU^{\prime}$. Under extra conditions (e.g., if $\sup_{n}\sup_{t=1,...,n}%
E|E(\varepsilon_{nt}^{2}-\sigma^{2}|\mathcal{F}_{n,t-i})|\rightarrow0$ as
$i\rightarrow\infty$ for some filtrations $\mathcal{F}_{n,t}$, $n\in\mathbb{N}
$, to which $\{\varepsilon_{nt}^{2}$\} is adapted), also $V\left( \cdot\right)
=\int_{0}^{\cdot}UU^{\prime}$ (see Theorem A.1 of Cavaliere and Taylor, 2009).
$\hfill\square$ \medskip{}
\end{remark}

The null asymptotic distribution of $\mathscr{F}{}_{n}$ under Assumption
$\mathcal{H}$ is provided in Hansen (2000, Theorem 2):
\begin{equation}
\mathscr{F}{}_{n}\overset{w}{\rightarrow}\sup_{r\in\lbrack\underline
{r},\overline{r}]}\{\tilde{N}(r)^{\prime}\tilde{M}\left(  r\right)
^{-1}\tilde{N}(r)\}\label{eq asy distr of supF}%
\end{equation}
with $\tilde{N}\left(  u\right)  :=N\left(  u\right)  -M\left(  u\right)
M\left(  1\right)  ^{-1}N\left(  1\right)  $ and $\tilde{M}\left(  r\right)
:=M\left(  r\right)  -M\left(  r\right)  M\left(  1\right)  ^{-1}M\left(
r\right)  $. In the case of (asymptotically)\ stationary regressors,
$\mathscr{F}{}_{n}$ converges to the supremum of a squared tied-down Bessell
process; see Andrews (1993). In the general case, however, since the
asymptotic distribution in (\ref{eq asy distr of supF}) depends on the joint
distribution of the limiting processes $M,N,V$, which is unspecified under
Assumption $\mathcal{H}$, asymptotic inference based on
(\ref{eq asy distr of supF}) is unfeasible. Simulation methods as the
bootstrap can therefore be appealing devices for computing \emph{p}-values
associated with $\mathscr{F}{}_{n}$.

\subsubsection{Bootstrap test and random limit bootstrap distribution}

Following Hansen (2000), we consider here a fixed-regressor wild bootstrap
introduced to accommodate possible conditional heteroskedasticity of
$\varepsilon_{nt}$. It is based on the residuals $\tilde{e}_{nt}$ from the OLS
regression of $y_{nt}$ on $x_{nt}$ and $x_{nt}\mathbb{I}_{\{t\geq\left\lfloor
\tilde{r}n\right\rfloor \}}$, where $\tilde{r}:=\arg\max_{r\in\lbrack
\underline{r},\overline{r}]}F_{\left\lfloor nr\right\rfloor }$ is the
estimated break fraction for the original sample. The bootstrap statistic is
\[
\mathscr{F}{}_{n}^{\ast}:=\max_{r\in\lbrack\underline{r},\overline{r}%
]}F_{\left\lfloor nr\right\rfloor }^{\ast}\text{,}%
\]
where $F_{\left\lfloor nr\right\rfloor }^{\ast}$ is the $F$ statistic for the
auxiliary null hypothesis that $\theta^{\ast}=0$ in the regression
\begin{equation}
y_{t}^{\ast}=\beta^{\ast\prime}x_{nt}+\theta^{\ast\prime}x_{nt}\mathbb{I}%
_{\{t\geq\left\lfloor rn\right\rfloor \}}+\text{error}_{nt}^{\ast
},\label{eq br}%
\end{equation}
with bootstrap data $y_{t}^{\ast}:=\tilde{e}_{nt}w_{t}^{\ast}$ for an i.i.d.
N(0,1) sequence of bootstrap multipliers $w_{t}^{\ast}$ independent of the data.

The weak limit of the bootstrap statistic $\mathscr{F}{}_{n}^{\ast}$ is stated
in the next theorem.

\begin{theorem}
\label{th h}Under Assumption $\mathcal{H}$ and under $\mathsf{H}_{0}$, it
holds that
\begin{equation}
\mathscr{F}{}_{n}^{\ast}\overset{w^{\ast}}{\rightarrow}_{w}\left.  \sup
_{r\in\lbrack\underline{r},\overline{r}]}\{\tilde{N}(r)^{\prime}\tilde
{M}\left(  r\right)  ^{-1}\tilde{N}(r)\}\right\vert (M,V),\label{eq d}%
\end{equation}
where $\tilde{M}\left(  r\right) $, $\tilde{N}\left(  r\right) $ are as in
(\ref{eq asy distr of supF}).
\end{theorem}

\begin{remark}
\label{Remark 4.4}Theorem \ref{th h} establishes that, in general, the weak
limit of the fixed-regressor bootstrap statistic is \emph{random}. In
particular, it is distinct from the limit in eq. (\ref{eq asy distr of supF})
and, as a result, the bootstrap does not estimate consistently the
unconditional limit distribution of the statistic $\mathscr{F}_{n}$ under
$\mathsf{H}_{0}$ (contrary to the claim in Theorem 6 of Hansen, 2000). To
illustrate the limiting randomness, consider the case $M=V$ with a scalar
regressor $x_{nt}\in\mathbb{R}$. By a change of variable (as in Theorem 3 of
Hansen, 2000), convergence (\ref{eq d}) reduces to
\[
\mathscr{F}{}_{n}^{\ast}\overset{w^{\ast}}{\rightarrow}_{w}\left.  \sup_{u\in
I(M,\underline{r},\bar{r})}\left\{  \frac{W(u)^{2}}{u(1-u)}\right\}
\right\vert M\text{\ \ \ $\text{for\ \ \ }$}I(M,\underline{r},\bar
{r}):=\left[  \tfrac{M(\underline{r})}{M(1)},\tfrac{M(\bar{r})}{M(1)}\right]
,
\]
where $W$ is a standard Brownian bridge on $[0,1]$, independent of $M$. As the
maximization interval $I(M,\underline{r},\bar{r})$ depends on $M$, so does the
supremum itself.$\hfill\square$ \medskip{}
\end{remark}

\subsubsection{Bootstrap validity}

Although under Assumption $\mathcal{H}$ the bootstrap does not replicate the
asymptotic (unconditional) distribution in (\ref{eq asy distr of supF}),
unconditional bootstrap validity can be established under no further
assumptions than Assumption $\mathcal{H}$, by using the results in Section
\ref{sec gen on unc val}. In contrast, despite fixing the regressors across
bootstrap samples, if interest is in achieving bootstrap validity conditional
on the regressors $X_{n}:=\{x_{nt}\}_{t=1}^{n}$, further conditions are
required; e.g., the following Assumption \textsc{$\mathcal{C}$}.

\bigskip

\noindent\textsc{Assumption $\mathcal{C}$}. \emph{Assumption }%
\textsc{\emph{$\mathcal{H}$}}\emph{\ holds and, as random measures on
}$\mathscr{D}{}_{m\times m}\times\mathscr{D}{}_{m\times m}\times
\mathscr{D}{}_{m}$%
\[
\left.  \left(  \tfrac{1}{n}\sum_{t=1}^{\left\lfloor n\cdot\right\rfloor
}x_{nt}x_{nt}^{\prime},\tfrac{1}{n\sigma^{2}}\sum_{t=1}^{\left\lfloor
n\cdot\right\rfloor }x_{nt}x_{nt}^{\prime}\varepsilon_{nt}^{2},\tfrac
{1}{n^{1/2}\sigma}\sum_{t=1}^{\left\lfloor n\cdot\right\rfloor }%
x_{nt}\varepsilon_{nt}\right)  \right\vert X_{n}\overset{w}{\rightarrow}%
_{w}\left(  M,V,N\right)  |(M,V)
\]
\emph{jointly with the convergence in Assumption }\textsc{\emph{$\mathcal{H}$%
}}\emph{(iii).}

\bigskip

The results on the validity of the bootstrap parameter constancy tests are
summarized in the following theorem.

\begin{theorem}
\label{th fb}Let the parameter constancy hypothesis $\mathsf{H}_{0}$ hold for
model (\ref{eq reg model}). Then, under Assumption $\mathcal{H}$, the
bootstrap based on $\tau_{n}=\mathscr{F}{}_{n}$ and $\tau_{n}^{\ast
}=\mathscr{F}{}_{n}^{\ast}$ is unconditionally valid. If Assumption
$\mathcal{C}$ holds, then the bootstrap based on $\mathscr{F}{}_{n}$ and
$\mathscr{F}{}_{n}^{\ast}$ is valid also conditionally on $X_{n}$.
\end{theorem}

\begin{remark}
\label{Remark 4.6}In the proof of Theorem \ref{th fb}, we refer to Theorem
\ref{th2} and Corollary \ref{c1}(a) for establishing respectively
unconditional and conditional bootstrap validity. Notice that Assumption
$\mathcal{C}$ is stronger than Assumption $\mathcal{H}$ due to the fact that
--\thinspace differently from the bootstrap variates $w_{t}^{\ast}$
--\thinspace the errors $\{\varepsilon_{nt}\}$ need not be independent of
$\{x_{nt}\}$. The third DGP of Section
\ref{sec uncond validity W/O conditional validity} could be used to construct
an example, with $x_{nt}:=n^{-1/2}x_{t}$ and $\varepsilon_{nt}:=\varepsilon
_{t}$, where Assumption $\mathcal{H}$(iii) holds but Assumption $\mathcal{C}$
does not.
\end{remark}

\begin{remark}
The meaning of `jointly' in Assumption $\mathcal{C}$ is given in eq.
(\ref{eq:wcrm1}). By Lemma \ref{le crpr}(b), the convergence will be
automatically joint if in $\mathscr{D}{}_{m\times m}$, $n^{-1}\sigma^{-2}%
\sum_{t=1}^{\left\lfloor n\cdot\right\rfloor }x_{nt}x_{nt}^{\prime
}(\varepsilon_{nt}^{2}-E(\varepsilon_{nt}^{2}|X_{n}))=o_{p}(1)$, such that
$n^{-1}\sigma^{-2}\sum_{t=1}^{\left\lfloor n\cdot\right\rfloor }x_{nt}%
x_{nt}^{\prime}\varepsilon_{nt}^{2}$\emph{\ }is asymptotically equivalent to
an\emph{\ }$X_{n}$-measurable process.$\hfill\square$
\end{remark}

\section{Conclusions}

\label{sec conclusion}

When the distribution of a bootstrap statistic conditional on the data is
random in the limit, the bootstrap fails to estimate consistently the
asymptotic distribution of the original statistic. In this case, the bootstrap
is usually regarded as invalid. Renormalization of the statistic of interest
cannot always be used to eliminate the limiting bootstrap randomness (e.g., it
cannot be used in any of the four applications discussed in Section
\ref{Section on Applications}). We have shown, however, that if
(asymptotic)\ bootstrap validity is defined as (large sample)\ control over
the frequency of correct inferences, then randomness of the limit bootstrap
distribution does not imply invalidity of the bootstrap, even without
renormalizing the original statistic.

For the asymptotic validity of bootstrap inference, in an unconditional or a
conditional sense, we have established sufficient conditions and strategies to
verify these conditions in specific applications. The conditions differ mainly
in their demands on the dependence structure of the data, and are more
restrictive for conditional validity to hold.

We have provided four applications to well-known econometric inference
problems which feature randomness of the limit bootstrap distribution. Among
the further applications where randomness of the limit bootstrap distribution
is likely to appear, and that could be analyzed using our approach, are
bootstrap inference in weakly or partially identified models, inference in
time series models with time-varying (stochastic) volatility, inference after
model selection, and the bootstrap in high-dimensional models. In addition,
the methods we provide for establishing conditional bootstrap validity could
be useful in problems involving nuisance parameters that are not consistently
estimable under the null hypothesis but where sufficient statistics are
available (with the bootstrap being potentially valid conditionally on such statistics).

An important issue not analyzed in the paper is whether the bootstrap can
deliver refinements over standard asymptotics in cases where the limit
bootstrap measure is random. We have seen in Sections 2 and
\ref{sec example CUSUM} that bootstrap inference in such cases could be exact
or close to exact. This seems to suggest that a potential for refinements
exists. Moreover, there is also a potential for the bootstrap to inherit the
finite-sample refinements offered by conditional asymptotic expansions (in
line with Barndorff-Nielsen's p{*}-formula, see Barndorff-Nielsen and Cox,
1994, Sec. 6.2), as has been established for some bootstrap procedures
(DiCiccio and Young, 2008) in the special case of correctly specified
parametric models. The study of such questions requires mathematical tools
different from those employed here, and is therefore left for further research.

\section*{References}

\begin{description}
\item \textsc{Andrews, D.W.K. }(1993): Tests for parameter instability and
structural change with unknown change point, \emph{Econometrica} 61, 821--856.

\item ------ (1997): A conditional Kolmogorov test, \emph{Econometrica} 65, 1097--1128.

\item ------ (1999): Estimation when a parameter is on a boundary,
\emph{Econometrica}, 67, 1341--1383.

\item ------ (2000): Inconsistency of the bootstrap when a parameter is on the
boundary of the parameter space, \emph{Econometrica}, 68, 399--405.

\item \textsc{Andrews, D.W.K. and W. Ploberger }(1994): Optimal tests when a
nuisance parameter is present only under the alternative, \emph{Econometrica}
62, 1383--1414.

\item \textsc{Athreya, K.B.}(1987): Bootstrap of the mean in the infinite
variance case, \textit{The Annals of Statistics} 15, 724-731.

\item \textsc{Aue, A., I. Berkes and L. Horv{\small \`{a}}th} (2008):
Selection from a stable box,\emph{Bernoulli }14, 125--139.

\item \textsc{O.E.Barndorff-Nielsen and D.R. Cox} (1994): \emph{Inference and
Asymptotics}, Chapman \& Hall.

\item \textsc{Basawa, I.V., A.K. Mallik, W.P. McCormick, J.H. Reeves, and R.L.
Taylor }(1991): Bootstrapping unstable first-order autoregressive processes,
\emph{The Annals of Statistics} 19, 1098--1101.

\item \textsc{Bentkus, V}. (2005): A Lyapunov-type bound in $R^{d}$,
$\emph{Theory\ of}$ \emph{Probability and Its Applications} 49(2), 311--323.

\item \textsc{Beran, R. }(1997): Diagnosing Bootstrap success, \emph{Annals of
the Institute of Statistical Mathematics }49, 1--24.

\item \textsc{Billingsley, P}. (1968): \emph{Convergence of Probability
Measures}, John Wiley \& Sons, NY.

\item \textsc{Cattaneo, M., M. Jansson and K. Nagasawa }(2017):
Bootstrap-based inference for cube root consistent estimators, \emph{arXiv}:1704.08066.

\item \textsc{Cavaliere, G. and I. Georgiev} (2019): Inference under random
limit bootstrap measures: supplemental material.

\item \textsc{Cavaliere, G., I. Georgiev and A.M.R. Taylor} (2016):
Sieve-based inference for infinite-variance linear processes, \emph{Annals of
Statistics }44, 1467--1494.

\item \textsc{Cavaliere, G., H.B. Nielsen and A.\ Rahbek }(2015): Bootstrap
testing of hypotheses on co-integration relations in vector autoregressive
models, \emph{Econometrica},\emph{\ }83, 813--831.

\item ------ (2017): On the consistency of bootstrap testing for a parameter
on the boundary of the parameter space, \emph{Journal of Time Series Analysis}
38, 513--534

\item \textsc{Cavaliere, G. and A.M.R. Taylor }(2009): Heteroskedastic time
series with a unit root, \emph{Econometric Theory} 25, 1228--1276.

\item \textsc{Chan, N.H. and C.Z. Wei} (1988): Limiting distributions of Least
Squares estimates of unstable autoregressive processes, \emph{Annals of
Statistics }16, 367--401.

\item \textsc{Chow G.} (1960): Tests of equality between sets of coefficients
in two linear regressions, \emph{Econometrica }28, 591--605.

\item \textsc{Crimaldi} \textsc{I. and L. Pratelli} (2005): Convergence
results for conditional expectations, \emph{Bernoulli} 11, 737--745.

\item \textsc{DasGupta A.} (2008): \emph{Asymptotic Theory of Statistics and
Probability}, Springer-Verlag: Berlin.

\item \textsc{DiCiccio T. and G.A. Young} (2008): Conditional properties of
unconditional parametric bootstrap procedures for inference in exponential
families, \emph{Biometrika} 95, 747--758.

\item \textsc{Fang Z. and A. Santos} (2019): Inference on directionally
differentiable functions, \emph{Review of Economic Studies} 86, 377--412.

\item \textsc{Georgiev, I., D. Harvey, S. Leybourne and A.M.R. Taylor }(2019):
A bootstrap stationarity test for predictive regression invalidity,
\emph{Journal of Business \& Economic Statistics} 37, 528--541.

\item \textsc{Hall, P.} (1992): \emph{The Bootstrap and Edgeworth Expansion},
Springer-Verlag: Berlin.

\item \textsc{Hansen} \textsc{B.E.} (2000): Testing for structural change in
conditional models. \emph{Journal of Econometrics} 97, 93--115.

\item \textsc{H\"{a}usler, E. and H. Luschgy} (2015): \emph{Stable Convergence
and Stable Limit Theorems}, Springer-Verlag: Berlin.

\item \textsc{Horowitz, J.L. }(2001): The bootstrap. In Heckman, J.J. and E.
Leamer (eds.) \emph{Handbook of Econometrics} 5, chapter 52, Elsevier: Amsterdam.

\item \textsc{Hounyo, U., S. Gon\c{c}alves and N. Meddahi} (2017):
Bootstrapping pre-averaged realized volatility under market microstructure
noise, \emph{Econometric Theory}\textit{\ }33, 791--838.

\item \textsc{Jacod, J., Y. Li, P. Mykland, M. Podolskij and M. Vetter}
(2009): Microstructure noise in the continuous case: The pre-averaging
approach. \textit{Stochastic Processes and Their Applications} 119, 2249-2276.

\item \textsc{Johansen} \textsc{S. and B. Nielsen} (2016): Analysis of the
Forward Search using some new results for martingales and empirical processes,
\emph{Bernoulli} 22, 1131--1183.

\item \textsc{Knight, K.} (1989): On the bootstrap of the sample mean in the
infinite variance case, \textit{The Annals of Statistics} 17, 1168-1175.

\item \textsc{Kallenberg O.} (1997): \textit{Foundations of Modern
Probability}, Springer-Verlag: Berlin.

\item \textsc{Kallenberg O.} (2017): \textit{Random measures: theory and
applications}, Springer-Verlag: Berlin.

\item \textsc{LePage, R. and K. Podg\'{o}rsky}: (1996): Resampling
permutations in regression without second moments, \emph{Journal of
Multivariate Analysis} 57, 119--141.

\item \textsc{LePage, R., K. Podg\'{o}rsky and M. Ryznar} (1997): Strong and
conditional invariance principles for samples attracted to stable laws,
\emph{Probability Theory and Related Fields} 108, 281--298.

\item \textsc{Linde, W. }(1989): Gaussian measure of translated balls in a
Banach space, \emph{Theory of Probabability and its Applications }34, 349--359.

\item \textsc{Lockhart} R. (2012): Conditional limit laws for goodness-of-fit
tests, \textit{Bernoulli} 18, 857--882.

\item \textsc{M\"{u}ller, U.K. and M. Watson} (2008):\ Testing models of low
frequency variability, \emph{Econometrica} 76, 979--1016.

\item \textsc{Nyblom} J., (1989): Testing for the constancy of parameters over
time, \emph{Journal of the American Statistical Association} 84, 223--230.

\item \textsc{Phillips, P.C.B}. (2014): On confidence intervals for
autoregressive roots and predictive regression, \emph{Econometrica} 82, 1177--1195.

\item \textsc{Politis, D., J. Romano and M. Wolf} (1999): \emph{Subsampling},
Springer, Berlin.

\item \textsc{Ploberger, W. and W. Kr{\small \"{a}}mer }(1992): The CUSUM test
with OLS residuals, \emph{Econometrica} 60, 271--285.

\item \textsc{Quandt} R. (1960): Tests of the hypothesis that a linear
regression system obeys two separate regimes, \textit{Journal of the American
Statistical Association} 55, 324--330.

\item \textsc{Reid N}. (1995): The roles of conditioning in inference,
\emph{Statistical Science} 10, 138--199.

\item \textsc{Rubshtein }B. (1996): A central limit theorem for conditional
distributions, In Bergelson V., P. March, J. Rosenblatt (eds.),
\textit{\ Convergence in Ergodic Theory and Probability}, De Gruyter: Berlin.

\item \textsc{Sen, B., M. Banerjee, and M. Woodroofe }(2010): Inconsistency of
Bootstrap: The Grenander Estimator, \emph{Annals of Statistics} 38, 1953--1977.

\item \textsc{Shao X. and D.N. Politis }(2013): Fixed \emph{b} subsampling and
the block bootstrap: improved confidence sets based on \emph{p}-value
calibration, \textit{Journal of the Royal Statistical Society B} 75, 161--184.

\item \textsc{Sweeting T.J. }(1989): On conditional weak convergence,
\textit{Journal of Theoretical Probability 2, 461--474.}

\item \textsc{Zhang, T. and W. Wu }(2012): Inference of time-varying
regression models, \textit{Annals of Statistics} 40, 1376--1402.
\end{description}

\appendix
%dummy comment inserted by tex2lyx to ensure that this paragraph is not empty

\section*{Appendices}

%\global\long

%\setcounter{equation}{0}\global\long

%\global\long

\label{Sec Appendix}

\setcounter{theorem}{0}\setcounter{equation}{0}\setcounter{lemma}{0}\renewcommand{\thelemma}{\thesection.\arabic{lemma}}\renewcommand{\thetheorem}{\thesection.\arabic{theorem}}\numberwithin{theorem}{section}
\numberwithin{lemma}{section} \numberwithin{remark}{section}

\section{Weak convergence in distribution}

\label{sec itere}

In this section we establish some properties of weak convergence in
distribution for random elements of Polish spaces. They are useful in
applications, in order to verify the high-level conditions of our main
theorems, as well as to prove these very theorems. Recall the convention that,
throughout, Polish spaces are equipped with their Borel sets. Finite $n
$-tuples of random elements defined on the same probability space are
considered as random elements of a product space with the product topology and
$\sigma$-algebra.

Let $(Z_{n},X_{n})$ and $(Z,X)$ be random elements such that $Z_{n}%
=(Z_{n}^{\prime},Z_{n}^{\prime\prime})$ and $Z=(Z^{\prime},Z^{\prime\prime})$
are $\mathcal{S}_{Z}^{\prime}\times\mathcal{S}_{Z}^{\prime\prime}$-valued,
whereas $X_{n}=(X_{n}^{\prime},X_{n}^{\prime\prime})$ and $X=(X^{\prime
},X^{\prime\prime})$ are resp. $\mathcal{S}^{\prime}\times\mathcal{S}%
^{\prime\prime}$-valued and $\mathcal{S}_{X}^{\prime}\times\mathcal{S}%
_{X}^{\prime\prime}$-valued ($n\in\mathbb{N}$). We say that $Z_{n}^{\prime
}|X_{n}^{\prime}\overset{w}{\rightarrow}_{w}Z^{\prime}|X^{\prime}$ and
$Z_{n}^{\prime\prime}|X_{n}^{\prime\prime}\overset{w}{\rightarrow}%
_{w}Z^{\prime\prime}|X^{\prime\prime}$ \textit{jointly} (denoted by
$(Z_{n}^{\prime}|X_{n}^{\prime},Z_{n}^{\prime\prime}|X_{n}^{\prime\prime
})\overset{w}{\rightarrow}_{w}(Z^{\prime}|X^{\prime},Z^{\prime\prime
}|X^{\prime\prime})$) if
\begin{equation}
\left(  E\{h^{\prime}(Z_{n}^{\prime})|X_{n}^{\prime}\},\,E\{h^{\prime\prime
}(Z_{n}^{\prime\prime})|X_{n}^{\prime\prime}\}\right)  \overset{w}%
{\rightarrow}\left(  E\{h^{\prime}(Z^{\prime})|X^{\prime}\},\,E\{h^{\prime
\prime}(Z^{\prime\prime})|X^{\prime\prime}\}\right) \label{eq:wcrm}%
\end{equation}
for all $h^{\prime}\in\mathcal{C}_{b}(\mathcal{S}_{Z}^{\prime})$ and
$h^{\prime\prime}\in\mathcal{C}_{b}(\mathcal{S}_{Z}^{\prime\prime})$. Even for
$X_{n}^{\prime}=X_{n}^{\prime\prime}$, this property is weaker than the
convergence $(Z_{n}^{\prime},Z_{n}^{\prime\prime})|X_{n}^{\prime}\overset
{w}{\rightarrow}_{w}(Z^{\prime},Z^{\prime\prime})|X$ defined by $E\{g(Z_{n}%
^{\prime},Z_{n}^{\prime\prime})|X_{n}^{\prime}\}\overset{w}{\rightarrow
}E\{g(Z^{\prime},Z^{\prime\prime})|X\}$ for all $g\in\mathcal{C}%
_{b}(\mathcal{S}_{Z}^{\prime}\times\mathcal{S}_{Z}^{\prime\prime})$. We notice
that for $Z_{n}^{\prime}=X_{n}^{\prime}$, (\ref{eq:wcrm}) reduces to
\begin{equation}
\left(  Z_{n}^{\prime},E\{h^{\prime\prime}(Z_{n}^{\prime\prime})|X_{n}%
^{\prime\prime}\}\right)  \overset{w}{\rightarrow}\left(  Z^{\prime
},E\{h^{\prime\prime}(Z^{\prime\prime})|X^{\prime\prime}\}\right)
\label{eq:wcrm1}%
\end{equation}
for all $h^{\prime\prime}\in\mathcal{C}_{b}(\mathcal{S}_{Z}^{\prime\prime})$
and in this case we write $(Z_{n}^{\prime},(Z_{n}^{\prime\prime}|X_{n}%
^{\prime\prime}))\overset{w}{\rightarrow}_{w}(Z^{\prime\prime},(Z^{\prime
}|X^{\prime}))$ (see Corollary \ref{co coeq}\ of the accompanying Supplement).

The first lemma given here is divided in two parts. In the first part, we
provide conditions for strengthening weak convergence in distribution to weak
convergence in probability. The second part, in its simplest form, provides
conditions such that the two convergence facts $\left(  Z_{n},X_{n}\right)
\overset{w}{\rightarrow}\left(  Z,X\right)  $ and $Z_{n}|X_{n}\overset
{w}{\rightarrow}_{w}Z|X$ imply the joint convergence $((Z_{n}|X_{n}%
),Z_{n},X_{n})\overset{w}{\rightarrow}_{w}((Z|X),Z,X)$.

\begin{lemma}
\label{le crpr}Let $\mathcal{S}_{Z},$ $\mathcal{S}_{Z}^{\prime},\mathcal{S}%
_{X}$ and $\mathcal{S}_{X}^{\prime}$ be Polish spaces. Consider the random
elements $Z_{n},Z$ ($\mathcal{S}_{Z}$-valued), $Z_{n}^{\prime},Z^{\prime}$
($\mathcal{S}_{Z}^{\prime}$-valued), $X_{n}$ ($\mathcal{S}_{X}$-valued) and
$X_{n}^{\prime}$, $X$ ($\mathcal{S}_{X}^{\prime}$-valued) for $n\in\mathbb{N}%
$. Assume that $X_{n}^{\prime}$ are $X_{n}$-measurable and $Z_{n}%
|X_{n}\overset{w}{\rightarrow}_{w}Z|X$.

(a) If all the considered random elements are defined on the same probability
space, $\left(  Z_{n},X_{n}^{\prime}\right)  \overset{w}{\rightarrow}\left(
Z,X\right)  $ and $X_{n}^{\prime}\overset{p}{\rightarrow}X$, then $Z_{n}%
|X_{n}\overset{w}{\rightarrow}_{p}Z|X$.

(b) If $\left(  Z_{n},X_{n}^{\prime},Z_{n}^{\prime}\right)  \overset
{w}{\rightarrow}\left(  Z,X,Z^{\prime}\right)  $, then the joint convergence
$((Z_{n}|X_{n}),Z_{n},X_{n}^{\prime},Z_{n}^{\prime})\overset{w}{\rightarrow
}_{w}((Z|X),Z,X,Z^{\prime})$ holds in the sense that, for all $h\in
\mathcal{C}_{b}(\mathcal{S}_{Z})$,%
\begin{equation}
(E\{h(Z_{n})|X_{n}\},Z_{n},X_{n}^{\prime},Z_{n}^{\prime})\overset
{w}{\rightarrow}(E\{h(Z)|X\},Z,X,Z^{\prime})\text{.}\label{ur crpr}%
\end{equation}

\end{lemma}

\noindent Notice that, by choosing $Z_{n}^{\prime}=Z^{\prime}=1$, a corollary
of Lemma \ref{le crpr}(b) not involving $Z_{n}^{\prime}$ and $Z^{\prime}$ is
obtained. It states that $Z_{n}|X_{n}\overset{w}{\rightarrow}_{w}Z|X$ and
$\left(  Z_{n},X_{n}^{\prime}\right)  \overset{w}{\rightarrow}\left(
Z,X\right)  $ together imply the joint convergence $((Z_{n}|X_{n}),Z_{n}%
,X_{n}^{\prime})\overset{w}{\rightarrow}_{w}((Z|X),Z,X) $, provided that
$X_{n}^{\prime}$ are $X_{n}$-measurable.

By means of eq. (\ref{eq:wcrm}) we defined \emph{joint }weak convergence in
distribution and denoted it by $(Z_{n}^{\prime}|X_{n}^{\prime},Z_{n}%
^{\prime\prime}|X_{n}^{\prime\prime})\overset{w}{\rightarrow}_{w}(Z^{\prime
}|X^{\prime},Z^{\prime\prime}|X^{\prime\prime})$. We now extend it to%
\begin{equation}
((Z_{n}^{\prime}|X_{n}^{\prime}),(Z_{n}^{\prime\prime}|X_{n}^{\prime\prime
}),Z_{n}^{\prime\prime\prime})\overset{w}{\rightarrow}_{w}((Z^{\prime
}|X^{\prime}),(Z^{\prime\prime}|X^{\prime\prime}),Z^{\prime\prime\prime
})\text{,}\label{ur j3}%
\end{equation}
defined to mean that%
\begin{equation}
(E\{h^{\prime}(Z_{n}^{\prime})|X_{n}^{\prime}\},E\{h^{\prime\prime}%
(Z_{n}^{\prime\prime})|X_{n}^{\prime\prime}\},Z_{n}^{\prime\prime\prime
})\overset{w}{\rightarrow}(E\{h^{\prime}(Z^{\prime})|X^{\prime}\},E\{h^{\prime
\prime}(Z^{\prime\prime})|X^{\prime\prime}\},Z^{\prime\prime\prime
})\label{ur j4}%
\end{equation}
for all continuous and bounded real $h^{\prime},h^{\prime\prime}$ with
matching domain. The natural equivalence of $((Z_{n}^{\prime}|X_{n}^{\prime
}),(Z_{n}^{\prime}|X_{n}^{\prime}),Z_{n}^{\prime\prime\prime})\overset
{w}{\rightarrow}_{w}((Z^{\prime}|X^{\prime}),(Z^{\prime}|X^{\prime}%
),Z^{\prime\prime\prime})$ and $((Z_{n}^{\prime}|X_{n}^{\prime}),Z_{n}%
^{\prime\prime\prime})\overset{w}{\rightarrow}_{w}((Z^{\prime}|X^{\prime
}),Z^{\prime\prime\prime})$ holds under separability of the space
$\mathcal{S}^{\prime\prime\prime}$ where $Z_{n}^{\prime\prime\prime}%
,Z^{\prime\prime\prime}$ take values (see Remark \ref{Remark A1}).

In Lemma \ref{le kal}(b) below we relate (\ref{ur j3}) to the joint weak
convergence of the respective conditional cdf's in the case of rv's
$Z_{n}^{\prime},Z_{n}^{\prime\prime},Z^{\prime}$ and $Z^{\prime\prime}$.
Before that, in Lemma \ref{le kal}(a) we show how joint weak convergence can
be strengthened to a.s. weak convergence on a special probability space. For a
single convergence $Z_{n}^{\prime}|X_{n}^{\prime}\overset{w}{\rightarrow}%
_{w}Z^{\prime}|X^{\prime}$, part (a) implies that there exists a Skorokhod
representation $(\tilde{Z}_{n}^{\prime},\tilde{X}_{n}^{\prime})\overset{d}%
{=}({Z}_{n}^{\prime},{X}_{n}^{\prime})$, $(\tilde{Z}^{\prime},\tilde
{X}^{\prime})\overset{d}{=}({Z}^{\prime},{X}^{\prime})$ such that $\tilde
{Z}_{n}^{\prime}|\tilde{X}_{n}^{\prime}\overset{w}{\rightarrow}_{a.s.}%
\tilde{Z}^{\prime}|\tilde{X}^{\prime}$.

\begin{lemma}
\label{le kal}Let $(Z_{n}^{\prime},Z_{n}^{\prime\prime},Z_{n}^{\prime
\prime\prime},X_{n}^{\prime},X_{n}^{\prime\prime})$ and $(Z^{\prime}%
,Z^{\prime\prime},Z^{\prime\prime\prime},X^{\prime},X^{\prime\prime})$ be
random elements of the same Polish product space, defined on possibly
different probability spaces ($n\in\mathbb{N}$).

(a) If (\ref{ur j3})-(\ref{ur j4}) hold, then there exist a probability space
$(\tilde{\Omega},\mathcal{\tilde{F}},\tilde{P})$ and random elements
$(\tilde{X}_{n}^{\prime},\tilde{X}_{n}^{\prime\prime},\tilde{Z}_{n}^{\prime
},\tilde{Z}_{n}^{\prime\prime},\tilde{Z}_{n}^{\prime\prime\prime})\overset
{d}{=}(X_{n}^{\prime},X_{n}^{\prime\prime},Z_{n}^{\prime},Z_{n}^{\prime\prime
},Z_{n}^{\prime\prime\prime})$, $(\tilde{X}^{\prime},\tilde{X}^{\prime\prime
},\tilde{Z}^{\prime},\tilde{Z}^{\prime\prime},\tilde{Z}^{\prime\prime\prime
})\overset{d}{=}(X^{\prime},X^{\prime\prime},Z^{\prime},Z^{\prime\prime
},Z^{\prime\prime\prime})$ defined on this space such that $\tilde{Z}%
_{n}^{\prime}|\tilde{X}_{n}^{\prime}\overset{w}{\rightarrow}_{a.s.}\tilde
{Z}^{\prime}|\tilde{X}^{\prime},\tilde{Z}_{n}^{\prime\prime}|\tilde{X}%
_{n}^{\prime\prime}\overset{w}{\rightarrow}_{a.s.}\tilde{Z}^{\prime\prime
}|\tilde{X}^{\prime\prime}\ $and $\tilde{Z}_{n}^{\prime\prime\prime}%
\overset{a.s.}{\rightarrow}\tilde{Z}^{\prime\prime\prime}$.

(b) Let $Z^{\prime},Z^{\prime\prime}$ be rv's and $Z^{\prime\prime\prime}$ be
$\mathcal{S}^{\prime\prime\prime}$-valued. If the conditional distributions
$Z^{\prime}|X^{\prime}$ and $Z^{\prime\prime}|X^{\prime\prime}$ are diffuse,
then (\ref{ur j3})-(\ref{ur j4}) is equivalent to the weak convergence of the
associated random cdf's:
\begin{equation}
\left(  P(Z_{n}^{\prime}\leq\cdot|X_{n}^{\prime}),\,P(Z_{n}^{\prime\prime}%
\leq\cdot|X_{n}^{\prime\prime}),Z_{n}^{\prime\prime\prime}\right)  \overset
{w}{\rightarrow}\left(  P(Z^{\prime}\leq\cdot|X^{\prime}),\,P(Z^{\prime\prime
}\leq\cdot|X^{\prime\prime}),Z^{\prime\prime\prime}\right) \label{or cokal}%
\end{equation}
as random elements of $\mathscr{D}(\mathbb{R})\times\mathscr{D}(\mathbb{R)}%
\times\mathcal{S}^{\prime\prime\prime}$.
\end{lemma}

The definition of the convergence $Z_{n}|X_{n}\overset{w}{\rightarrow}_{w}Z|X$
implies that $h\left(  Z_{n}\right)  |X_{n}\overset{w}{\rightarrow}_{w}h(Z)|X$
for any continuous $h:\mathcal{S}_{Z}\rightarrow\mathcal{S}_{Z}^{\prime}$
between Polish spaces. A generalization for functions $h$ with a negligible
set of discontinuities is provided in the following CMT (for weak convergence
a.s. and weak convergence in probability, see Theorem 10 of Sweeting, 1989).

\begin{theorem}
\label{th cmt}Let $\mathcal{S}_{Z},\mathcal{S}_{Z}^{\prime},\mathcal{S}_{X}$
and $\mathcal{S}_{X}^{\prime}$ be Polish spaces and the random elements
$Z_{n},Z$ be $\mathcal{S}_{Z}$-valued, $X_{n}$ be $\mathcal{S}_{X}$-valued and
$X$ be $\mathcal{S}_{X}^{\prime}$-valued. If $Z_{n}|X_{n}\overset
{w}{\rightarrow}_{w}Z|X$ and $h:\mathcal{S}_{Z}\rightarrow\mathcal{S}%
_{Z}^{\prime}$ has its set of discontinuity points $D_{h}$ with $P(Z\in
D_{h}|X)=0$ a.s., then $h\left(  Z_{n}\right)  |X_{n}\overset{w}{\rightarrow
}_{w}h(Z)|X$.
\end{theorem}

Next, we prove in Theorem \ref{p citex} a weak convergence result for iterated
conditional expectations. The theorem provides conditions under which the
convergence $E(z_{n}|X_{n})\overset{w}{\rightarrow}E(z|X^{\prime}%
,X^{\prime\prime})$ implies, upon iteration of the expectations, that
$E\{E(z_{n}|X_{n})|X_{n}^{\prime}\}\overset{w}{\rightarrow}E\{E(z|X^{\prime
},X^{\prime\prime})|X^{\prime}\}$ for rv's $z_{n},z$ and for $X_{n}%
$-measurable $X_{n}^{\prime}$. In terms of weak convergence in distribution,
the result allows to pass from $Z_{n}|X_{n}\overset{w}{\rightarrow}%
_{w}Z|(X^{\prime},X^{\prime\prime})$ to $Z_{n}|X_{n}^{\prime}\overset
{w}{\rightarrow}_{w}Z|X^{\prime}$. The most transparent conclusions from
Theorem \ref{p citex} are that, (i), $(E(z_{n}|X_{n}),X_{n}^{\prime}%
)\overset{w}{\rightarrow}(E(z|X^{\prime}),X^{\prime})$ implies $E(z_{n}%
|X_{n}^{\prime})$ $\overset{w}{\rightarrow}$ $E(z|X^{\prime})$, and (ii), that
$((Z_{n}|X_{n}),X_{n}^{\prime})\overset{w}{\rightarrow}_{w}((Z|X^{\prime
}),X^{\prime})$ implies $Z_{n}|X_{n}^{\prime}$ $\overset{w}{\rightarrow}_{w}$
$Z|X^{\prime}$. We need, however, a more elaborate version for joint weak convergence.

\begin{theorem}
\label{p citex}For $n\in\mathbb{N}$, let $z_{n}$ be integrable rv's, $X_{n}%
$,$Y_{n}$ and $(X_{n}^{\prime},X_{n}^{\prime\prime})$ be random elements of
Polish spaces (say, $\mathcal{S}_{X},$ $\mathcal{S}_{Y}$ and $\mathcal{S}%
_{X}^{\prime}$), defined on the probability spaces $\left(  \Omega
_{n},\mathcal{F}_{n},P_{n}\right)  $ and such that $(X_{n}^{\prime}%
,X_{n}^{\prime\prime})$ are $X_{n}$-measurable $(n\in\mathbb{N})$. Let also
$z$ be an integrable rv and $Y,(X^{\prime},X^{\prime\prime})$ be random
elements of the Polish spaces $\mathcal{S}_{Y}$, $\mathcal{S}_{X}^{\prime}$
defined on a probability space $\left(  \Omega,\mathcal{F},P\right)  $. If
\begin{equation}
(E(z_{n}|X_{n}),X_{n}^{\prime},X_{n}^{\prime\prime},Y_{n})\overset
{w}{\rightarrow}(E(z|X^{\prime},X^{\prime\prime}),X^{\prime},X^{\prime\prime
},Y)\label{ur cocon}%
\end{equation}
and $X_{n}^{\prime\prime}|X_{n}^{\prime}\overset{w}{\rightarrow}_{w}%
X^{\prime\prime}|X^{\prime}$, then $E(z_{n}|X_{n}^{\prime})\overset
{w}{\rightarrow}E(z|X^{\prime})$ jointly with (\ref{ur cocon}).

Moreover, let $Z_{n}$, $Z$ be random elements of a Polish space $\mathcal{S}%
_{Z}$ defined resp. on $\left(  \Omega_{n},\mathcal{F}_{n},P_{n}\right)  $ and
$\left(  \Omega,\mathcal{F},P\right)  $. If
\begin{equation}
((Z_{n}|X_{n}),X_{n}^{\prime},X_{n}^{\prime\prime},Y_{n})\overset
{w}{\rightarrow}_{w}((Z|(X^{\prime},X^{\prime\prime})),X^{\prime}%
,X^{\prime\prime},Y)\label{ur j5}%
\end{equation}
and $X_{n}^{\prime\prime}|X_{n}^{\prime}\overset{w}{\rightarrow}_{w}%
X^{\prime\prime}|X^{\prime}$, then%
\begin{equation}
((Z_{n}|X_{n}^{\prime}),(Z_{n}|X_{n}),X_{n}^{\prime},X_{n}^{\prime\prime
},Y_{n})\overset{w}{\rightarrow}_{w}((Z|X^{\prime}),(Z|(X^{\prime}%
,X^{\prime\prime})),X^{\prime},X^{\prime\prime},Y).\label{ur j6}%
\end{equation}

\end{theorem}

\begin{remark}
\label{Remark 3.42}A special case with $X_{n}^{\prime\prime}=X^{\prime\prime
}=1$ and $Y_{n}=Y=1$ is that where $(E(z_{n}|X_{n}),X_{n}^{\prime})\overset
{w}{\rightarrow}(E(z|X^{\prime}),X^{\prime})$ such that $X_{n}^{\prime\prime
}|X_{n}^{\prime}\overset{w}{\rightarrow}_{w}X^{\prime\prime}|X^{\prime}$ is
trivial, and thus, as a result, $E(z_{n}|X_{n}^{\prime})\overset
{w}{\rightarrow}E(z|X^{\prime})$. In terms of conditional distributions,\ the
joint convergence $((Z_{n}|X_{n}),X_{n}^{\prime})\overset{w}{\rightarrow}%
_{w}((Z|X^{\prime}),X^{\prime})$ implies that $((Z_{n}|X_{n}^{\prime}%
),(Z_{n}|X_{n}),X_{n}^{\prime})\overset{w}{\rightarrow}_{w}((Z|X^{\prime
}),(Z|X^{\prime}),X^{\prime})$.
\end{remark}

\begin{remark}
\label{Remark bpv}Theorem \ref{p citex} can be applied to the bootstrap
\emph{p}-value. Let (\ref{ur cocon}) hold for $z_{n}=p_{n}^{\ast}$ and
$Y_{n}=Y=1$, and let $G^{\ast}$ be the conditional cdf of $p^{\ast}%
|(X^{\prime},X^{\prime\prime})$. If $E(G^{\ast}|X^{\prime})$ equals pointwise
the cdf of the $U(0,1)$ distribution, then the convergence $p_{n}^{\ast}%
|X_{n}^{\prime}\overset{w}{\rightarrow}_{w}p^{\ast}|X^{\prime}$ implied by
Theorem \ref{p citex} under the condition $X_{n}^{\prime\prime}|X_{n}^{\prime
}\overset{w}{\rightarrow}_{w}X^{\prime\prime}|X^{\prime}$ becomes $p_{n}%
^{\ast}|X_{n}^{\prime}\overset{w}{\rightarrow}_{p}U(0,1)$.$\hfill\square$
\end{remark}

We conclude the section with a result that is used for establishing the joint
convergence of original and bootstrap quantities as an implication of a
marginal and a conditional convergence.

\begin{lemma}
\label{t cmj}Let $(\Omega\times\Omega^{\ast},\mathcal{F}\times\mathcal{F}%
^{\ast},P\times P^{b})$ be a product probability space. Let $D_{n}%
:\Omega\rightarrow\mathcal{S}_{D}$, $W_{n}^{\ast}:\Omega^{\ast}\rightarrow
\mathcal{S}_{W}$, $X:\Omega\rightarrow\mathcal{S}_{X}$ and $Z^{\ast}%
:\Omega\times\Omega^{\ast}\rightarrow\mathcal{S}_{Z}$ ($n\in\mathbb{N} $) be
random elements of the Polish spaces $\mathcal{S}_{D}$, $\mathcal{S}_{W}$,
$\mathcal{S}_{X}=\mathcal{S}_{X}^{\prime}\times\mathcal{S}_{X}^{\prime\prime}$
and $\mathcal{S}_{Z}$. Assume further that $X_{n}$ are $D_{n}$-measurable
random elements of $\mathcal{S}_{X}$ and $Z_{n}$ are $(D_{n},W_{n}^{\ast}%
)$-measurable random elements of $\mathcal{S}_{Z}$ $(n\in\mathbb{N})$. If
$X_{n}\overset{p}{\rightarrow}X=(X^{\prime},X^{\prime\prime})$ and
$Z_{n}^{\ast}|D_{n}\overset{w}{\rightarrow}_{p}Z^{\ast}|X^{\prime}$, then
$(Z_{n}^{\ast},X_{n})\overset{w}{\rightarrow}(Z,X)$ and $(Z_{n}^{\ast}%
,X_{n})|D_{n}\overset{w}{\rightarrow}_{p}(Z,X)|X$ on $\mathcal{S}_{Z}%
\times\mathcal{S}_{X}$, where $Z$ is a random element of $\mathcal{S}_{Z}$
such that $Z^{\ast}|X^{\prime}\overset{d}{=}Z|X^{\prime}\overset{d}{=}Z|X$.
\end{lemma}

The existence of a random element $Z$ with the specified properties, possibly
on an extension of the original probability space, is ensured by Lemma 5.9 of
Kallenberg (1997). A well-known special case of Lemma \ref{t cmj} is that
where $X_{n}=(1,X_{n}^{\prime\prime})$, $X=(1,X^{\prime\prime})$,
$X_{n}^{\prime\prime}\overset{p}{\rightarrow}X^{\prime\prime}$ and
$Z_{n}^{\ast}|X_{n}^{\prime\prime}\overset{w}{\rightarrow}_{p}X^{\prime\prime
}$ (such that $D_{n}=X_{n}^{\prime\prime}$). Then $(Z_{n}^{\ast},X_{n}%
^{\prime\prime})\overset{w}{\rightarrow}(Z,X^{\prime\prime})$ with
$Z|X^{\prime\prime}\overset{d}{=}Z\overset{d}{=}X^{\prime\prime}$ reducing to
the condition that $X^{\prime\prime}$ and $Z$ are independent and distributed
like $X^{\prime\prime}$ (DasGupta, 2008, p.475).

\section{Proofs}

\label{sec all proofs}

\subsection{Proofs of the results in Section \ref{sec example}}

\label{sec Proofs of the results in Section <ref>sec g</ref>}

\noindent\textsc{Proof of Eq. (\ref{eq:xit})}. Let $\mathring{\varepsilon}%
_{t}:=\varepsilon_{t}-E(\varepsilon_{t}|\eta_{t})$, $t\in\mathbb{N}$. Then
$(\mathring{\varepsilon}_{t},E(\varepsilon_{t}|\eta_{t}),\eta_{t})^{\prime}$
is an i.i.d. sequence with diagonal covariance matrix $\operatorname*{diag}%
(\omega_{\varepsilon|\eta},1-\omega_{\varepsilon|\eta},1) $, $\omega
_{\varepsilon|\eta}:=E\{Var(\varepsilon_{t}|\eta_{t})\}\in(0,1)$, and it is a
standard fact that
\begin{equation}
n^{-1/2}(%
%TCIMACRO{\tsum _{t=1}^{\left\lfloor n\cdot\right\rfloor }}%
%BeginExpansion
{\textstyle\sum_{t=1}^{\left\lfloor n\cdot\right\rfloor }}
%EndExpansion
\mathring{\varepsilon}_{t},%
%TCIMACRO{\tsum _{t=1}^{\left\lfloor n\cdot\right\rfloor }}%
%BeginExpansion
{\textstyle\sum_{t=1}^{\left\lfloor n\cdot\right\rfloor }}
%EndExpansion
E(\varepsilon_{t}|\eta_{t}),%
%TCIMACRO{\tsum _{t=1}^{\left\lfloor n\cdot\right\rfloor }}%
%BeginExpansion
{\textstyle\sum_{t=1}^{\left\lfloor n\cdot\right\rfloor }}
%EndExpansion
\eta_{t})\overset{w}{\rightarrow}(\omega_{\varepsilon|\eta}^{1/2}%
B_{y1},(1-\omega_{\varepsilon|\eta})^{1/2}B_{y2},B_{\eta})\label{eq a}%
\end{equation}
in $\mathscr{D}_{3}$, where $(B_{y1},B_{y2},B_{\eta})$ is a standard Brownian
motion in $\mathbb{R}^{3}$. Further, by the conditional invariance principle
of Rubshtein (1996),
\begin{equation}
\left.  n^{-1/2}%
%TCIMACRO{\tsum _{t=1}^{\left\lfloor n\cdot\right\rfloor }}%
%BeginExpansion
{\textstyle\sum_{t=1}^{\left\lfloor n\cdot\right\rfloor }}
%EndExpansion
\mathring{\varepsilon}_{t}\right\vert
%TCIMACRO{\tsum _{t=1}^{\left\lfloor n\cdot\right\rfloor }}%
%BeginExpansion
{\textstyle\sum_{t=1}^{\left\lfloor n\cdot\right\rfloor }}
%EndExpansion
\eta_{t}\overset{w}{\rightarrow}_{p}\omega_{\varepsilon|\eta}^{1/2}%
B_{y1}\overset{d}{=}\left.  \omega_{\varepsilon|\eta}^{1/2}B_{y1}\right\vert
(B_{y2},B_{\eta})\label{eq b}%
\end{equation}
as a convergence of random measures on $\mathscr{D}$. Since $\sigma(\sum
_{t=1}^{\left\lfloor n\cdot\right\rfloor }\eta_{t})=\sigma(\sum_{t=1}%
^{\left\lfloor n\cdot\right\rfloor }E(\varepsilon_{t}|\eta_{t}),$ $\sum
_{t=1}^{\left\lfloor n\cdot\right\rfloor }\eta_{t})=\sigma(X_{n})$, the
convergence
\[
\left.  n^{-1/2}\left(
%TCIMACRO{\tsum _{t=1}^{\left\lfloor n\cdot\right\rfloor }}%
%BeginExpansion
{\textstyle\sum_{t=1}^{\left\lfloor n\cdot\right\rfloor }}
%EndExpansion
\mathring{\varepsilon}_{t},%
%TCIMACRO{\tsum _{t=1}^{\left\lfloor n\cdot\right\rfloor }}%
%BeginExpansion
{\textstyle\sum_{t=1}^{\left\lfloor n\cdot\right\rfloor }}
%EndExpansion
E(\varepsilon_{t}|\eta_{t}),%
%TCIMACRO{\tsum _{t=1}^{\left\lfloor n\cdot\right\rfloor }}%
%BeginExpansion
{\textstyle\sum_{t=1}^{\left\lfloor n\cdot\right\rfloor }}
%EndExpansion
\eta_{t}\right)  \right\vert X_{n}\overset{w}{\rightarrow}_{w}\left.
(\omega_{\varepsilon|\eta}^{1/2}B_{y1},(1-\omega_{\varepsilon|\eta}%
)^{1/2}B_{y2},B_{\eta})\right\vert (B_{y2},B_{\eta})
\]
follows from (\ref{eq a}) and (\ref{eq b})\ by Theorem 2.1\ of Crimaldi and
Pratelli (2005), for random measures on $\mathscr{D}_{3}$. Notice that
$(n^{-1}\sum_{t=1}^{n}\eta_{t}\mathring{\varepsilon}_{t},n^{-1}\sum_{t=1}%
^{n}\eta_{t}E(\varepsilon_{t}|\eta_{t}))\overset{p}{\rightarrow}0$ and the
convergence is preserved upon conditioning on $X_{n}$. Then, by using
conditional convergence to stochastic integrals (Theorem 3 of Georgiev et al.,
2018), it further follows that
\begin{align*}
& \left.  \left(  n^{-2}M_{n},n^{-1}%
%TCIMACRO{\tsum _{t=1}^{\left\lfloor n\cdot\right\rfloor }}%
%BeginExpansion
{\textstyle\sum_{t=1}^{\left\lfloor n\cdot\right\rfloor }}
%EndExpansion
x_{t}\mathring{\varepsilon}_{t},n^{-1}%
%TCIMACRO{\tsum _{t=1}^{\left\lfloor n\cdot\right\rfloor }}%
%BeginExpansion
{\textstyle\sum_{t=1}^{\left\lfloor n\cdot\right\rfloor }}
%EndExpansion
x_{t}E(\varepsilon_{t}|\eta_{t})\right)  \right\vert X_{n}\overset{}{=}\\
& \hspace{1.5cm}\left.  \left(  n^{-2}M_{n},n^{-1}%
%TCIMACRO{\tsum _{t=1}^{\left\lfloor n\cdot\right\rfloor }}%
%BeginExpansion
{\textstyle\sum_{t=1}^{\left\lfloor n\cdot\right\rfloor }}
%EndExpansion
x_{t-1}\mathring{\varepsilon}_{t}+o_{p}(1),n^{-1}%
%TCIMACRO{\tsum _{t=1}^{\left\lfloor n\cdot\right\rfloor }}%
%BeginExpansion
{\textstyle\sum_{t=1}^{\left\lfloor n\cdot\right\rfloor }}
%EndExpansion
x_{t-1}E(\varepsilon_{t}|\eta_{t})+o_{p}(1)\right)  \right\vert X_{n}\\
& \hspace{1.5cm}\overset{w}{\rightarrow}_{w}(M,\omega_{\varepsilon|\eta}%
^{1/2}M^{1/2}\xi_{1},(1-\omega_{\varepsilon|\eta})^{1/2}M^{1/2}\xi
_{2})|(B_{y2},B_{\eta})
\end{align*}
with $M:=\int B_{\eta}^{2}$, $\xi_{1}:=(\int B_{\eta}^{2})^{-1/2}\int B_{\eta
}dB_{y1}$, $\xi_{2}:=(\int B_{\eta}^{2})^{-1/2}$ $\int B_{\eta}dB_{y2}$
jointly independent and $\xi_{i}\sim N(0,1)$,$\,i=1,2$. Then, by Theorem
\ref{th cmt}, $\tau_{n}$ of (\textsc{\ref{eq:ucl}}) satisfies%
\begin{equation}
\left.  \left(  \tau_{n},n^{-2}M_{n},n^{-1}%
%TCIMACRO{\tsum _{t=1}^{n}}%
%BeginExpansion
{\textstyle\sum_{t=1}^{n}}
%EndExpansion
x_{t}E(\varepsilon_{t}|\eta_{t})\right)  \right\vert X_{n}\overset
{w}{\rightarrow}_{w}(\tau,M,(1-\omega_{\varepsilon|\eta})^{1/2}M^{1/2}\xi
_{2})|(B_{y2},B_{\eta})\label{ur samt}%
\end{equation}
with $\tau:=M^{-1/2}(\omega_{\varepsilon|\eta}^{1/2}\xi_{1}+(1-\omega
_{\varepsilon|\eta})^{1/2}\xi_{2})$. This yields (\ref{eq:xit}). The
bootstrap, instead of estimating consistently the limiting conditional
distribution of $\tau_{n}$ given $X_{n}$, estimates the random distribution
obtained by averaging this limit over $\xi_{2}$. As a result, conditionally on
$X_{n}$ the bootstrap \emph{p}-value is not asymptotically uniformly
distributed:
\begin{equation}
p_{n}^{\ast}|X_{n}=\left.  \Phi(\hat{\omega}_{\varepsilon}^{-1/2}M_{n}%
^{1/2}(\hat{\beta}-\beta))\right\vert X_{n}\overset{w}{\rightarrow}_{w}\left.
\Phi(\omega_{\varepsilon|\eta}^{1/2}\xi_{1}+(1-\omega_{\varepsilon|\eta
})^{1/2}\xi_{2})\right\vert \xi_{2},\label{eq:phi}%
\end{equation}
which is not the cdf of a $U(0,1)$ rv. $\hfill\square$

\subsection{Proofs of the results in Section \ref{sec g}}

\label{sec prg}

\noindent\textsc{Proof of Theorem \ref{th2}.} The result is a special case of
Theorem \ref{th2a} (proved later in an independent manner) with $\tau^{\ast
}=\tau$ and $F^{\ast}=F$. $\hfill\square$

\bigskip

\noindent\textsc{Proof of Theorem \ref{p2 copy(2)}.} The result (as well as
Corollary \ref{p2 part 2}) follows from Theorem \ref{p2}, which is proved
below in an independent manner. Specifically, as the conditions of Theorem
\ref{p2} are satisfied, it holds that $P(p_{n}^{\ast}\leq q|X_{n})\overset
{w}{\rightarrow}F( F^{\ast-1}(q)) $. Let $g(\cdot)=\min\{\cdot,1\}\mathbb{I}%
_{\{\cdot\geq0\}}$. By the definition of weak convergence,
\begin{align*}
P\left(  p_{n}^{\ast}\leq q\right)   & =E\{g(P(p_{n}^{\ast}\leq q|X_{n}%
))\}\overset{w}{\rightarrow}E\{g(F(F^{\ast-1}(q)))\}\\
& =E\{F(F^{\ast-1}(q))\}=E\{E[F(F^{\ast-1}(q))|F^{\ast}\}=E\{F^{\ast}%
(F^{\ast-1}(q))\}=q
\end{align*}
using for the penultimate equality the $F^{\ast}$-measurability of $F^{\ast
-1}(q)$ and the relation $E(F(\gamma)|F^{\ast})=F^{\ast}(\gamma)$ for
$F^{\ast}$-measurable rv's $\gamma$. Thus, $P(p_{n}^{\ast}\leq q)\overset
{}{\rightarrow}q$ for almost all $q\in\left(  0,1\right)  $, which proves that
$p_{n}^{\ast}\overset{w}{\rightarrow}U\left(  0,1\right)  $. $\hfill\square$

\bigskip

\noindent\textsc{Details of Remark }\ref{Remark exp}. We justify an assertion
of Remark \ref{Remark exp} regarding condition ($\dagger$). Let $\tilde{\tau
}_{n}^{\ast}$ be a measurable transformation of $X_{n}$ and $W_{n}^{\ast}$
such that the expansion $\tau_{n}^{\ast}=\tilde{\tau}_{n}^{\ast}+o_{p}(1)$
holds w.r.t. the probability measure on the space where $D_{n}$ and
$W_{n}^{\ast}$ are jointly defined. Then condition ($\dagger$) is satisfied
with $X_{n}^{\prime}:=P(\left.  \tilde{\tau}_{n}\leq\cdot\right\vert X_{n}%
)\in\mathscr{D}(\mathbb{R})$ and $X=F^{\ast}$, provided that $F^{\ast}$ is
sample-path continuous. In fact, under the assumed expansion, it holds that
$(\tau_{n}^{\ast}-\tilde{\tau}_{n}^{\ast})|D_{n}\overset{w}{\rightarrow}_{p}0$
because convergence in probability to zero is preserved upon conditioning. As
$\tilde{\tau}_{n}^{\ast}|D_{n}\overset{d}{=}\tilde{\tau}_{n}^{\ast}|X_{n}$, it
follows that the L\'{e}vy distance between $F_{n}^{\ast}(\cdot):=P\left(
\left.  \tau_{n}^{\ast}\leq\cdot\right\vert D_{n}\right)  $ and $X_{n}%
^{\prime}:=P(\left.  \tilde{\tau}_{n}\leq\cdot\right\vert X_{n})$ is
$o_{p}\left(  1\right)  $, and since the weak limit $F^{\ast}$ of $F_{n}%
^{\ast}$ is sample-path continuous, $F_{n}^{\ast}=X_{n}^{\prime}+o_{p}\left(
1\right)  $ in the uniform distance. Thus, also $X_{n}^{\prime}\overset
{w}{\rightarrow}F^{\ast}$.$\hfill\square$

\bigskip

\noindent\textsc{Proof of Corollary \ref{p2 part 2}.} Convergence
(\ref{eq coroll joint conv}) implies condition (\ref{eq joint conv}) with the
specified $F$, $F^{\ast}$ by Lemma \ref{le kal}, since the limit random
measures are diffuse. Part (a) follows from Theorem \ref{p2 copy(2)} with
$F=F^{\ast}$ (see Remark \ref{Remark exp}), and part (b)\ from Theorem
\ref{p2 copy(2)} with $F^{\ast}(u)=E(F(u)|X^{\prime})$, $u\in\mathbb{R}%
$.$\hfill\square$

\bigskip

\noindent\textsc{Proof of Theorem \ref{p2}.} Introduce $F_{n}\left(
\cdot\right)  :=P\left(  \left.  \tau_{n}\leq\cdot\right\vert X_{n}\right)  $,
$F_{n}^{\ast}\left(  \cdot\right)  :=P\left(  \left.  \tau_{n}^{\ast}\leq
\cdot\right\vert D_{n}\right)  $ and $\tilde{F}_{n}^{\ast}(\cdot):=P\left(
\left.  \tau_{n}^{\ast}\leq\cdot\right\vert X_{n}\right)  $ as random elements
of $\mathscr{D}(\mathbb{R})$. On the probability space of $X^{\prime}$,
possibly upon extending it, define $\tau^{\ast}:=F^{\ast-1}(\zeta)$ for a rv
$\zeta\sim U(0,1)$ which is independent of $X^{\prime}$. Then the convergence
$\left(  F_{n}^{\ast},X_{n}^{\prime}\right)  \overset{w}{\rightarrow}(F^{\ast
},X^{\prime})$, where $F^{\ast}$ is $X^{\prime}$-measurable and sample-path
continuous, implies that $\left(  (\tau_{n}^{\ast}|D_{n}),X_{n}^{\prime
}\right)  \overset{w}{\rightarrow}_{w}((\tau^{\ast}|X^{\prime}),X^{\prime})$
by Lemma \ref{le kal}(b). Since $X_{n}^{\prime}$ is $D_{n}$-measurable, by
Theorem \ref{p citex} (see also Remark \ref{Remark 3.42})\ it follows that
$(\tau_{n}^{\ast}|D_{n},~\tau_{n}^{\ast}|X_{n}^{\prime})\overset
{w}{\rightarrow}_{w}(\tau^{\ast}|X^{\prime},~\tau^{\ast}|X^{\prime})$. Since
the conditional cdf $F^{\ast}$ of $\tau^{\ast}|X^{\prime}$ is sample-path
continuous, for $r_{n}(\cdot):=F_{n}^{\ast}(\cdot)-P(\tau_{n}^{\ast}\leq
\cdot|X_{n}^{\prime})$ it follows that $\sup_{x\in\mathbb{R}}|r_{n}%
(x)|\overset{p}{\rightarrow}0$, by using Lemma \ref{le kal}(b). Then the
$D_{n}$-measurability of $X_{n}$, the $X_{n}$-measurability of $X_{n}^{\prime
}$ and Jensen's inequality yield%
\[
\left\vert \tilde{F}_{n}^{\ast}(u)-P(\tau_{n}^{\ast}\leq u|X_{n}^{\prime
})\right\vert =|E\{r_{n}(u)|X_{n}\}|\leq E\{|r_{n}(u)||X_{n}\}\leq
E\{\sup\nolimits_{x\in\mathbb{R}}|r_{n}(x)||X_{n}\}
\]
for every $u\in\mathbb{R}$, and further,%
\[
\sup\nolimits_{\mathbb{R}}\left\vert \tilde{F}_{n}^{\ast}-P(\tau_{n}^{\ast
}\leq\cdot|X_{n}^{\prime})\right\vert \leq E\{\sup\nolimits_{x\in\mathbb{R}%
}|r_{n}(x)||X_{n}\}\overset{p}{\rightarrow}0
\]
because the $o_{p}\left(  1\right)  $ property of $\sup_{x\in\mathbb{R}}%
|r_{n}(x)|$ is preserved upon conditioning and because $\sup_{x\in\mathbb{R}%
}|r_{n}(x)|$ is bounded. Therefore, $F_{n}^{\ast}=P(\tau_{n}^{\ast}\leq
\cdot|X_{n}^{\prime})+r_{n}=\tilde{F}_{n}^{\ast}+o_{p}\left(  1\right)  $
uniformly. Then the convergence $(F_{n},F_{n}^{\ast})\overset{w}{\rightarrow
}(F,F^{\ast})$ in $\mathscr{D}(\mathbb{R})^{\times2}$ extends to $(F_{n}%
,F_{n}^{\ast},\tilde{F}_{n}^{\ast})\overset{w}{\rightarrow}(F,F^{\ast}%
,F^{\ast})$ in $\mathscr{D}(\mathbb{R})^{\times3}$.

Fix a $q\in\left(  0,1\right)  $ at which $F^{\ast-1}$ is a.s. continuous;
such $q$ are all but countably many because $F^{\ast-1}$ is c\`{a}dl\`{a}g.
Here $F^{\ast-1}$ stands for the right-continuous generalized inverse of
$F^{\ast}$, and similarly for other cdf's. It follows from the CMT that
$(F_{n},F_{n}^{\ast-1}(q),\tilde{F}_{n}^{\ast-1}(q))\overset{w}{\rightarrow
}(F,F^{\ast-1}(q),F^{\ast-1}(q))$ in $\mathscr{D}\left(  \mathbb{R}\right)
\times\mathbb{R}^{2}$. Hence, $F_{n}^{\ast-1}(q)=\tilde{F}_{n}^{\ast
-1}(q)+o_{p}(1)$ such that $P(|F_{n}^{\ast-1}(q)-\tilde{F}_{n}^{\ast
-1}(q)|<\epsilon)\rightarrow1$ for all $\epsilon>0$. With $\mathbb{I}%
_{n,\epsilon}:=\mathbb{I}_{\{|F_{n}^{\ast-1}(q)-\tilde{F}_{n}^{\ast
-1}(q)|<\epsilon\}}=1+o_{p}(1)$, it holds that
\begin{align*}
|P(\tau_{n} &  \leq F_{n}^{\ast-1}(q)|X_{n})-P(\tau_{n}\leq\tilde{F}_{n}%
^{\ast-1}(q)+\epsilon|X_{n})|\\
&  \leq\mathbb{I}_{n,\epsilon}|P(\tau_{n}\leq\tilde{F}_{n}^{\ast
-1}(q)+\epsilon|X_{n})-P(\tau_{n}\leq\tilde{F}_{n}^{\ast-1}(q)-\epsilon
|X_{n})|+(1-\mathbb{I}_{n,\epsilon})\\
&  =\mathbb{I}_{n,\epsilon}|F_{n}(\tilde{F}_{n}^{\ast-1}(q)+\epsilon
)-F_{n}(\tilde{F}_{n}^{\ast-1}(q)-\epsilon)|+(1-\mathbb{I}_{n,\epsilon}),
\end{align*}
the equality because $\tilde{F}_{n}^{\ast-1}(q)\pm\epsilon$ are $X_{n}%
$-measurable. Using the continuity of $F$ and the CMT, we conclude that the
upper bound in the previous display converges weakly to $|F(F^{\ast
-1}(q)+\epsilon)-F(F^{\ast-1}(q)-\epsilon)|$, which in its turn converges in
probability to zero as $\epsilon\rightarrow0^{+}$ again by the continuity of
$F$. Therefore,
\[
\lim_{\epsilon\rightarrow0^{+}}\limsup_{n\rightarrow\infty}P\left(
|P(\tau_{n}\leq F_{n}^{\ast-1}(q)|X_{n})-P(\tau_{n}\leq\tilde{F}_{n}^{\ast
-1}(q)+\epsilon|X_{n})|>\eta\right)  =0
\]
for every $\eta>0$. On the other hand, as it was already used, by the $X_{n} $
measurability of $\tilde{F}_{n}^{\ast-1}(q)+\epsilon$ and the CMT,
\[
P(\tau_{n}\leq\tilde{F}_{n}^{\ast-1}(q)+\epsilon|X_{n})=F_{n}(\tilde{F}%
_{n}^{\ast-1}(q)+\epsilon)\underset{n\rightarrow\infty}{\overset
{w}{\longrightarrow}}F(F^{\ast-1}(q)+\epsilon)\underset{\epsilon
\rightarrow0^{+}}{\overset{w}{\longrightarrow}}F(F^{\ast-1}(q)).
\]
Theorem 4.2 of Billingsley (1968) thus yields $P(\tau_{n}\leq F_{n}^{\ast
-1}(q)|X_{n})\overset{w}{\rightarrow}F(F^{\ast-1}(q))$. The proof of
(\ref{eq:pis}) is concluded by noting that $P(p_{n}^{\ast}\leq q|X_{n})$
differs from $P(\tau_{n}\leq F_{n}^{\ast-1}(q)|X_{n})$ by no more than the
largest jump of $F_{n}^{\ast}$, which tends in probability to zero because the
weak limit of $F_{n}^{\ast}$ is continuous.

Asymptotic validity of the bootstrap conditional on $X_{n}$ requires that $F(
F^{\ast-1}(q)) =q$ for almost all $q\in(0,1),$ which by the continuity of $F$
and $F^{\ast}$ reduces to $F=F^{\ast}$.$\hfill\square$

\bigskip

\noindent\textsc{Proof of Corollary \ref{c1}}. Part (a) follows from Theorem
\ref{p2} with $F=F^{\ast}$ and Polya's theorem. Regarding part (b),
$((\tau_{n}|X_{n}),(\tau_{n}^{\ast}|D_{n}),X_{n}^{\prime},X_{n}^{\prime\prime
})\overset{w}{\rightarrow}_{w}(\tau|(X^{\prime},X^{\prime\prime}),$
$(\tau|X^{\prime}),X^{\prime},$ $X^{\prime\prime})$ and $X_{n}^{\prime\prime
}|X_{n}^{\prime}\overset{w}{\rightarrow}_{w}X^{\prime\prime}|X^{\prime}$
imply, by Theorem \ref{p citex} with $Y_{n}=E\{g(\tau_{n}^{\ast})|D_{n}\}$,
$Y=E\{g(\tau)|X^{\prime}\}$ and an arbitrary $g\in\mathcal{C}_{b}(\mathbb{R}%
)$, that $(\tau_{n}|X_{n}^{\prime},$ $\tau_{n}^{\ast}|D_{n})\overset
{w}{\rightarrow}_{w}(\tau|X^{\prime},$ $\tau|X^{\prime})$. As the conditional
distribution $\tau|X^{\prime}$ is diffuse, the proof is completed as in part
(a). $\hfill\square$

\bigskip

\noindent\textsc{Details of Remark \ref{Remark remalt}.} By extended Skorokhod
coupling (Corollary 5.12 of Kallenberg, 1997), consider a Skorokhod
representation of $D_{n}$ and $\left(  \tau,\tau^{\ast},X,X^{\prime}\right)  $
such that $(\tau_{n},\tau_{n}^{\ast},\phi_{n}(X_{n}),\psi_{n}(D_{n}%
))\overset{a.s.}{\rightarrow}(\tau,\tau^{\ast},X,$ $X^{\prime})$. Then, by
Lemma \ref{le crpr}(a), on the Skorokhod-representation space it holds that
$\tau_{n}|X_{n}\overset{w}{\rightarrow}_{p}\tau|X$ and $\tau_{n}^{\ast}%
|D_{n}\overset{w}{\rightarrow}_{p}\tau^{\ast}|X^{\prime}$, such that on a
general probability space $(\tau_{n}|X_{n},\tau_{n}^{\ast}|D_{n})\overset
{w}{\rightarrow}_{w}(\tau|X,\tau^{\ast}|X^{\prime})$. $\hfill\square$

\bigskip

\noindent\textsc{Details of Remark \ref{Remark 3.45}. }With $(X_{n}^{\prime
},X_{n}^{\prime\prime})$ and $(X^{\prime},X^{\prime\prime})$ as in Remark
\ref{Remark 3.45}, and with the notation of Section
\ref{sec Proofs of the results in Section <ref>sec g</ref>}, we argue next
that the weak convergence of $\tau_{n}|X_{n}$, $\tau_{n}^{\ast}|D_{n}$ and
$\left(  X_{n}^{\prime},X_{n}^{\prime\prime}\right)  $ is joint. Consider a
Skorokhod representation of $D_{n}$ and $(M,\xi_{1},\xi_{2})$ on a probability
space where convergence (\ref{ur samt}) is strengthened to $(\tau_{n}%
,X_{n}^{\prime},(X_{n}^{\prime})^{1/2}X_{n}^{\prime\prime})|X_{n}$
$\overset{w}{\rightarrow}_{a.s.}(\tau,M,(1-\omega_{\varepsilon|\eta}%
)M^{1/2}\xi_{2})|(M,\xi_{2})$ (by Lemma \ref{le kal}(a)), and $\hat{\omega
}_{\varepsilon}\overset{a.s.}{\rightarrow}\omega_{\varepsilon}$. Thus, on this
space, $\tau_{n}|X_{n}\overset{w}{\rightarrow}_{a.s.}\tau|(M,\xi_{2})$,
$\left(  X_{n}^{\prime},X_{n}^{\prime\prime}\right)  \overset{a.s.}%
{\rightarrow}(M,(1-\omega_{\varepsilon|\eta})\xi_{2})$, and by
(\ref{eq limit of BS measure - example}), also $P^{\ast}(\tau_{n}^{\ast}\leq
u)\overset{a.s.}{\rightarrow}\Phi(\omega_{\varepsilon}^{-1/2}M^{1/2}u)$,
$u\in\mathbb{R}$, such that $\tau_{n}^{\ast}|D_{n}\overset{w}{\rightarrow
}_{a.s.}\tau|M$. It follows that on a general probability space $((\tau
_{n}|X_{n}),(\tau_{n}^{\ast}|D_{n}),X_{n}^{\prime},X_{n}^{\prime\prime
})\overset{w}{\rightarrow}_{w}(\tau|(M,\xi_{2}),(\tau|M),M,(1-\omega
_{\varepsilon|\eta})\xi_{2}) $.$\hfill\square$

\bigskip

\noindent\textsc{Proof of Theorem \ref{th2a}}.\textsc{\ }The random element
$\left(  \tau,F^{\ast}\right)  $ of $\mathbb{R}\times\mathscr{D}(\mathbb{R}) $
is a measurable function of $\left(  \tau,X\right)  $ determined up to
indistinguishability by the joint distribution of $\left(  \tau,\tau^{\ast
},X\right)  $. By extended Skorokhod coupling (Corollary 5.12 of Kallenberg,
1997), we can regard the data and $(\tau,\tau^{\ast},X)$ as defined on a
special probability space where $\left(  \tau_{n},F_{n}^{\ast}\right)
\rightarrow\left(  \tau,F^{\ast}\right)  $ a.s. in $\mathbb{R}\times
\mathscr{D}(\mathbb{R})$ and $F^{\ast}(\cdot)=P(\tau^{\ast}\leq\cdot|X) $
still holds. We can also replace the redefined $F^{\ast}$ by a sample-path
continuous random cdf that it is indistinguishable from it (and maintain the
notation $F^{\ast}$).

Since $F^{\ast}$ is sample-path continuous and $F_{n}^{\ast},F^{\ast}$ are
(random) cdf's, $F_{n}^{\ast}\overset{a.s.}{\rightarrow}F^{\ast}$ in
$\mathscr{D}(\mathbb{R})$ implies that $\sup_{u\in\mathbb{R}}|F_{n}^{\ast
}(u)-F^{\ast}(u)|$$\overset{a.s.}{\rightarrow}0$. Therefore, $F_{n}^{\ast
}(\tau_{n})-F^{\ast}(\tau_{n})\overset{a.s.}{\rightarrow}0$. Since $\tau
_{n}\overset{a.s.}{\rightarrow}\tau$ and $F^{\ast}$ is sample-path continuous,
it holds further that $F^{\ast}(\tau_{n})\overset{a.s.}{\rightarrow}F^{\ast
}(\tau)$, so also $F_{n}^{\ast}(\tau_{n})\overset{a.s.}{\rightarrow}F^{\ast
}(\tau)$ on the special probability space. Hence, in general, $F_{n}^{\ast
}(\tau_{n})\overset{w}{\rightarrow}F^{\ast}(\tau)$.

We now find the cdf of $F^{\ast}(\tau)$. Again by continuity of $F^{\ast}$ and
by the choice of $F^{\ast-1}$ as the right-continuous inverse, the equality of
events $\{F^{\ast}(u)\leq q\}=\{u\leq F^{\ast-1}(q)\}$, $q\in(0,1)$, holds and
implies that
\[
P\left(  \left.  F^{\ast}(u)|_{u=\tau}\leq q\right\vert X\right)  =P\left(
\left.  \tau\leq F^{\ast-1}(q)\right\vert X\right)  =F\left(  F^{\ast
-1}(q)\right)  ,
\]
the latter equality because $F^{\ast-1}(q)$ is $X$-measurable. We conclude
that $F^{\ast}(\tau)$ has cdf $E\{F( F^{\ast-1}(q)) \}$, $q\in(0,1)$, as
asserted.$\hfill\square$

\subsection{Proofs of the results in Section \ref{Section on Applications}}

\label{sec Proofs of the results in Section applications}

\noindent\textsc{Proof of Eq. (\ref{eq:jbb})}. By extended Skorokhod coupling
(Corollary 5.12 of Kallenberg, 1997), we can regard the data and $X$ as
defined on a single probability space such that $n^{-\alpha/2}x_{\left\lfloor
n\cdot\right\rfloor }\overset{a.s.}{\rightarrow}X$ in $\mathscr{D}$. Then, by
a product-space construction, we can extend this space to define also an
i.i.d. standard Gaussian sequence $\{\varepsilon_{t}^{\ast}\}$ independent of
the data and (by Lemma 5.9 of Kallenberg, 1997), a random element $(W,b)$ of
$\mathscr{D}$$\times\mathbb{R}$ such that $(W,b)|X$ has the conditional
distribution specified in the text. Consider outcomes (say $\omega$) in the
factor-space of $n^{-\alpha/2}x_{\left\lfloor n\cdot\right\rfloor }$ such that
$(\text{$n^{-\alpha-1}M_{n}(\omega),n^{-\alpha/2-1}\xi_{n}(\omega))$%
}\rightarrow(M(\omega),\xi(\omega))$, $n^{-(\alpha+1)/2}\sup|x_{\left\lfloor
n\cdot\right\rfloor }(\omega)|\rightarrow0$ and $M(\omega)>0$; such outcomes
have probability one. For every such outcome, $(n^{1/2}W_{n}^{\ast}%
,M_{n}^{1/2}\hat{\beta}^{\ast})$ is tight in $\mathscr{D}$$\times\mathbb{R}$
because $n^{1/2}W_{n}^{\ast}$ and $M_{n}^{1/2}\hat{\beta}^{\ast}$ are tight in
$\mathscr{D}$ and $\mathbb{R}$ resp., and its finite-dimensional distributions
converge, by the multivariate Lyapunov CLT (Bentkus, 2005), to those of
$(W^{\omega},b^{\omega})$, where $W^{\omega}$ and $b^{\omega}$ are resp. a
standard Brownian bridge and a standard Gaussian rv with $\operatorname*{Cov}%
(W^{\omega}(u),b^{\omega})=M(\omega)^{-1/2}\xi(\omega)\psi(u)$, $u\in
\lbrack0,1]$. It follows by disintegration (Theorem 5.4 of Kallenberg, 1997)
that $(n^{1/2}W_{n}^{\ast},M_{n}^{1/2}\hat{\beta}^{\ast})|x_{\left\lfloor
n\cdot\right\rfloor }$$\overset{w}{\rightarrow}_{a.s.}(W,b)|X\overset{d}%
{=}(W,b)|(M,\xi)$, and further, that
\[
(n^{1/2}W_{n}^{\ast},M_{n}^{1/2}\hat{\beta}^{\ast},n^{-\alpha-1}%
M_{n},n^{-\alpha/2-1}\xi_{n})|x_{\left\lfloor n\cdot\right\rfloor }\overset
{w}{\rightarrow}_{a.s.}(W,b,M,\xi)|(M,\xi)
\]
by Lemma \ref{t cmj}, since $(\text{$n^{-\alpha-1}M_{n},n^{-\alpha/2-1}\xi
_{n})$ are }x_{\left\lfloor n\cdot\right\rfloor }$-measurable. Still further,
by a CMT for a.s. weak convergence (Theorem 10 of Sweeting, 1989),
\[
(n^{1/2}W_{n}^{\ast},\,n^{(\alpha+1)/2}\hat{\beta}^{\ast},\,n^{-\alpha/2-1}%
\xi_{n})|x_{\left\lfloor n\cdot\right\rfloor }\overset{w}{\rightarrow}%
_{a.s.}(W,M^{-1/2}b,\xi)|(M,\xi)
\]
on the special probability space. This implies (\ref{eq:jbb}) on a general
probability space.$\hfill\square$

\bigskip{}

\noindent\textsc{Proof of Eq. (\ref{eq tri})}. Under $\mathsf{H}_{0}$, by
extended Skorokhod coupling (Corollary 5.12 of Kallenberg, 1997), we regard
the data and $\left(  X,\tau\right)  $ as defined on a single probability
space such that $(n^{-\alpha/2}x_{\left\lfloor n\cdot\right\rfloor },\tau
_{n})\overset{a.s.}{\rightarrow}(X,\tau)$ in $\mathscr{D}\times\mathbb{R}$,
and which is extended to support the independent bootstrap sequence
$\{\varepsilon_{t}^{\ast}\}$ and $\left(  W,b\right)  $ such that $\left(
W,b\right)  |X$ has the conditional distribution specified in the text. We
have by the same argument as for eq. (\ref{eq:tst}) that, on this space,
$\tau_{n}^{\ast}\overset{w^{\ast}}{\rightarrow}_{p}\tau|(M,\xi)$ so that
$F_{n}^{\ast}(\cdot):=P(\tau_{n}^{\ast}\leq\cdot|D_{n})\overset{p}%
{\rightarrow}F(\cdot):=P(\tau\leq\cdot|M,\xi)$ in $\mathscr{D}(\mathbb{R})$,
because $F$ is sample-path continuous (\textit{e.g.}, by Proposition 3.2 of
Linde, 1989, applied conditionally on $M,\xi$). As further $\tau_{n}%
\overset{a.s.}{\rightarrow}\tau$ on this space, we can collect the previous
convergence facts into $(\tau_{n},F_{n}^{\ast})\overset{p}{\rightarrow}%
(\tau,F)$, which proves that on a general probability space eq. (\ref{eq tri}) holds.

Notice that (i) $(\tau_{n},n^{-\alpha-1}M_{n},n^{-\alpha/2-1}\xi_{n}%
)\overset{a.s.}{\rightarrow}\left(  \tau,M,\xi\right)  $, as implied by the
a.s. convergence of $\tau_{n}$ and $n^{-\alpha/2}x_{\left\lfloor
n\cdot\right\rfloor }$, and (ii) $\tau_{n}^{\ast}\overset{w^{\ast}%
}{\rightarrow}_{p}\tau|(M,\xi)$, jointly imply, by Lemma \ref{t cmj}, that
$(\tau_{n},\tau_{n}^{\ast},n^{-\alpha-1}M_{n},n^{-\alpha/2-1}\xi_{n}%
)\overset{w}{\rightarrow}\left(  \tau,\tau^{\ast},M,\xi\right)  $ with
$\tau|(M,\xi)\overset{d}{=}\tau^{\ast}|(M,\xi)$. The latter convergence
remains valid on general probability spaces and is used in the discussion of
conditional bootstrap validity.$\hfill\square$

\bigskip{}

\noindent\textsc{Proof of theorem \ref{Lemma bootstrap with boundary}}.
Introduce $\tilde{x}_{t}:=(1,x_{n,t-1})^{\prime}$. Let $\mu_{n}:=n^{1/2}%
(\hat{\theta}-\theta_{0})$ and $N_{n}^{\ast}:=n^{-1/2}\sum_{t=1}%
^{n}\varepsilon_{t}^{\ast}\tilde{x}_{t}$. Moreover, let the normalized
bootstrap estimator be denoted by $\mu_{n}^{\ast}:=n^{1/2}(\hat{\theta}^{\ast
}-\hat{\theta})$; similarly, $\tilde{\mu}_{n}^{\ast}:=n^{1/2}(\tilde{\theta
}^{\ast}-\hat{\theta})$, where $\tilde{\theta}^{\ast}$is the unrestricted
(OLS) bootstrap estimator.

As in Theorem 3 of Georgiev \textit{et al.} (2018), it follows,
\textit{mutatis mutandis}, that $(\mu_{n},M_{n},N_{n}^{\ast})\overset{w^{\ast
}}{\rightarrow}_{w}(\ell(\theta_{0}),M,M^{1/2}\xi^{\ast})|(M,\ell(\theta
_{0}))$ in $\mathbb{R}{}^{2\times4}$, where $M$ is of full rank with
probability one, $\xi^{\ast}|(M,\ell(\theta_{0}))\sim N(0,\sigma_{e}^{2}%
I_{2})$ and $\sigma_{e}^{2}$ denotes the variance of $\varepsilon_{t}$
corrected for $\Delta x_{n,t}$. To derive the main result, we analyze the
properties of $\mu_{n}^{\ast}$ on a special probability space where $(\mu
_{n},M_{n},N_{n}^{\ast})$ given the data converge weakly a.s. rather than
weakly in distribution. Specifically, by Lemma \ref{le kal}(a) we can consider
a probability space (where $\ell(\theta_{0}),M$ and, for every $n\in
\mathbb{N}$, also the original data and the bootstrap sample can be redefined,
maintaining their distribution), such that
\begin{equation}
\mu_{n}\overset{a.s.}{\rightarrow}\ell(\theta_{0})\text{, }M_{n}\overset
{a.s.}{\rightarrow}M\text{, }N_{n}^{\ast}\overset{w^{\ast}}{\rightarrow
}_{a.s.}M^{1/2}\xi^{\ast}|(M,\ell(\theta_{0}))\overset{d}{=}M^{1/2}\xi^{\ast
}|M\text{.}\label{eq weak conv in prob}%
\end{equation}

Let $q_{n}^{\ast}\left(  \theta\right)  :=n^{-1}\sum_{t=1}^{n}(y_{t}^{\ast
}-\theta^{\prime}\tilde{x}_{t})^{2}$ and notice that $\tilde{\theta}^{\ast
}:=\arg\min_{\theta\in\mathbb{R}^{2}}q_{n}^{\ast}\left(  \theta\right)  $ is
such that $\tilde{\mu}_{n}^{\ast}:=n^{1/2}(\tilde{\theta}^{\ast}-\hat{\theta
})=M_{n}^{-1}N_{n}^{\ast}$. On the special probability space, the asymptotic
distribution of $\tilde{\mu}_{n}^{\ast}$ follows from
(\ref{eq weak conv in prob}) and a CMT (Theorem 10 of Sweeting, 1989) as%
\begin{equation}
\tilde{\mu}_{n}^{\ast}\overset{w^{\ast}}{\rightarrow}_{a.s.}\tilde{\ell}%
^{\ast}|(M,\ell(\theta_{0}))\overset{d}{=}\tilde{\ell}^{\ast}|M\text{, }%
\tilde{\ell}^{\ast}:=\sigma_{e}^{2}M^{-1/2}\xi^{\ast}\text{.}%
\label{eq separata}%
\end{equation}

Let us turn to the bootstrap estimator $\hat{\theta}^{\ast}$. If $g(\theta
_{0})>g^{\ast}(\theta_{0})$, then the consistency facts $\hat{\theta}%
\overset{a.s.}{\rightarrow}\theta_{0}$ (from (\ref{eq weak conv in prob})) and
$\tilde{\theta}^{\ast}\overset{w^{\ast}}{\rightarrow}_{a.s.}\theta_{0}$ (from
(\ref{eq weak conv in prob})-(\ref{eq separata})), jointly with the continuity
of $g,g^{\ast}$ at $\theta_{0}$, imply that $P^{\ast}(g(\tilde{\theta}^{\ast
})\geq g^{\ast}(\hat{\theta}))\overset{a.s.}{\rightarrow}1$. Hence, $P^{\ast
}(\hat{\theta}^{\ast}=\tilde{\theta}^{\ast})\overset{a.s.}{\rightarrow}1$ and
$P^{\ast}(\mu_{n}^{\ast}=\tilde{\mu}_{n}^{\ast})\overset{a.s.}{\rightarrow}1$.
Using also (\ref{eq separata}), it follows that ${\mu}_{n}^{\ast}\overset
{w}{\rightarrow}_{a.s.}\tilde{\ell}^{\ast}|M$ on the special probability
space, and since $\mu_{n}\overset{a.s.}{\rightarrow}\ell(\theta_{0})$ on this
space, it follows further that $(\mu_{n},({\mu}_{n}^{\ast}|D_{n}))\overset
{w}{\rightarrow}_{w}(\ell(\theta_{0}),(\tilde{\ell}^{\ast}|M))$ on a general
probability space, as asserted in (\ref{eq semno}).

In the case $g^{\ast}(\theta_{0})=g(\theta_{0})$, it still holds that
$\hat{\theta}^{\ast}=\tilde{\theta}^{\ast}$ whenever $g(\tilde{\theta}^{\ast
})\geq g^{\ast}(\hat{\theta})$. However, the probability of the latter event
no longer tends to one. Whenever $g(\tilde{\theta}^{\ast})<g^{\ast}%
(\hat{\theta})$, the estimator $\hat{\theta}^{\ast}$ exists if and only if the
bootstrap estimator restricted to the bootstrap boundary, say $\check{\theta
}^{\ast}$, exists. Let $\mathbb{I}_{n}^{\ast}:=\mathbb{I}_{\mathbb{\{}%
h(\tilde{\theta}_{n}^{\ast})\geq0\}}$ with $h(\theta):=g(\theta)-g^{\ast}%
(\hat{\theta})$. In order to justify the equality
\begin{equation}
\hat{\theta}^{\ast}=\tilde{\theta}^{\ast}\mathbb{I}_{n}^{\ast}+\check{\theta
}^{\ast}(1-\mathbb{I}_{n}^{\ast}%
)\label{eq decomposition for the restricted estimator}%
\end{equation}
in an appropriate sense, we show in Section \ref{se wedge} of the accompanying
Supplement that $\check{\theta}^{\ast}(1-\mathbb{I}_{n}^{\ast})$ is
well-defined with $P^{\ast}$-probability approaching one a.s. We also show
there that $\Vert\check{\theta}^{\ast}-\hat{\theta}\Vert(1-\mathbb{I}%
_{n}^{\ast})=O_{p^{\ast}}(n^{-1/2})$ a.s., and as a result, $\Vert\hat{\theta
}^{\ast}-\hat{\theta}\Vert=O_{p^{\ast}}(n^{-1/2})$ a.s., using also
(\ref{eq separata}). We do not discuss the uniqueness of $\check{\theta}%
^{\ast}$ but we show instead that the measurable minimizers of $q_{n}^{\ast
}\left(  \theta\right)  $ over the bootstrap boundary are asymptotically
equivalent, as they give rise to the same asymptotic distribution for
$\hat{\theta}^{\ast}$.

To see this, we further establish in the Supplement that $\check{\theta}%
^{\ast}$ satisfies a first-order condition [foc] with $P^{\ast}$-probability
approaching one a.s. Let dots over function names denote differentiation
w.r.t. $\theta$ (\textit{e.g.}, $\dot{q}_{n}^{\ast}(\theta):=(\partial\dot
{q}_{n}^{\ast}/\partial\theta^{\prime})(\theta)$, a column vector). Then the
foc takes the form
\[
\{\dot{q}_{n}^{\ast}(\check{\theta}^{\ast})+\check{\delta}_{n}\dot{h}%
(\check{\theta}^{\ast})\}(1-\mathbb{I}_{n}^{\ast})=\{\dot{q}_{n}^{\ast}%
(\check{\theta}^{\ast})+\check{\delta}_{n}\dot{g}(\check{\theta}^{\ast
})\}(1-\mathbb{I}_{n}^{\ast})=0\text{, }h(\check{\theta}^{\ast})=0,
\]
where $\check{\delta}_{n}\in\mathbb{R}$. The foc implies, by means of a
standard argument, that we can choose a measurable $\bar{\theta}^{\ast}$
between $\check{\theta}^{\ast}$ and $\hat{\theta}$ such that%
\[
\{n^{1/2}(\check{\theta}^{\ast}-\hat{\theta})-(I_{2}-A_{n}^{\ast}\dot{g}%
(\bar{\theta}^{\ast})^{\prime})\tilde{\mu}_{n}^{\ast}-A_{n}^{\ast}%
n^{1/2}h(\hat{\theta})\}(1-\mathbb{I}_{n}^{\ast})=0,
\]
where $A_{n}^{\ast}:=M_{n}^{-1}\dot{g}(\check{\theta}^{\ast})[\dot{g}%
(\bar{\theta}^{\ast})^{\prime}M_{n}^{-1}\dot{g}(\check{\theta}^{\ast})]^{-1}$.
As further $\Vert\check{\theta}^{\ast}-\hat{\theta}\Vert(1-\mathbb{I}%
_{n}^{\ast})=O_{p^{\ast}}(n^{-1/2})$ a.s., $\Vert\bar{\theta}^{\ast}%
-\hat{\theta}\Vert(1-\mathbb{I}_{n}^{\ast})=O_{p^{\ast}}(n^{-1/2})$ a.s. and
$\hat{\theta}-\theta_{0}=O(n^{-1/2})$ a.s., using the continuity of $\dot
{g}(\theta)$ at $\theta_{0}$ it follows that
\[
\{n^{1/2}(\check{\theta}^{\ast}-\hat{\theta})-[(I_{2}-A^{\ast}\dot{g}^{\prime
})\tilde{\mu}_{n}^{\ast}-A^{\ast}(\dot{g}-\dot{g}^{\ast})^{\prime}n^{1/2}%
(\hat{\theta}-\theta_{0})]\}(1-\mathbb{I}_{n}^{\ast})=o_{p^{\ast}%
}(1)~~\text{a.s.,}%
\]
where $A^{\ast}:=M^{-1}\dot{g}[\dot{g}^{\prime}M^{-1}\dot{g}]^{-1}$ and
$P^{\ast}(|o_{p^{\ast}}(1)|>\eta)\overset{a.s.}{\rightarrow}1$ for all
$\eta>0$.

Returning to (\ref{eq decomposition for the restricted estimator}), we
conclude that%
\begin{equation}
n^{1/2}(\hat{\theta}^{\ast}-\hat{\theta})=\tilde{\mu}_{n}^{\ast}\mathbb{I}%
_{n}^{\ast}+\{(I_{2}-A^{\ast}\dot{g}^{\prime})\tilde{\mu}_{n}^{\ast}-A^{\ast
}(\dot{g}-\dot{g}^{\ast})^{\prime}n^{1/2}(\hat{\theta}-\theta_{0}%
)\}(1-\mathbb{I}_{n}^{\ast})+o_{p^{\ast}}(1)~~\text{a.s.}\label{eq dithe}%
\end{equation}
Consider the event indicated by $\mathbb{I}_{n}^{\ast}$. As $\Vert\hat{\theta
}^{\ast}-\hat{\theta}\Vert=O_{p^{\ast}}(n^{-1/2})$ a.s. and $\hat{\theta
}-\theta_{0}=O(n^{-1/2})$ a.s., by the mean value theorem and the continuous
differentiability of $g$, $g^{\ast}$ it holds that%
\[
n^{1/2}h(\tilde{\theta}^{\ast})=\dot{g}^{\prime}\tilde{\mu}_{n}^{\ast}%
+(\dot{g}-\dot{g}^{\ast})^{\prime}\mu_{n}+c_{n},
\]
where $c_{n}=o_{p^{\ast}}(1)$ a.s. Then $\mathbb{I}_{n}\overset{w^{\ast}%
}{\rightarrow}_{a.s.}\mathbb{I}_{\infty}|(M,\ell(\theta_{0}))$ with
$\mathbb{I}_{\infty}:=\mathbb{I}_{\{\dot{g}^{\prime}\tilde{\ell}^{\ast}%
\geq(\dot{g}^{\ast}-\dot{g})^{\prime}\ell(\theta_{0})\}}$, by
(\ref{eq weak conv in prob})-(\ref{eq separata}) and the CMT for weak a.s.
convergence (Theorem 10 of Sweeting, 1989), as the probability of the limiting
discontinuities is $0$: $P(\dot{g}^{\prime}\tilde{\ell}^{\ast}=(\dot{g}^{\ast
}-\dot{g})^{\prime}\ell(\theta_{0})|(M,\ell(\theta_{0})))=0$ a.s. By exactly
the same facts, passage to the limit directly in (\ref{eq dithe}) yields%
\[
n^{1/2}(\hat{\theta}^{\ast}-\hat{\theta})\overset{w^{\ast}}{\rightarrow
}_{a.s.}\{\tilde{\ell}^{\ast}\mathbb{I}_{\infty}+\check{\ell}^{\ast
}(1-\mathbb{I}_{\infty})\}|(M,\ell(\theta_{0}))\text{, }\check{\ell}^{\ast
}:=(I_{2}-A^{\ast}\dot{g}^{\prime})\tilde{\ell}^{\ast}-A^{\ast}(\dot{g}%
-\dot{g}^{\ast})^{\prime}\ell
\]
on the special probability space, where also $\mu_{n}\overset{a.s.}%
{\rightarrow}\ell(\theta_{0})$ by (\ref{eq weak conv in prob}). Therefore, on
a general probability space it holds that%
\[
(\mu_{n},(n^{1/2}(\hat{\theta}^{\ast}-\hat{\theta})|D_{n}))\overset{w^{\ast}%
}{\rightarrow}_{w}(\ell(\theta_{0}),[\{\tilde{\ell}^{\ast}\mathbb{I}_{\infty
}+\check{\ell}^{\ast}(1-\mathbb{I}_{\infty})\}|(M,\ell(\theta_{0}))]).
\]
The proof is completed by checking directly that, as asserted in
(\ref{eq amans}), $\tilde{\ell}^{\ast}\mathbb{I}_{\infty}+\check{\ell}^{\ast
}(1-\mathbb{I}_{\infty})$ equals $\arg\min_{\{\dot{g}^{\prime}\lambda\geq
(\dot{g}^{\ast}-\dot{g})^{\prime}\ell\}}||\lambda-\tilde{\ell}^{\ast}||_{M}$
a.s.$\hfill\square$

\bigskip

{}

\noindent\textsc{Proof of Theorem \ref{th h}}. Let $(M_{n},\tilde{V}%
_{n}):=(n^{-1}\sum_{t=1}^{\left\lfloor n\cdot\right\rfloor }x_{nt}%
x_{nt}^{\prime},n^{-1}\sigma^{-2}\sum_{t=1}^{\left\lfloor n\cdot\right\rfloor
}x_{nt}x_{nt}^{\prime}\tilde{e}_{nt}^{2})$. As $\tilde{V}_{n}=n^{-1}%
\sigma^{-2}\sum_{t=1}^{\left\lfloor n\cdot\right\rfloor }x_{nt}x_{nt}^{\prime
}\varepsilon_{nt}^{2}+o_{p}(1)$ under $\mathsf{H_{0}}$ and Assumption
$\mathcal{H}$, it further holds that $(M_{n},\tilde{V}_{n})\overset
{w}{\rightarrow}(M,V)$ in $\mathscr{D}{}_{m\times m}\mathscr{\times
D}{}_{m\times m}$. The data $D_{n}:=\{x_{nt},y_{nt}\}_{t=1}^{n}$ and the
bootstrap multipliers $\{w_{t}^{\ast}\}_{t\in\mathbb{N}}$ can be regarded
(upon padding with zeroes) as defined on the Polish space $(\mathbb{R^{\infty
}})^{k+2}$. Therefore, by Corollary 5.12 of Kallenberg (1997), there exists a
special probability space where $(M,V)$, and for every $n\in\mathbb{N}$, also
the original and the bootstrap data can be redefined, maintaining their
distribution (we also maintain the notation), such that $(M_{n},\tilde{V}%
_{n})$$\overset{a.s.}{\rightarrow}$$(M,V)$.

Consider $N_{n}^{\ast}:=n^{-1/2}\sigma^{-1}\sum_{t=1}^{\left\lfloor
n\cdot\right\rfloor }x_{nt}y_{t}^{\ast}$. As $N_{n}^{\ast}$ conditionally on
the data is a zero-mean Gaussian process with independent increments and
variance function $\tilde{V}_{n}$, the argument for Theorem 5 of Hansen (2000)
yields the conditional convergence
\begin{equation}
\left.  N_{n}^{\ast}\right\vert D_{n}\overset{d}{=}\left.  N_{n}^{\ast
}\right\vert (M_{n},\tilde{V}_{n})\overset{w}{\rightarrow}_{a.s.}%
N|(M,V)\label{eq x}%
\end{equation}
on the special probability space. The marginal convergence $(M_{n},\tilde
{V}_{n})\overset{a.s.}{\rightarrow}(M,V)$ in $(\mathscr{D}{}_{m\times m})^{2}$
and (\ref{eq x}) jointly imply, by Lemma \ref{t cmj}, that
\begin{equation}
\left.  (M_{n},\tilde{V}_{n},N_{n}^{\ast})\right\vert D_{n}\overset{d}%
{=}\left.  (M_{n},\tilde{V}_{n},N_{n}^{\ast})\right\vert (M_{n},\tilde{V}%
_{n})\overset{w}{\rightarrow}_{p}\left.  \left(  M,V,N\right)  \right\vert
(M,V)\label{eq ch}%
\end{equation}
as a convergence of random measures on $(\mathscr{D}{}_{m\times m})^{2}%
\times\mathscr{D}{}_{m}$.

The proof is completed as in Theorems 5 and 6 of Hansen (2000), by using the
following uniform expansion in $r\in\lbrack\underline{r},\overline{r}]:$
$F_{\left\lfloor nr\right\rfloor }^{\ast}=\tilde{F}_{n}(r)$$+o_{p}(1)$ with
\[
\tilde{F}_{n}(r)=\left\Vert (M_{n}(r)-M_{n}(r)M_{n}(1)^{-1}M_{n}%
(r))^{-1/2}(N^{\ast}(r)-M_{n}(r)M_{n}(1)^{-1}N_{n}^{\ast}(1))\right\Vert ^{2}%
\]
and where convergence is w.r.t. the joint measure over the original and the
bootstrap data. As $\tilde{F}_{n}(r)$ depends on the data only through
$M_{n},\tilde{V}_{n}$, it follows that
\[
P^{\ast}(\max_{r\in\lbrack\underline{r},\overline{r}]}\tilde{F}_{n}%
(r)\leq\cdot)=P(\max_{r\in\lbrack\underline{r},\overline{r}]}\tilde{F}%
_{n}(r)\leq\cdot|M_{n},\tilde{V}_{n}),
\]
and since $\{\max_{r\in\lbrack\underline{r},\overline{r}]}\tilde{F}%
_{n}(r)\}|(M_{n},\tilde{V}_{n})\overset{w}{\rightarrow}_{p}\mathscr{F}|(M,V) $
by (\ref{eq ch}) and a CMT for weak convergence in probability (Theorem 10 of
Sweeting, 1989), with
\[
\mathscr{F}:=\sup_{r\in\lbrack\underline{r},\overline{r}]}\{\tilde
{N}(r)^{\prime}\tilde{M}\left(  1\right)  ^{-1}\tilde{N}(r)\}
\]
as in of eq. (\ref{eq asy distr of supF}), also\ $\max_{r\in\lbrack
\underline{r},\overline{r}]}\tilde{F}_{n}(r)\overset{w^{\ast}}{\rightarrow
}_{p}\mathscr{F}|(M,V)$. Finally, as $\mathscr{F}_{n}^{\ast}:=\max
_{r\in\lbrack\underline{r},\overline{r}]}F_{\left\lfloor nr\right\rfloor
}^{\ast} $ $=\max_{r\in\lbrack\underline{r},\overline{r}]}\tilde{F}%
_{n}(r)+o_{p}(1)$ and `$(\cdot)\overset{p}{\rightarrow}0$' becomes
`$(\cdot)\overset{w^{\ast}}{\rightarrow}_{p}0$' upon conditioning on the data,
we conclude that $\mathscr{F}_{n}^{\ast}\overset{w^{\ast}}{\rightarrow}%
_{p}\mathscr{F}|(M,V)$ on the special probability space. Then $\mathscr{F}_{n}%
^{\ast}\overset{w^{\ast}}{\rightarrow}_{w}\mathscr{F}|(M,V) $ in
general.$\hfill\square$

\bigskip{}

\noindent\textsc{Proof of Theorem \ref{th fb}}. Additionally to the notation
introduced in the proof of Theorem \ref{th h}, let $V_{n}:=n^{-1}\sigma
^{-2}\sum_{t=1}^{\left\lfloor n\cdot\right\rfloor }x_{nt}x_{nt}^{\prime
}\varepsilon_{nt}^{2}$ and $X_{n}:=\{x_{nt}\}_{t=1}^{n}.$ Under Assumption
$\mathcal{H}$, by Corollary 5.12 of Kallenberg (1997), consider a single
probability space where, for every $n\in\mathbb{N},$ the original and the
bootstrap data are redefined together with $(M,V,N)$, maintaining their
distribution (we also maintain the notation), such that
\begin{equation}
\left(  M_{n},V_{n},\tilde{V}_{n},\tfrac{1}{n^{1/2}\sigma}%
%TCIMACRO{\tsum _{t=1}^{\left\lfloor n\cdot\right\rfloor }}%
%BeginExpansion
{\textstyle\sum_{t=1}^{\left\lfloor n\cdot\right\rfloor }}
%EndExpansion
x_{nt}\varepsilon_{nt},\mathscr{F}_{n}\right)  \overset{a.s.}{\rightarrow
}\left(  M,V,V,N,\mathscr{F}\right) \label{eq mc}%
\end{equation}
in $(\mathscr{D}{}_{m\times m})^{3}\times\mathscr{D}{}_{m}\times\mathbb{R}$,
with $\mathscr{F}:=\sup_{r\in\lbrack\underline{r},\overline{r}]}\{\tilde
{N}(r)^{\prime}\tilde{M}\left(  r\right)  ^{-1}\tilde{N}(r)\}$ of eq.
(\ref{eq asy distr of supF}). On this space also $\mathscr{F}_{n}^{\ast
}\overset{w^{\ast}}{\rightarrow}_{p}\mathscr{F}|(M,V)$ holds, by the proof of
Theorem \ref{th h}, or equivalently, $P^{\ast}(\mathscr{F}_{n}^{\ast}\leq
\cdot)\overset{p}{\rightarrow}P(\mathscr{F}\leq\cdot|M,V)$ in
$\mathscr{D}(\mathbb{R})$, given that sample-path continuity of the
conditional cdf $P(\mathscr{F}\leq\cdot|M,V)$ is guaranteed by Proposition 3.2
of Linde (1989) applied conditionally on $M,V$. We see that $(\mathscr{F}_{n}%
,P(\mathscr{F}_{n}^{\ast}\leq\cdot|D_{n}))\overset{p}{\rightarrow
}(\mathscr{F}_{{}},P(\mathscr{F}\leq\cdot|M,V))$ on the special probability
space. This implies that $(\mathscr{F}_{n},P^{\ast}(\mathscr{F}_{n}^{\ast}%
\leq\cdot))\overset{w}{\rightarrow}(\mathscr{F},P(\mathscr{F}\leq\cdot|M,V))$
in $\mathbb{R\times}\mathscr{D}(\mathbb{R})$ on general probability spaces.
Theorem \ref{th2} becomes applicable and the conclusion of Theorem \ref{th h}
about unconditional validity of the bootstrap follows.

Let now Assumption $\mathcal{C}$ hold. Let the original and the bootstrap data
be redefined on another probability space where, by Lemma \ref{le kal}(a),
(\ref{eq mc}) holds (and thus, $\mathscr{F}_{n}^{\ast}\overset{w^{\ast}%
}{\rightarrow}_{p}\mathscr{F}|(M,V)$ by the proof of Theorem \ref{th h}), and
additionally, the convergence in Assumption $\mathcal{C}$ holds as an a.s.
convergence of random probability measures:
\[
\left.  \left(  M_{n},V_{n},\tfrac{1}{n^{1/2}\sigma}%
%TCIMACRO{\tsum _{t=1}^{\left\lfloor n\cdot\right\rfloor }}%
%BeginExpansion
{\textstyle\sum_{t=1}^{\left\lfloor n\cdot\right\rfloor }}
%EndExpansion
x_{nt}\varepsilon_{nt}\right)  \right\vert X_{n}\overset{w}{\rightarrow
}_{a.s.}\left(  M,V,N\right)  |(M,V).
\]
By expanding $F_{\left\lfloor nr\right\rfloor }$ similarly to $F_{\left\lfloor
nr\right\rfloor }^{\ast}$ in the proof of Theorem \ref{th h} and applying the
CMT\ of Sweeting (1989, Theorem 10), we can conclude that $\mathscr{F}_{n}%
|X_{n}\overset{w}{\rightarrow}_{p}\mathscr{F}|(M,V)$. Recalling that also
$\mathscr{F}_{n}^{\ast}\overset{w^{\ast}}{\rightarrow}_{p}\mathscr{F}|(M,V)$,
it follows that on a general probability space $(\mathscr{F}_{n}%
|X_{n},\mathscr{F}_{n}^{\ast}|D_{n})\overset{w}{\rightarrow}_{w}%
(\mathscr{F}|(M,V),\mathscr{F}|(M,V))$.

As previously, the continuity requirement of Corollary \ref{c1}(a) (with
$\tau:=\mathscr{F}$ and $X=X^{\prime}:=(M,V)$) is satisfied by Proposition 3.2
of Linde (1989) applied conditionally. The bootstrap based on $\mathscr{F}_{n}%
$ and $\mathscr{F}_{n}^{\ast}$ is then concluded to be valid conditionally on
$X_{n}$.$\hfill\square$\newpage\appendix
%dummy comment inserted by tex2lyx to ensure that this paragraph is not empty
%

%TCIMACRO{\TeXButton{\appendix special}{\setcounter{section}{0}
%\def\thesection{S.\arabic{section}}
%\def\thesubsection{S.\arabic{section}.\arabic{subsection}}}}%
%BeginExpansion
\setcounter{section}{0}
\def\thesection{S.\arabic{section}}
\def\thesubsection{S.\arabic{section}.\arabic{subsection}}%
%EndExpansion
%

%TCIMACRO{\TeXButton{Lemma numbering}{\renewcommand{\thelemma}{S.\arabic
%{lemma}}
%\setcounter{lemma}{0}}}%
%BeginExpansion
\renewcommand{\thelemma}{S.\arabic{lemma}}
\setcounter{lemma}{0}%
%EndExpansion
%

%TCIMACRO{\TeXButton{Equation numbering}{\renewcommand{\theequation}%
%{S.\arabic{equation}}
%\setcounter{equation}{0}}}%
%BeginExpansion
\renewcommand{\theequation}{S.\arabic{equation}}
\setcounter{equation}{0}%
%EndExpansion
%

%TCIMACRO{\TeXButton{Table numbering}{\renewcommand{\thetable}{S.\arabic
%{table}}
%\setcounter{table}{0}}}%
%BeginExpansion
\renewcommand{\thetable}{S.\arabic{table}}
\setcounter{table}{0}%
%EndExpansion
%

%TCIMACRO{\TeXButton{Figure numbering}{\renewcommand{\thefigure}{S.\arabic
%{figure}}
%\setcounter{figure}{0}}}%
%BeginExpansion
\renewcommand{\thefigure}{S.\arabic{figure}}
\setcounter{figure}{0}%
%EndExpansion
%

%TCIMACRO{\TeXButton{%
%\setcounter{page}{1}%
%}{\setcounter{page}{1}}}%
%BeginExpansion
\setcounter{page}{1}%
%EndExpansion
\newpage

\section*{Inference under Random Limit Bootstrap Measures: \newline
Supplemental Material}

\begin{center}
\bigskip

\textsc{Giuseppe Cavaliere and Iliyan Georgiev}

\vspace{1cm}

\bigskip

{\small November 30, 2019}

\vspace{2cm}

\renewcommand{\thesection}{S.\arabic{section}}
\end{center}

\section{Introduction}

This supplement contains in Section \ref{sec proofs of WC in D} the proofs of
the results formulated in Appendix \ref{sec itere} of Cavaliere and Georgiev
(2019), CG hereafter. In Section \ref{se wedge}, it makes available some
technical details referred to in the proof of Theorem
\ref{Lemma bootstrap with boundary} in CG. In Section \ref{Appendix MC}, the
supplement provides a description of the Monte Carlo simulation design used in
Section \ref{sec example} of CG. For notation, see CG. Unless differently
specified, all references are to sections, equations and results in CG.
Additional references are reported at the end of this document.

\section{Weak convergence in distribution: proofs}

\label{sec proofs of WC in D}

Throughout this supplement, references to extended Skorokhod coupling are
based on Corollary 5.12 of Kallenberg (1997). The exposition could sometimes
be shortened by explicitly considering the random measures of interest as
random elements of a Polish space of measures. To avoid an extra level of
abstraction, we do not adopt this perspective.

\noindent\textsc{Proof of Lemma \ref{le crpr}.} The proof of part (a) is a
straightforward modification of step 1 in the proof of Theorem 2.1 in Crimaldi
and Pratelli (2005), where $X_{n}^{\prime}=X_{n}$ is considered. For part (b),
let $X_{n}^{\prime}=\phi_{n}\left(  X_{n}\right)  $ ($n\in\mathbb{N}$) for
some measurable functions $\phi_{n}$. Without loss of generality, we can
consider that $\mathcal{S}_{X}=\mathcal{S}_{X}^{\prime}$, for otherwise we
could identify $X_{n}$ and $X$ with some random elements $\left(  X_{n}%
,Y_{n}\right)  $ and $\left(  Y,X\right)  $ of $\mathcal{S}_{X}\times
\mathcal{S}_{X}^{\prime}$ for arbitrary constant random elements $Y_{n}$
($\mathcal{S}_{X}^{\prime}$-valued) and $Y$ ($\mathcal{S}_{X}$-valued) defined
on the probability spaces of resp. $X_{n}$ and $X$. Then, by extended
Skorokhod coupling, consider a single probability space supporting $(\tilde
{Z}_{n},\tilde{X}_{n},\tilde{Z}_{n}^{\prime})\overset{d}{=}(Z_{n},X_{n}%
,Z_{n}^{\prime})$ and $(\tilde{Z},\tilde{X},\tilde{Z}^{\prime})\overset{d}%
{=}(Z,X,Z^{\prime})$ with the respective $\tilde{X}_{n}^{\prime}:=\phi
_{n}(\tilde{X}_{n})$ such that $(\tilde{Z}_{n},\tilde{X}_{n}^{\prime}%
,\tilde{Z}_{n}^{\prime})\overset{a.s.}{\rightarrow}(\tilde{Z},\tilde{X}%
,\tilde{Z}^{\prime})$. Then also $\tilde{Z}_{n}|\tilde{X}_{n}\overset
{w}{\rightarrow}_{w}\tilde{Z}|\tilde{X}$ because weak convergence in
distribution is property of the distributions of $(\tilde{Z}_{n},\tilde{X}%
_{n})$ and $(\tilde{Z},\tilde{X})$. From part (a) it follows that $\tilde
{Z}_{n}|\tilde{X}_{n}\overset{w}{\rightarrow}_{p}\tilde{Z}|\tilde{X}$ such
that $E\{h(\tilde{Z}_{n})|\tilde{X}_{n}\}\overset{p}{\rightarrow}%
E\{h(\tilde{Z})|\tilde{X}\}$ for every $h\in\mathcal{C}_{b}(\mathcal{S}_{Z})$.
The convergence $(E\{h(\tilde{Z}_{n})|\tilde{X}_{n}\},\tilde{Z}_{n},\tilde
{X}_{n}^{\prime},\tilde{Z}_{n}^{\prime})\overset{p}{\rightarrow}%
(E\{h(\tilde{Z})|\tilde{X}\},\tilde{Z},\tilde{X},\tilde{Z}^{\prime})$ for such
$h$ implies (\ref{ur crpr}) on a general probability space.$\hfill
\square\medskip$

\begin{remark}
\label{Remark A1}We establish here the `natural' fact that the convergence
\[
((Z_{n}^{\prime}|X_{n}^{\prime}),(Z_{n}^{\prime}|X_{n}^{\prime}),Z_{n}%
^{\prime\prime\prime})\overset{w}{\rightarrow}_{w}((Z^{\prime}|X^{\prime
}),(Z^{\prime}|X^{\prime}),Z^{\prime\prime\prime})
\]
is equivalent to $((Z_{n}^{\prime}|X_{n}^{\prime}),Z_{n}^{\prime\prime\prime
})\overset{w}{\rightarrow}_{w}((Z^{\prime}|X^{\prime}),Z^{\prime\prime\prime
})$ under separability of the space $\mathcal{S}^{\prime\prime\prime}$ where
$Z_{n}^{\prime\prime\prime},Z^{\prime\prime\prime}$ take values. In fact, in
this case, (\ref{ur j4}) with $Z_{n}^{\prime}=Z_{n}^{\prime\prime}$ and
$X_{n}^{\prime}=X_{n}^{\prime\prime}$ is equivalent to
\begin{equation}
(E\{h(Z_{n}^{\prime})|X_{n}^{\prime}\},Z_{n}^{\prime\prime\prime})\overset
{w}{\rightarrow}(E\{h(Z^{\prime})|X^{\prime}\},Z^{\prime\prime\prime
})\label{or ussh}%
\end{equation}
for all continuous and bounded real $h$ with matching domain, since both are
equivalent to $uE\{h(Z_{n}^{\prime})|X_{n}^{\prime}\}+v\mathbb{I}%
_{\{Z_{n}^{\prime\prime\prime}\in A\}}\overset{w}{\rightarrow}uE\{h(Z^{\prime
})|X^{\prime}\}+v\mathbb{I}_{\{Z^{\prime\prime\prime}\in A\}}$ for all such
$h$, all continuity sets $A$ of the distribution of $Z^{\prime\prime\prime}$
and all $(u,v)\in\mathbb{R}^{2}$, by Theorem 3.1 of Billingsley (1968) and the
Cram\'{e}r-Wold theorem.
\end{remark}

\noindent\textsc{Proof of Lemma \ref{le kal}(}a\textsc{).} Let (\ref{ur j3}%
)-(\ref{ur j4}) hold. Then $Z_{n}^{\prime}\overset{w}{\rightarrow}Z^{\prime}$
such that the sequence of probability measures $\{P_{n}^{\prime}\}$ induced by
$Z_{n}^{\prime}$ is tight. The sequence of conditional measures $Z_{n}%
^{\prime}|X_{n}^{\prime}$ has the tight sequence $\{P_{n}^{\prime}\}$ as its
sequence of average measures. As a result, there exists a countable set of
continuous and bounded real functions, say $\{h_{i}^{\prime}\}_{i\in
\mathbb{N}}$, such that the convergence $E\{h_{i}^{\prime}\left(
Z_{n}^{\prime}\right)  |X_{n}^{\prime}\}\overset{a.s.}{\rightarrow}%
E\{h_{i}^{\prime}\left(  Z^{\prime}\right)  |X^{\prime}\} $, were it to hold
for all $i\in\mathbb{N}$, would imply $E\{h^{\prime}(Z_{n}^{\prime}%
)|X_{n}^{\prime}\}\overset{a.s.}{\rightarrow}E\{h^{\prime}\left(  Z^{\prime
}\right)  |X^{\prime}\}$ for all continuous and bounded real $h^{\prime}$ with
the domain of $h_{i}^{\prime}$ (by Theorem 2.2 of Berti, Pratelli and Rigo,
2006). Similarly, there exists a sequence of continuous and bounded real
functions $\{h_{i}^{\prime\prime}\}_{i\in\mathbb{N}}$, such that the
convergence $E\{h_{i}^{\prime\prime}\left(  Z_{n}^{\prime\prime}\right)
|X_{n}^{\prime\prime}\}\overset{a.s.}{\rightarrow}E\{h_{i}^{\prime\prime
}\left(  Z^{\prime\prime}\right)  |X^{\prime\prime}\}$ for all $i\in
\mathbb{N}$ would imply $E\{h^{\prime\prime}(Z_{n}^{\prime\prime}%
)|X_{n}^{\prime\prime}\}\overset{a.s.}{\rightarrow}E\{h^{\prime\prime}\left(
Z^{\prime\prime}\right)  |X^{\prime\prime}\}$ for all continuous and bounded
real $h^{\prime\prime}$ with the domain of $h_{i}^{\prime\prime}$.

Consider the measurable functions $H_{n}$ with values in $\mathcal{S}%
^{\prime\prime\prime}\times\mathbb{R}^{\infty}$ defined by%
\[
H_{n}(X_{n}^{\prime},X_{n}^{\prime\prime},Z_{n}^{\prime\prime\prime}%
)=(Z_{n}^{\prime\prime\prime},\phi_{n1}^{\prime}(X_{n}^{\prime}),\phi
_{n1}^{\prime\prime}(X_{n}^{\prime\prime}),\phi_{n2}^{\prime}(X_{n}^{\prime
}),\phi_{n2}^{\prime\prime}(X_{n}^{\prime\prime}),...)
\]
such that a version $\phi_{ni}^{\prime}(X_{n}^{\prime})$ of $E\{h_{i}^{\prime
}(Z_{n}^{\prime})|X_{n}^{\prime}\}$ and a version $\phi_{ni}^{\prime\prime
}(X_{n}^{\prime\prime})$ of $E\{h_{i}^{\prime\prime}(Z_{n}^{\prime\prime
})|X_{n}^{\prime\prime}\}$ appear resp. at positions $2i$ and $2i+1$, and the
analogous%
\[
H(X^{\prime},X^{\prime\prime},Z^{\prime\prime\prime})=(Z^{\prime\prime\prime
},\phi_{1}^{\prime}(X^{\prime}),\phi_{1}^{\prime\prime}(X^{\prime\prime}%
),\phi_{2}^{\prime}(X^{\prime}),\phi_{2}^{\prime\prime}(X^{\prime\prime
}),...),
\]
where $\phi_{i}^{\prime}(X^{\prime})$ and $\phi_{i}^{\prime\prime}%
(X^{\prime\prime})$ are versions of resp. $E\{h_{i}^{\prime}(Z^{\prime
})|X^{\prime}\}$ and $E\{h_{i}^{\prime\prime}(Z^{\prime\prime})|X^{\prime
\prime}\}$ ($i\in\mathbb{N}$). By separability and Theorem 3.1 of Billingsley
(1968), $H_{n}(X_{n}^{\prime},X_{n}^{\prime\prime},Z_{n}^{\prime\prime\prime
})\overset{w}{\rightarrow}H(X^{\prime},X^{\prime\prime},$ $Z^{\prime
\prime\prime})$ in $\mathcal{S}^{\prime\prime\prime}\times\mathbb{R}^{\infty}$
would follow if%
\begin{align*}
& (\mathbb{I}_{\{Z_{n}^{\prime\prime\prime}\in A\}},\phi_{n1}^{\prime}%
(X_{n}^{\prime}),\phi_{n1}^{\prime\prime}(X_{n}^{\prime\prime}),\phi
_{n2}^{\prime}(X_{n}^{\prime}),\phi_{n2}^{\prime\prime}(X_{n}^{\prime\prime
}),...)\\
& \hspace{2cm}\overset{w}{\rightarrow}(\mathbb{I}_{\{Z^{\prime\prime\prime}\in
A\}},\phi_{1}^{\prime}(X^{\prime}),\phi_{1}^{\prime\prime}(X^{\prime\prime
}),\phi_{2}^{\prime}(X^{\prime}),\phi_{2}^{\prime\prime}(X^{\prime\prime
}),...)
\end{align*}
in $\mathbb{R}^{\infty}$ for every continuity set $A$ of the distribution of
$Z^{\prime\prime\prime}$. The previous is equivalent to weak convergence of
the finite-dimensional distributions of the considered sequences for every
such $A$ (Billingsley, 1968, p.19). Any linear combination of finitely many
functions among $\{h_{i}^{\prime}\}_{i\in\mathbb{N}}$ is a bounded and
continuous real function, and so for $\{h_{i}^{\prime\prime}\}_{i\in
\mathbb{N}}$ , and for any such two linear combinations (say $h^{\prime}%
=\sum_{s=1}^{m}u_{s}h_{i_{s}}^{\prime}$ and $h^{\prime\prime}=\sum_{s=1}%
^{l}v_{s}h_{j_{s}}^{\prime\prime}$), it holds that $\sum_{s=1}^{m}u_{s}%
\phi_{ni_{s}}^{\prime}(X_{n}^{\prime})$ and $\sum_{s=1}^{l}v_{s}\phi_{nj_{s}%
}^{\prime\prime}(X_{n}^{\prime\prime})$ are versions of resp. $E\{h^{\prime
}(Z_{n}^{\prime})|X_{n}^{\prime}\}\ $and $E\{h^{\prime\prime}(Z_{n}%
^{\prime\prime})|X_{n}^{\prime\prime}\}$, whereas $\sum_{s=1}^{m}u_{s}%
\phi_{_{i_{s}}}^{\prime}(X^{\prime})$ and $\sum_{s=1}^{l}v_{s}\phi_{j_{s}%
}^{\prime\prime}(X^{\prime\prime})$ are versions of resp. $E\{h^{\prime
}(Z^{\prime})|X^{\prime}\}\ $and $E\{h^{\prime\prime}(Z^{\prime\prime
})|X^{\prime\prime}\}$;\ therefore,%
\begin{align*}
& \left(  \mathbb{I}_{\{Z_{n}^{\prime\prime\prime}\in A\}},\sum_{s=1}^{m}%
u_{s}\phi_{ni_{s}}^{\prime}(X_{n}^{\prime}),\sum_{s=1}^{l}v_{s}\phi_{nj_{s}%
}^{\prime\prime}(X_{n}^{\prime\prime})\right) \\
& \hspace{2cm}\overset{w}{\rightarrow}\left(  \mathbb{I}_{\{Z^{\prime
\prime\prime}\in A\}},\sum_{s=1}^{m}u_{s}\phi_{i_{s}}^{\prime}(X^{\prime
}),\sum_{s=1}^{l}v_{s}\phi_{j_{s}}^{\prime\prime}(X^{\prime\prime})\right)
\end{align*}
by (\ref{ur j4}) and Theorem 3.1 of Billingsley (1968). By the Cram\'{e}r-Wold
theorem, this implies that the finite-dimensional distributions of
$H_{n}(X_{n}^{\prime},X_{n}^{\prime\prime},Z_{n}^{\prime\prime\prime})$ weakly
converge to those of $H(Z^{\prime},X^{\prime},X^{\prime\prime})$. As a result,
$H_{n}(X_{n}^{\prime},X_{n}^{\prime\prime},Z_{n}^{\prime\prime\prime}%
)\overset{w}{\rightarrow}H(X^{\prime},X^{\prime\prime},Z^{\prime\prime\prime
})$ in $\mathcal{S}^{\prime\prime\prime}\times\mathbb{R}^{\infty}$.

By extended Skorokhod coupling, $(X_{n}^{\prime},X_{n}^{\prime\prime}%
,Z_{n}^{\prime},Z_{n}^{\prime\prime},Z_{n}^{\prime\prime\prime})$ and
$(X^{\prime},X^{\prime\prime},Z^{\prime},Z^{\prime\prime},Z^{\prime
\prime\prime})$ can be redefined, maintaining their distribution, on a new
probability space where $H_{n}(X_{n}^{\prime},X_{n}^{\prime\prime}%
,Z_{n}^{\prime\prime\prime})\overset{a.s.}{\rightarrow}H(X^{\prime}%
,X^{\prime\prime},Z^{\prime\prime\prime})$ in $\mathcal{S}^{\prime\prime
\prime}\times\mathbb{R}^{\infty}$ (we subsume the \symbol{126}-notation for
the redefined variables). On the new probability space, the relevant
components of $H_{n}$, $H$ are still versions of the conditional expectations
for the redefined variables, for conditional expectations are determined up to
equivalence by the underlying joint distributions. As a result, $Z_{n}%
^{\prime\prime\prime}\overset{a.s.}{\rightarrow}Z^{\prime\prime\prime}$,
$E\{h_{i}^{\prime}\left(  Z_{n}^{\prime}\right)  |X_{n}^{\prime}%
\}\overset{a.s.}{\rightarrow}E\{h_{i}^{\prime}\left(  Z^{\prime}\right)
|X^{\prime}\}$ for all $i\in\mathbb{N}$, and similarly for $h_{i}%
^{\prime\prime}$. By the choice of $\{h_{i}^{\prime}\}_{i\in\mathbb{N}}$ and
$\{h_{i}^{\prime\prime}\}_{i\in\mathbb{N}}$, on this space $Z_{n}^{\prime
}|X_{n}^{\prime}\overset{w}{\rightarrow}_{a.s.}Z^{\prime}|X^{\prime}$ and
$Z_{n}^{\prime\prime}|X_{n}^{\prime\prime}\overset{w}{\rightarrow}%
_{a.s.}Z^{\prime\prime}|X^{\prime\prime} $. \medskip

\noindent\textsc{Proof of Lemma \ref{le kal}(}b\textsc{).} Let $Z_{n}^{\prime
},Z_{n}^{\prime\prime},Z^{\prime},Z^{\prime\prime}$ ($n\in\mathbb{N}$) be
rv's. By the proof of Kallenberg (2017, Theorem 4.20), on the
Skorokhod-coupling space considered in the proof of part (a) it holds that
$P(Z_{n}^{\prime}\leq\cdot|X_{n}^{\prime})\overset{a.s.}{\rightarrow
}P(Z^{\prime}\leq\cdot|X^{\prime})$ in $\mathscr{D}(\mathbb{R)}$ and
$P(Z_{n}^{\prime\prime}\leq\cdot|X_{n}^{\prime\prime})\overset{a.s.}%
{\rightarrow}P(Z^{\prime\prime}\leq\cdot|X^{\prime\prime})$ in
$\mathscr{D}(\mathbb{R)}$. Since on this space also $Z_{n}^{\prime\prime
\prime}\overset{a.s.}{\rightarrow}Z^{\prime\prime\prime}$, (\ref{or cokal})
follows on a general probability space.

Conversely, let (\ref{or cokal}) hold. Notice that $P(Z_{n}^{\prime}\leq
\cdot|X_{n}^{\prime})$, $P(Z_{n}^{\prime\prime}\leq\cdot|X_{n}^{\prime\prime
})$, $P(Z^{\prime}\leq\cdot|X^{\prime})$ and $P(Z^{\prime\prime}\leq
\cdot|X^{\prime\prime})$ as random elements of $\mathscr{D}(\mathbb{R)}$ are
measurable transformations of resp. $X_{n}^{\prime}$, $X_{n}^{\prime\prime}$,
$X^{\prime}$ and $X^{\prime\prime}$ that are determined up to
indistinguishability by the joint distributions of resp. $\left(
Z_{n}^{\prime},X_{n}^{\prime}\right)  $, $\left(  Z_{n}^{\prime\prime}%
,X_{n}^{\prime\prime}\right)  $, $\left(  Z^{\prime},X^{\prime}\right)  $ and
$\left(  Z^{\prime\prime},X^{\prime\prime}\right)  $. By extended Skorokhod
coupling, $(X_{n}^{\prime},X_{n}^{\prime\prime},Z_{n}^{\prime},Z_{n}%
^{\prime\prime},Z_{n}^{\prime\prime\prime})$ and $(X^{\prime},X^{\prime\prime
},Z^{\prime},Z^{\prime\prime},Z^{\prime\prime\prime})$ can be redefined,
maintaining their distribution, on a new probability space where
$Z_{n}^{\prime\prime\prime}\overset{a.s.}{\rightarrow}Z^{\prime\prime\prime}$,
$P(Z_{n}^{\prime}\leq\cdot|X_{n}^{\prime})\overset{a.s.}{\rightarrow
}P(Z^{\prime}\leq\cdot|X^{\prime})$ in $\mathscr{D}(\mathbb{R)}$ and
$P(Z_{n}^{\prime\prime}\leq\cdot|X_{n}^{\prime\prime})\overset{a.s.}%
{\rightarrow}P(Z^{\prime\prime}\leq\cdot|X^{\prime\prime})$ in
$\mathscr{D}(\mathbb{R)}$. By the proof of Kallenberg (2017, Theorem 4.20), on
this space
\[
\left(  E\{h^{\prime}(Z_{n}^{\prime})|X_{n}^{\prime}\},\,E\{h^{\prime\prime
}(Z_{n}^{\prime\prime})|X_{n}^{\prime\prime}\},Z_{n}^{\prime\prime\prime
}\right)  \overset{a.s.}{\rightarrow}\left(  E\{h^{\prime}(Z^{\prime
})|X^{\prime}\},\,E\{h^{\prime\prime}(Z^{\prime\prime})|X^{\prime\prime
}\},Z^{\prime\prime\prime}\right)
\]
for all $h^{\prime},h^{\prime\prime}\in\mathcal{C}_{b}(\mathbb{R})$, and
therefore, (\ref{ur j4}) holds on a general probability space.$\hfill
\square\medskip$

The following corollary, in its simplest version, established the `natural'
equivalence of $Z_{n}\overset{w}{\rightarrow}Z$ and $Z_{n}|Z_{n}\overset
{w}{\rightarrow}_{w}Z|Z$ for random elements $Z_{n},Z$ of a Polish space.

\begin{corollary}
\label{co coeq}Let $(Z_{n},X_{n})$ and $(Z,X)$ be random elements such that
$Z_{n}=(Z_{n}^{\prime},Z_{n}^{\prime\prime})$ and $Z=(Z^{\prime}%
,Z^{\prime\prime})$ are $\mathcal{S}_{Z}^{\prime}\times\mathcal{S}_{Z}%
^{\prime\prime}$-valued, whereas $X_{n}$ and $X$ are resp. $\mathcal{S}
$-valued and $\mathcal{S}_{X}$-valued ($n\in\mathbb{N}$), with all the
mentioned spaces being Polish metric spaces. Then the convergence
$(Z_{n}^{\prime}|Z_{n}^{\prime},Z_{n}^{\prime\prime}|X_{n})\overset
{w}{\rightarrow}_{w}(Z^{\prime}|Z^{\prime},Z^{\prime\prime}|X)$ in the sense
of (\ref{eq:wcrm}) is equivalent to the convergence $(Z_{n}^{\prime}%
,(Z_{n}^{\prime\prime}|X_{n}))\overset{w}{\rightarrow}_{w}(Z^{\prime
},(Z^{\prime\prime}|X))$ in the sense of (\ref{eq:wcrm1}). If additionally
$\mathcal{S}_{Z}^{\prime\prime}=\mathbb{R}$ and the conditional distribution
$Z^{\prime\prime}|X$ is diffuse, then both convergence facts are equivalent to
$(Z_{n}^{\prime},P(Z_{n}^{\prime\prime}\leq\cdot|X_{n}^{\prime\prime
}))\overset{w}{\rightarrow}_{w}(Z^{\prime},P(Z^{\prime\prime}\leq
\cdot|X^{\prime\prime}))$ as random elements of $\mathcal{S}_{Z}^{\prime
}\times\mathscr{D}(\mathbb{R)}.$
\end{corollary}

\noindent\textsc{Proof of Corollary \ref{co coeq}}. As in the proof of Lemma
\ref{le crpr}, there is no loss of generality in assuming the equality
$\mathcal{S}_{X}=\mathcal{S}$. First, let $(Z_{n}^{\prime}|Z_{n}^{\prime
},Z_{n}^{\prime\prime}|X_{n})\overset{w}{\rightarrow}_{w}(Z^{\prime}%
|Z^{\prime},Z^{\prime\prime}|X)$. By Lemma \ref{le kal}(a), consider a
Skorokhod representation such that $\tilde{Z}_{n}^{\prime}|\tilde{Z}%
_{n}^{\prime}\overset{w}{\rightarrow}_{a.s.}\tilde{Z}^{\prime}|\tilde
{Z}^{\prime}$ and $\tilde{Z}_{n}^{\prime\prime}|\tilde{X}_{n}\overset
{w}{\rightarrow}_{a.s.}\tilde{Z}^{\prime\prime}|\tilde{X}$. Then $h^{\prime
}(\tilde{Z}_{n}^{\prime})\overset{a.s.}{\rightarrow}h^{\prime}(\tilde
{Z}^{\prime})$ for every $h^{\prime}\in\mathcal{C}_{b}(\mathcal{S}_{Z}%
^{\prime})$. This implies that $\tilde{Z}_{n}^{\prime}\overset{p}{\rightarrow
}\tilde{Z}^{\prime}$ by the proof of Proposition 4.3(i) of Crimaldi and
Pratelli (2005). As further $E\{h^{\prime\prime}(\tilde{Z}_{n}^{\prime\prime
})|\tilde{X}_{n}\}\overset{a.s.}{\rightarrow}E\{h^{\prime\prime}(\tilde
{Z}^{\prime\prime})|\tilde{X}\}$ for every $h^{\prime\prime}\in\mathcal{C}%
_{b}(\mathcal{S}_{Z}^{\prime\prime})$, the convergence $(\tilde{Z}_{n}%
^{\prime},E\{h^{\prime\prime}(\tilde{Z}_{n}^{\prime\prime})|\tilde{X}%
_{n}\})\overset{p}{\rightarrow}(\tilde{Z}^{\prime},E\{h^{\prime\prime}%
(\tilde{Z}^{\prime\prime})|\tilde{X}\})$ implies that $(Z_{n}^{\prime
},E\{h^{\prime\prime}(Z_{n}^{\prime\prime})|X_{n}\})\overset{w}{\rightarrow}$
\linebreak$(Z^{\prime},E\{h^{\prime\prime}(Z^{\prime\prime})|X\})$ for every
such $h^{\prime\prime}$, which is (\ref{eq:wcrm1}). Second, let $(Z_{n}%
^{\prime},E\{h^{\prime\prime}(Z_{n}^{\prime\prime})|X_{n}\})\overset
{w}{\rightarrow}(Z^{\prime},E\{h^{\prime\prime}(Z^{\prime\prime})|X\})$ hold
for every $h^{\prime\prime}\in\mathcal{C}_{b}(\mathcal{S}_{Z}^{\prime\prime}%
)$. Then $(h^{\prime}(Z_{n}^{\prime}),E\{h^{\prime\prime}(Z_{n}^{\prime\prime
})|X_{n}\})\overset{w}{\rightarrow}(h^{\prime}(Z^{\prime}),E\{h^{\prime\prime
}(Z^{\prime\prime})|X\})$ for every $h^{\prime}\in\mathcal{C}_{b}%
(\mathcal{S}_{Z}^{\prime})$, by the CMT. The latter statement is equivalent to
(\ref{eq:wcrm}) with $Z_{n}^{\prime}=X_{n}^{\prime}$. Finally, equivalence to
the convergence involving the random cdf $P(Z^{\prime\prime}\leq
\cdot|X^{\prime\prime})$ follows from Lemma \ref{le kal}(b);\ see also
\ref{Remark A1}.$\hfill\square\medskip$

\noindent\textsc{Proof of Theorem \ref{th cmt}}. As in the proof of Lemma
\ref{le crpr}, without loss of generality, we can assume that $\mathcal{S}%
_{X}=\mathcal{S}_{X}^{\prime}$. By Lemma \ref{le kal}(a), consider a Skorokhod
representation such that $\tilde{Z}_{n}|\tilde{X}_{n}\overset{w}{\rightarrow
}_{a.s.}\tilde{Z}|\tilde{X}$. Then $h(\tilde{Z}_{n})|\tilde{X}_{n}\overset
{w}{\rightarrow}_{a.s.}h(\tilde{Z})|\tilde{X}$ on the Skorokhod-representation
space by Theorems 8(i) and 10 of Sweeting (1989). Therefore, $h(Z_{n})|X_{n}$
$\overset{w}{\rightarrow}_{w}$ $h(Z)|X$ on a general probability
space.$\hfill\square\medskip$

\noindent\textsc{Proof of Theorem \ref{p citex}}. In terms of conditional
expectations, the first part of the theorem asserts that if (\ref{ur cocon})
holds and $E\{h(X_{n}^{\prime\prime})|X_{n}^{\prime}\}\overset{w}{\rightarrow
}E\{h\left(  X^{\prime\prime}\right)  |X^{\prime}\}$ for all continuous and
bounded real functions $h$ with matching domain, where $\left(  X_{n}^{\prime
},X_{n}^{\prime\prime}\right)  $ are $X_{n}$-measurable, then the iterated
expectations
\[
E(z_{n}|X_{n}^{\prime})=E\{E(z_{n}|X_{n})|X_{n}^{\prime}\}\text{ and
}E(z|X^{\prime})=E\{E(z|X^{\prime},X^{\prime\prime})|X^{\prime}\}
\]
satisfy the convergence
\begin{equation}
(E(z_{n}|X_{n}^{\prime}),E(z_{n}|X_{n}),X_{n}^{\prime},X_{n}^{\prime\prime
},Y_{n})\overset{w}{\rightarrow}(E(z|X^{\prime}),E(z|X^{\prime},X^{\prime
\prime}),X^{\prime},X^{\prime\prime},Y).\label{conitex}%
\end{equation}
We set up the proof in these terms.

By Theorem 2.1 of Crimaldi and Pratelli (2005), $(X_{n}^{\prime},X_{n}%
^{\prime\prime})\overset{w}{\rightarrow}(X^{\prime},X^{\prime\prime})$ and
$X_{n}^{\prime\prime}|X_{n}^{\prime}$ $\overset{w}{\rightarrow}_{w}$
$X^{\prime\prime}|X^{\prime}$ imply $(X_{n}^{\prime},X_{n}^{\prime\prime
})|X_{n}^{\prime}$ $\overset{w}{\rightarrow}_{w}$ $(X^{\prime},X^{\prime
\prime})|X^{\prime}$; i.e., for all $h\in\mathcal{C}_{b}(\mathcal{S}%
_{X}^{\prime})$, it holds that $E\{h(X_{n}^{\prime},X_{n}^{\prime\prime
})|X_{n}^{\prime}\}\overset{w}{\rightarrow}E\{h\left(  X^{\prime}%
,X^{\prime\prime}\right)  |X^{\prime}\}$.

Let $\phi_{n}$ and $\phi$ be measurable real functions such that $\phi
_{n}\left(  X_{n}\right)  $ and $\phi(X^{\prime},X^{\prime\prime})$ are
versions respectively of the conditional expectations $E(z_{n}|X_{n})$ and
$E(z|X^{\prime},X^{\prime\prime})$. We proceed in two steps. First, we argue
that we can redefine $(X_{n},Y_{n})$ and $\left(  X^{\prime},X^{\prime\prime
},Y\right)  $, maintaining their distribution, on a new probability space
where $\left(  \phi_{n}(X_{n}),X_{n}^{\prime},X_{n}^{\prime\prime}%
,Y_{n}\right)  \overset{a.s.}{\rightarrow}\left(  \phi(X^{\prime}%
,X^{\prime\prime}),X^{\prime},X^{\prime\prime},Y\right)  $ and $E\{h(X_{n}%
^{\prime},X_{n}^{\prime\prime})|X_{n}^{\prime}\}\overset{p}{\rightarrow
}E\{h\left(  X^{\prime},X^{\prime\prime}\right)  |X^{\prime}\}$ for all
$h\in\mathcal{C}_{b}(\mathcal{S}_{X}^{\prime})$. Second, we show that on this
space $E\{\phi_{n}(X_{n})|X_{n}^{\prime}\}\overset{p}{\rightarrow}%
E\{\phi(X^{\prime},X^{\prime\prime})|X^{\prime}\}$, which implies convergence
(\ref{conitex}) on a general probability space.\medskip{}

\noindent\textsc{Step 1.} Let the measurable function $\psi_{n}$ be such that
$(X_{n}^{\prime},X_{n}^{\prime\prime})=\psi_{n}(X_{n})$, thus $\left(
\phi_{n}(X_{n}),\psi_{n}(X_{n}),Y_{n}\right)  \overset{w}{\rightarrow}\left(
\phi(X^{\prime},X^{\prime\prime}),X^{\prime},X^{\prime\prime},Y\right)  $. By
extended Skorokhod coupling, there exist a probability space and random
elements $(\tilde{X}_{n},\tilde{Y}_{n})\overset{d}{=}(X_{n},Y_{n})$,
$(\tilde{X}^{\prime},\tilde{X}^{\prime\prime},\tilde{Y})\overset{d}{=}\left(
X^{\prime},X^{\prime\prime},Y\right)  $ defined on this space such that
$(\phi_{n}(\tilde{X}_{n}),\psi_{n}(\tilde{X}_{n}),\tilde{Y}_{n})\overset
{a.s.}{\rightarrow}(\phi(\tilde{X}^{\prime},\tilde{X}^{\prime\prime}%
),\tilde{X}^{\prime},\tilde{X}^{\prime\prime},$ $\tilde{Y})$. On this space it
also holds that $E\{h(\tilde{X}_{n}^{\prime},\tilde{X}_{n}^{\prime\prime
})|\tilde{X}_{n}^{\prime}\}\overset{w}{\rightarrow}E\{h(\tilde{X}^{\prime
},\tilde{X}^{\prime\prime})|\tilde{X}^{\prime}\}$ for all $h\in\mathcal{C}%
_{b}(\mathcal{S}_{X}^{\prime})$ and $(\tilde{X}_{n}^{\prime},\tilde{X}%
_{n}^{\prime\prime}):=\psi_{n}(\tilde{X}_{n})$, as a consequence of the
distributional equalities $(\tilde{X}_{n}^{\prime},\tilde{X}_{n}^{\prime
\prime})\overset{d}{=}\left(  X_{n}^{\prime},X_{n}^{\prime\prime}\right)  $
and $(\tilde{X}^{\prime},\tilde{X}^{\prime\prime})\overset{d}{=}\left(
X^{\prime},X^{\prime\prime}\right)  $. Moreover, this convergence can be
strengthened to $E\{h(\tilde{X}_{n}^{\prime},\tilde{X}_{n}^{\prime\prime
})|\tilde{X}_{n}^{\prime}\}\overset{p}{\rightarrow}E\{h(\tilde{X}^{\prime
},\tilde{X}^{\prime\prime})|\tilde{X}^{\prime}\}$ for all $h\in\mathcal{C}%
_{b}(\mathcal{S}_{X}^{\prime})$ by Lemma \ref{le crpr}(a). The next step of
the proof takes place in this special probability space (we subsume the
$\symbol{126}$-notation).\medskip{}

\noindent\textsc{Step 2. }As $\mathcal{C}_{b}(\mathcal{S}_{X}^{\prime})$ is
dense in the real functions on $\mathcal{S}_{X}^{\prime}$ that are integrable
w.r.t. the probability measure induced by $\left(  X^{\prime},X^{\prime\prime
}\right)  $, it follows that for every $\varepsilon\in\left(  0,1\right)  $
there exists a $\phi_{\varepsilon}\in\mathcal{C}_{b}(\mathcal{S}_{X}^{\prime
})$ such that $E\left\vert \phi_{\varepsilon}\left(  X^{\prime},X^{\prime
\prime}\right)  -\phi(X^{\prime},X^{\prime\prime})\right\vert <(\varepsilon
/5)^{2}$. We decompose
\begin{align*}
& \left\vert E\{\phi_{n}(X_{n})|X_{n}^{\prime}\}-E\{\phi(X^{\prime}%
,X^{\prime\prime})|X^{\prime}\}\right\vert \overset{}{\leq}E\left\{  \left.
\left\vert \phi_{n}(X_{n})-\phi(X^{\prime},X^{\prime\prime})\right\vert
\right\vert X_{n}^{\prime}\right\} \\
& \hspace{6.3cm}+\left\vert E\left\{  \phi(X^{\prime},X^{\prime\prime}%
)-\phi_{\varepsilon}(X^{\prime},X^{\prime\prime})|X_{n}^{\prime}\right\}
\right\vert \\
& \hspace{6.3cm}+\left\vert E\left\{  \phi_{\varepsilon}(X^{\prime}%
,X^{\prime\prime})-\phi_{\varepsilon}(X_{n}^{\prime},X_{n}^{\prime\prime
})|X_{n}^{\prime}\right\}  \right\vert \\
& \hspace{6.3cm}+\left\vert E\left\{  \phi_{\varepsilon}(X_{n}^{\prime}%
,X_{n}^{\prime\prime})|X_{n}^{\prime}\right\}  -E\{\phi_{\varepsilon
}(X^{\prime},X^{\prime\prime})|X^{\prime}\}\right\vert \\
& \hspace{6.3cm}+\left\vert E\{\phi_{\varepsilon}(X^{\prime},X^{\prime\prime
})-\phi(X^{\prime},X^{\prime\prime})|X^{\prime}\}\right\vert
\end{align*}
and label the addends on the right-hand side $\rho_{i}$, $i=1,...,5$, in order
of appearance. The term $\rho_{1}$ is $o_{p}\left(  1\right)  $ because
$\left\vert \phi_{n}(X_{n})-\phi(X^{\prime},X^{\prime\prime})\right\vert
\overset{p}{\rightarrow}0$ and the $o_{p}\left(  1\right)  $ property is
preserved upon conditioning. By Markov's inequality and the choice of
$\phi_{\varepsilon}$, it follows that $P\left(  \rho_{2}\geq\varepsilon
/5\right)  \leq\varepsilon/5$. Since $\left(  X_{n}^{\prime},X_{n}%
^{\prime\prime}\right)  \overset{a.s.}{\rightarrow}\left(  X^{\prime
},X^{\prime\prime}\right)  $ and $\phi_{\varepsilon}$ is continuous, it holds
that $\left\vert \phi_{\varepsilon}(X^{\prime},X^{\prime\prime})-\phi
_{\varepsilon}\left(  X_{n}^{\prime},X_{n}^{\prime\prime}\right)  \right\vert
\overset{a.s.}{\rightarrow}0$ and $\rho_{3}$ is $o_{p}\left(  1\right)  $
similarly to $\rho_{1}$. For $h=\phi_{\varepsilon}$ it holds that
$|E\{h(X_{n}^{\prime},X_{n}^{\prime\prime})|X_{n}^{\prime}\}-E\{h\left(
X^{\prime},X^{\prime\prime}\right)  |X^{\prime}\}|\overset{p}{\rightarrow}0$
such that $\rho_{4}=o_{p}(1)$. Finally, $P\left(  \rho_{5}\geq\varepsilon
/5\right)  \leq\varepsilon/5$ by Markov's inequality, similarly to $\rho_{2}$.
By combining these results, it follows that
\[
P\left(  \left\vert E\{\phi_{n}(X_{n})|X_{n}^{\prime}\}-E\{\phi(X^{\prime
},X^{\prime\prime})|X^{\prime}\}\right\vert \geq\varepsilon\right)
<\varepsilon
\]
for large enough $n$. This proves $E(z_{n}|X_{n}^{\prime})=E\{\phi_{n}%
(X_{n})|X_{n}^{\prime}\}\overset{p}{\rightarrow}E\{\phi(X^{\prime}%
,X^{\prime\prime})|X^{\prime}\}=E(z|X^{\prime})$ on the special probability
space, and since also $\left(  E(z_{n}|X_{n}),X_{n}^{\prime},X_{n}%
^{\prime\prime},Y_{n}\right)  \overset{a.s.}{\rightarrow}\left(
E(z|X^{\prime}),X^{\prime},X^{\prime\prime},Y\right)  $ on that space,
(\ref{conitex}) follows on the original probability spaces.

To prove the second part of the theorem, let $h^{\prime},h^{\prime\prime}%
\in\mathcal{C}_{b}( \mathcal{S}_{Z})$ be arbitrary. By Remark \ref{Remark A1},
(\ref{ur j5}) implies (\ref{ur cocon}) with $z_{n}=h^{\prime}\left(
Z_{n}\right)  $, $Y_{n}=E\{h^{\prime\prime}(Z_{n})|X_{n}\},$ $z=h^{\prime
}\left(  Z\right)  $, $Y=E\{h^{\prime\prime}(Z)|X^{\prime},X^{\prime\prime}%
\}$. By the first part of the theorem and the arbitrariness of $h^{\prime
},h^{\prime\prime}$, (\ref{ur j6}) follows.$\hfill\square$

\medskip

\noindent\textsc{Proof of Lemma \ref{t cmj}}. Let $\{f_{j}\}_{j\in\mathbb{N}}$
be a convergence-determining countable set of bounded Lipschitz functions
$\mathcal{S}_{X}\times\mathcal{S}_{Z}\rightarrow\mathbb{R}$ such that
$(X_{n},Z_{n}^{\ast})\overset{w}{\rightarrow}(X,Z)$ is implied by the
convergence
\begin{equation}
Ef_{j}(X_{n},Z_{n}^{\ast})\rightarrow Ef_{j}(X,Z)\text{ as }n\rightarrow
\infty\text{ for all }j\in\mathbb{N}.\label{ur cdem}%
\end{equation}
The existence of such $\{f_{j}\}_{j\in\mathbb{N}}$ follows from the proof of
Proposition 3.4.4 of Ethier and Kurtz (2005);\ see also Proposition 2.2 of
Worm and Hille (2011). If $\{n_{m}\}$ is an arbitrary subsequence of the
naturals, there exists a further subsequence $\{n_{m_{k}}\}$ such that
$X_{n_{m_{k}}}\overset{a.s.}{\rightarrow}X$ and the (random) conditional
distribution of $Z_{n_{m_{k}}}^{\ast}$ given $D_{n_{m_{k}}}$ a.s. converges to
the (random) conditional distribution of $Z^{\ast}$ given $X^{\prime}$ (the
latter by Corollary 2.4 of Berti, Pratelli and Rigo, 2006). In particular,
$E\{h(Z_{n_{m_{k}}}^{\ast})|D_{n_{m_{k}}}\}\overset{a.s.}{\rightarrow
}E\{h(Z_{{}}^{\ast})|X^{\prime}\}$ as $k\rightarrow\infty$ for every
$h\in\mathcal{C}_{b}(\mathcal{S}_{Z})$. If we show that $Ef_{j}(X_{n_{m_{k}}%
},Z_{n_{m_{k}}}^{\ast})\rightarrow Ef_{j}(X,Z)$ as $k\rightarrow\infty$ for
all $j\in\mathbb{N}$, (\ref{ur cdem}) will follow. Hence, without loss of
generality we can take $X_{n}\overset{a.s.}{\rightarrow}X$ and $E\{h(Z_{n}%
^{\ast})|D_{n}\}\overset{a.s.}{\rightarrow}E\{h(Z^{\ast})|X^{\prime}\}$ for
every $h\in\mathcal{C}_{b}(\mathcal{S}_{Z})$, and prove that, as a result,
(\ref{ur cdem}) holds.

Write $Z_{n}^{\ast}=$ $Z_{n}^{\ast}(D_{n},W_{n}^{\ast})$ and define the
measurable functions $\phi_{nj}:\mathcal{S}_{X}\times\mathcal{S}%
_{D}\rightarrow\mathbb{R}$ and $\phi_{j}:\mathcal{S}_{X}\times\Omega
\rightarrow\mathbb{R}$ by%
\[
\phi_{nj}(x,d):=E_{P^{b}}\{f_{j}(x,Z_{n}^{\ast}(d,W_{n}^{\ast}))\}\text{ and
}\phi_{j}(x,\omega):=\int_{\mathcal{S}_{Z}}f_{j}(x,z)\nu(dz,X^{\prime}%
(\omega)),
\]
where $\nu$ is a regular conditional distribution of $Z^{\ast}$ given
$X^{\prime}$. First, we show that there exists an event $A\in\mathcal{F}$ with
$P(A)=1$ such that%
\begin{equation}
\phi_{nj}(x,D_{n}(\omega))\rightarrow\phi_{j}(x,\omega)\text{ for all }%
j\in\mathbb{N},x\in\mathcal{S}_{X},\omega\in A\text{.}\label{ur confj}%
\end{equation}
Second, we conclude that $Ef_{j}(X,Z_{n}^{\ast})\rightarrow Ef_{j}(X,Z)$ as
$n\rightarrow\infty$ for all $j\in\mathbb{N}$, and then we obtain
(\ref{ur cdem}).

Let $\{x_{i}\}_{i\in\mathbb{N}}$ be a countable dense subset of $\mathcal{S}%
_{X}$. As $f_{j}(x_{i},\cdot)\in\mathcal{C}_{b}(\mathcal{S}_{Z})$, it holds
that $E\{f_{j}(x_{i},Z_{n}^{\ast})|D_{n}\}\overset{a.s.}{\rightarrow}%
E\{f_{j}(x_{i},Z^{\ast})|X^{\prime}\}$ (take $h=f_{j}(x_{i},\cdot)$). Since
$\phi_{nj}(x_{i},D_{n})$ and $\phi_{j}(x_{i},\omega)$ are versions of
$E\{f_{j}(x_{i},Z_{n}^{\ast})|D_{n}\}$ and $E\{f_{j}(x_{i},Z^{\ast}%
)|X^{\prime}\}$ respectively (see Ex. 10.1.9 of Dudley, 2004, p.341, for the
former and Theorem 5.4 of Kallenberg, 1997, for both or the latter), there
exist sets $A_{ij}\in\mathcal{F}$ with $P(A_{ij})=1$ such that $\phi
_{nj}(x_{i},D_{n}(\omega))\rightarrow\phi_{j}(x,\omega)$ for all $\omega\in
A_{ij}$ and every $i,j\in\mathbb{N}$. Define $A:=\cap_{i,j\in\mathbb{N}}%
A_{ij}$ with $P\left(  A\right)  =1$. It then holds that%
\[
\phi_{nj}(x_{i},D_{n}(\omega))\rightarrow\phi_{j}(x_{i},\omega)\text{ for all
}i,j\in\mathbb{N},\omega\in A\text{.}%
\]
Since, for every $x\in\mathcal{S}_{X}$ and $j\in\mathbb{N}$, $|f_{j}%
(x_{i},\cdot)-f_{j}(x,\cdot)|\leq L_{j}\{\rho_{X}\left(  x_{i},x\right)
\wedge1\}$ can be made arbitrarily small by an appropriate choice of $x_{i}$,
where $\rho_{X}$ is the metric on $\mathcal{S}_{X}$ and $L_{j}\in\mathbb{R}$
are Lipschitz constants for $f_{j}$, from the definitions of $\phi_{nj}$ and
$\phi_{j}$ it follows that $|\phi_{nj}(x_{i},D_{n}(\omega))-\phi_{nj}%
(x,D_{n}(\omega))|$ and $|\phi_{j}(x_{i},\omega)-\phi_{j}(x,\omega)|$ for
every fixed $x,j$ can be made arbitrarily small uniformly over $n,\omega$.
From this fact and from the previous display, (\ref{ur confj}) follows for
$A=\cap_{i,j\in\mathbb{N}}A_{ij}$.

For arbitrary $j,n\in\mathbb{N}$, (\ref{ur confj}) ensures that $\phi
_{nj}(X(\omega),D_{n}(\omega))\rightarrow\phi_{j}(X(\omega),\omega)$ for all
$\omega\in A$, and thus, a.s. Since $\phi_{nj}(X,D_{n})$ is a version of
$E\{f_{j}(X,Z_{n}^{\ast})|D_{n},X\}\ $(by the product structure of the
probability space; see Ex. 10.1.9 of Dudley, 2004, p.341), it follows that
\begin{align*}
E\{f_{j}(X,Z_{n}^{\ast})|D_{n},X\}\overset{a.s.}{\rightarrow}\phi_{j}%
(X,\omega)  & =\left.  E\{f_{j}(x,Z^{\ast})|X^{\prime}\}\right\vert
_{x=X}\overset{(1)}{=}\left.  E\{f_{j}(x,Z)|X^{\prime}\}\right\vert _{x=X}\\
& \overset{(2)}{=}\left.  E\{f_{j}(x,Z)|X\}\right\vert _{x=X}\overset{(3)}%
{=}E\{f_{j}(X,Z)|X\}\text{ a.s.,}%
\end{align*}
equalities (1) and (2) from $Z^{\ast}|X^{\prime}\overset{d}{=}Z|X^{\prime
}\overset{d}{=}Z|X$, and equality (3) because for $X$-measurable $\xi$'s,
$\left.  E\{f_{j}(x,Z)|X\}\right\vert _{x=\xi}=E\{f_{j}(\xi,Z)|X\}$. By the
bounded convergence theorem, $Ef_{j}(X,Z_{n}^{\ast})\rightarrow Ef_{j}(X,Z)$.

Next, $|Ef_{j}(X_{n},Z_{n}^{\ast})-Ef_{j}(X,Z_{n}^{\ast})|\leq L_{j}%
E\{\rho_{X}\left(  X_{n},X\right)  \wedge1\}\rightarrow0$ for every
$j\in\mathbb{N}$, again by the bounded convergence theorem, as $X_{n}%
\overset{a.s.}{\rightarrow}X$. Thus, $Ef_{j}(X_{n},Z_{n}^{\ast})=Ef_{j}%
(X,Z_{n}^{\ast})+o\left(  1\right)  \rightarrow Ef_{j}(X,Z)$ and
(\ref{ur cdem}) is proved. This establishes the convergence $(X_{n}%
,Z_{n}^{\ast})\overset{w}{\rightarrow}(X,Z)$.

Finally, $(X_{n},Z_{n}^{\ast})\overset{w}{\rightarrow}(X,Z)$ and $Z_{n}^{\ast
}|D_{n}\overset{w}{\rightarrow}_{p}Z|X$, where $X_{n}$ are $D_{n}$-measurable,
imply that $(\{(X_{n},Z_{n}^{\ast})|D_{n}\},X_{n})\overset{w}{\rightarrow}%
_{w}(\{(X,Z)|X\},X)$, by a straightforward modification of the proof of
Theorem 2.1 in Crimaldi and Pratelli (2005). By Theorem \ref{p citex} (see
also Remark \ref{Remark 3.42}), the latter convergence implies that
\begin{equation}
\left(  (X_{n},Z_{n}^{\ast})|D_{n},~(X_{n},Z_{n}^{\ast})|X_{n}\right)
\overset{w}{\rightarrow}_{w}\left(  (X,Z)|X,~(X,Z)|X\right)  .\label{eq predi}%
\end{equation}
In their turn, $X_{n}\overset{p}{\rightarrow}X$ and $(X_{n},Z_{n}^{\ast
})|X_{n}\overset{w}{\rightarrow}_{w}(X,Z)|X$ imply, by Corollary 4.4 of
Crimaldi and Pratelli (2005), that $(X_{n},Z_{n}^{\ast})|X_{n}\overset
{w}{\rightarrow}_{p}(X,Z)|X$, which jointly with (\ref{eq predi}) yields the
convergence $(X_{n},Z_{n}^{\ast})|D_{n}\overset{w}{\rightarrow}_{p}%
(X,Z)|X$.$\hfill\square$

\section{Details of the proof of Theorem \ref{Lemma bootstrap with boundary}}

\label{se wedge}

Here we establish the well-definition of $\check{\theta}^{\ast}$ in
(\ref{eq decomposition for the restricted estimator}) and its consistency at
the $n^{-1/2}$ rate in the sense that $\Vert\check{\theta}^{\ast}-\hat{\theta
}\Vert(1-\mathbb{I}_{n}^{\ast})=O_{p^{\ast}}(n^{-1/2})$ a.s.\medskip

STEP 1. \textit{Existence of a minimizer of } $q_{n}^{\ast}$ \textit{\ over a
portion of the bootstrap boundary close to} $\theta_{0}$. The choice
$(\theta,c)=(\theta_{0},g(\theta_{0}))$ trivially satisfies the equation
$g(\theta)=c$. Since $g$ is continuously differentiable in a neighborhood of
$\theta_{0}$ and $\dot{g}=(\dot{g}_{1}(\theta_{0}),\dot{g}_{2}(\theta
_{0}))^{\prime}\neq0$ (say that $\dot{g}_{1}(\theta_{0})\neq0$, with the
subscript denoting partial differentiation), by the implicit function theorem
there exist an $r>0$ and a unique function $\gamma:[\theta_{2,0}%
-r,\theta_{2,0}+r]\times\lbrack g(\theta_{0})-r,g(\theta_{0})+r]\rightarrow
\lbrack\theta_{1,0}-r,\theta_{1,0}+r]$ such that $\gamma(\theta_{2,0}%
,0)=\theta_{1,0}$, $g(\gamma(\theta_{2},c),\theta_{2})=c$;\ moreover, $\gamma$
is continuously differentiable. For outcomes such that $|g^{\ast}(\hat{\theta
})-g(\theta_{0})|\leq r$, the (non-empty) portion of the bootstrap boundary
$\partial\Theta^{\ast}=\{\theta\in\mathbb{R}^{2}:g(\theta)=g^{\ast}%
(\hat{\theta})\}$ contained in the rectangle $\Pi:=[\theta_{1,0}%
-r,\theta_{1,0}+r]\times\lbrack\theta_{2,0}-r,\theta_{2,0}+r]$ can be
parameterized as $\theta_{1}=\gamma({\theta_{2},g^{\ast}(\hat{\theta}))}$,
$\theta_{2}\in\lbrack\theta_{2,0}-r,\theta_{2,0}+r]$. Define $\check{\theta
}^{\ast}:=(\gamma(\check{\theta}_{2}^{\ast},g^{\ast}(\hat{\theta}^{r}%
)),\check{\theta}_{2}^{\ast})^{\prime}$, where $\check{\theta}_{2}^{\ast}$ is
a measurable minimizer of the continuous function $q_{n}^{\ast}(\gamma
(\theta_{2},g^{\ast}(\hat{\theta}^{r})),\theta_{2})$ over $\theta_{2}%
\in\lbrack\theta_{2,0}-r,\theta_{2,0}+r]$, with $\hat{\theta}^{r}:=\hat
{\theta}\mathbb{I}_{\{|g^{\ast}(\hat{\theta})-g(\theta_{0})|\leq r\}}%
+\theta_{0}\mathbb{I}_{\{|g^{\ast}(\hat{\theta})-g(\theta_{0})|>r\}}$. Since
$\mathbb{I}_{\{|g^{\ast}(\hat{\theta})-g(\theta_{0})|\leq r\}}\overset
{a.s.}{\rightarrow}1$ under $g^{\ast}(\theta_{0})=g(\theta_{0})$, it follows
that $\check{\theta}^{\ast}$ minimizes $q_{n}^{\ast}$ over $\partial
\Theta^{\ast}\cap\Pi$ with $P^{\ast}$-probability approaching one a.s.\medskip

\noindent STEP 2. \textit{Minimization of }$q_{n}^{\ast}$\textit{\ over the
entire bootstrap boundary.} For outcomes in
\[
A_{n}:=\{|g^{\ast}(\hat{\theta})-g(\theta_{0})|\leq r\}\cap\{g(\tilde{\theta
}^{\ast})<g^{\ast}(\hat{\theta})\}\cap\{\Vert\hat{\theta}-\theta_{0}%
\Vert+\Vert\tilde{\theta}^{\ast}-\hat{\theta}\Vert\leq\tfrac{r}{2}\},
\]
the minimum of $q_{n}^{\ast}$ over the entire bootstrap boundary
$\partial\Theta^{\ast}$ exists and is attained only in $\Pi$ (\textit{e.g.},
at the bootstrap estimator $\check{\theta}^{\ast}$ defined in Step 1),
provided that%
\[
\alpha_{n}:=\lambda_{\min}(M_{n})\tfrac{r^{2}}{4}-\lambda_{\max}(M_{n}%
)\Vert\tilde{\theta}^{\ast}-\hat{\theta}\Vert^{2}>0.
\]
To see this, consider $\theta^{c}:=c\hat{\theta}+(1-c)\tilde{\theta}^{\ast} $
where $c:=\inf\{a\in\lbrack0,1]:h(a\hat{\theta}+(1-a)\tilde{\theta}^{\ast
})=0\}$; $\theta^{c}$ is well-defined whenever $g(\tilde{\theta}^{\ast
})<g^{\ast}(\hat{\theta})$ because $g(\hat{\theta})\geq g^{\ast}(\hat{\theta
})$ and $h$ is continuous. Moreover, $\theta^{c}\in\Pi$ for outcomes in
$A_{n}$ because $\Vert\theta^{c}-\theta_{0}\Vert\leq\Vert\hat{\theta}%
-\theta_{0}\Vert+\Vert\tilde{\theta}^{\ast}-\hat{\theta}\Vert\leq\frac{r}{2}$
and, hence, $q_{n}^{\ast}(\theta^{c})\geq q_{n}^{\ast}(\check{\theta}^{\ast})$
for outcomes in $A_{n}$, by the minimizing property of $\check{\theta}^{\ast}$
on $\partial\Theta^{\ast}\cap\Pi$. For any $\theta\not \in \Pi$ and outcomes
in $A_{n}$, we therefore find that%
\begin{align*}
q_{n}^{\ast}(\theta)-q_{n}^{\ast}(\check{\theta}^{\ast})  & \geq q_{n}^{\ast
}(\theta)-q_{n}^{\ast}(\theta^{c})=q_{n}^{\ast}(\theta)-q_{n}^{\ast}%
(\tilde{\theta}^{\ast})+q_{n}^{\ast}(\tilde{\theta}^{\ast})-q_{n}^{\ast
}(\theta^{c})\\
& \geq\lambda_{\min}(M_{n})\Vert\theta-\tilde{\theta}^{\ast}\Vert^{2}%
-\lambda_{\max}(M_{n})\Vert\tilde{\theta}^{\ast}-\theta^{c}\Vert^{2}\\
& \geq\lambda_{\min}(M_{n})\{\Vert\theta-\theta_{0}\Vert-\Vert\tilde{\theta
}^{\ast}-\theta_{0}\Vert\}^{2}-\lambda_{\max}(M_{n})\Vert\tilde{\theta}^{\ast
}-\hat{\theta}\Vert^{2}\\
& \geq\lambda_{\min}(M_{n})\tfrac{r^{2}}{4}-\lambda_{\max}(M_{n})\Vert
\tilde{\theta}^{\ast}-\hat{\theta}\Vert^{2}=\alpha_{n}.
\end{align*}
Thus, for outcomes in $A_{n}\cap\{\alpha_{n}>0\}$, $q_{n}^{\ast}$ out of $\Pi$
is larger than $\min_{\theta\in\partial\Theta^{\ast}\cap\Pi}q_{n}^{\ast
}(\theta)$, such that $\check{\theta}^{\ast}$ minimizes $q_{n}^{\ast}$ over
the entire bootstrap boundary.

We find the associated probability%
\begin{align*}
& P^{\ast}\left(  (1-\mathbb{I}_{n}^{\ast})q_{n}^{\ast}(\check{\theta}^{\ast
})\leq(1-\mathbb{I}_{n}^{\ast})q_{n}(\theta)~~\forall\theta:h\left(
\theta\right)  =0\right) \\
& \hspace{0.5in}\overset{}{\geq}P^{\ast}\left(  |g^{\ast}(\hat{\theta
})-g(\theta_{0})|\leq r,\Vert\hat{\theta}-\theta_{0}\Vert\leq\tfrac{r}%
{4},\Vert\tilde{\theta}^{\ast}-\hat{\theta}\Vert\leq\frac{r}{4},\alpha_{n}%
\geq0\right) \\
& \hspace{0.5in}\overset{}{=}\mathbb{I}_{\{|g^{\ast}(\hat{\theta}%
)-g(\theta_{0})|\leq r\}\cap\{\Vert\hat{\theta}-\theta_{0}\Vert\leq
r/4\}}P^{\ast}\left(  \Vert\tilde{\theta}^{\ast}-\hat{\theta}\Vert\leq
\tfrac{r}{4},\alpha_{n}\geq0\right)  \overset{a.s.}{\rightarrow}1
\end{align*}
because $g(\hat{\theta})\overset{a.s.}{\rightarrow}g(\theta_{0})$,
$\lambda_{\min}(M_{n})\rightarrow\lambda_{\min}(M)>0$ a.s., $\lambda_{\max
}(M_{n})\rightarrow\lambda_{\max}(M)<\infty$ a.s. and $\Vert\tilde{\theta
}^{\ast}-\hat{\theta}\Vert\overset{w^{\ast}}{\rightarrow}_{a.s.}0$. This
establishes the fact that $\hat{\theta}^{\ast}$ of
(\ref{eq decomposition for the restricted estimator}), with $\check{\theta
}^{\ast}$ as defined in Step 1, minimizes $q_{n}^{\ast}$ over the bootstrap
parameter space $\Theta^{\ast}$ with $P^{\ast}$-probability approaching one
a.s.\medskip

\noindent STEP 3. \textit{Consistency rate of} $\check{\theta}^{\ast}$.
Similarly to Step 2, for outcomes in $A_{n}$,
\[
0\geq q_{n}^{\ast}(\check{\theta}^{\ast})-q_{n}^{\ast}(\theta^{c})\geq
\lambda_{\min}(M_{n})\Vert\check{\theta}^{\ast}-\tilde{\theta}^{\ast}\Vert
^{2}-\lambda_{\max}(M_{n})\Vert\tilde{\theta}^{\ast}-\hat{\theta}\Vert
^{2}\text{,}%
\]
the first inequality by the minimizing property of $\check{\theta}^{\ast}$
over $\partial\Theta^{\ast}\cap\Pi$. Therefore,%
\begin{align*}
& P^{\ast}\left(  (1-\mathbb{I}_{n}^{\ast})\Vert\check{\theta}^{\ast}%
-\tilde{\theta}^{\ast}\Vert^{2}\leq(1-\mathbb{I}_{n}^{\ast})\frac
{\lambda_{\max}(M_{n})}{\lambda_{\min}(M_{n})}\Vert\tilde{\theta}^{\ast}%
-\hat{\theta}\Vert^{2}\right) \\
& \geq P^{\ast}\left(  |g^{\ast}(\hat{\theta})-g(\theta_{0})|\leq r,\Vert
\hat{\theta}-\theta_{0}\Vert\leq\tfrac{r}{4},\Vert\tilde{\theta}^{\ast}%
-\hat{\theta}\Vert\leq\tfrac{r}{4}\right) \\
& =\mathbb{I}_{\{|g^{\ast}(\hat{\theta})-g(\theta_{0})|\leq r\}\cap\{\Vert
\hat{\theta}-\theta_{0}\Vert\leq r/4\}}P^{\ast}\left(  \Vert\tilde{\theta
}^{\ast}-\hat{\theta}\Vert\leq\tfrac{r}{4}\right)  \overset{a.s.}{\rightarrow
}1.
\end{align*}
As $\lambda_{\max}(M_{n})/\lambda_{\min}(M_{n})\overset{a.s.}{\rightarrow
}\lambda_{\max}(M)/\lambda_{\min}(M)$ and $\Vert\tilde{\theta}^{\ast}%
-\hat{\theta}\Vert=O_{p^{\ast}}(n^{-1/2})$ $P$-a.s. (the latter, by
(\ref{eq separata})), it follows that $(1-\mathbb{I}_{n}^{\ast})\Vert
\check{\theta}^{\ast}-\tilde{\theta}^{\ast}\Vert=O_{p^{\ast}}(n^{-1/2})$
$P$-a.s. and $\Vert\hat{\theta}^{\ast}-\tilde{\theta}^{\ast}\Vert=O_{p^{\ast}%
}(n^{-1/2})$ $P$-a.s. for $\hat{\theta}^{\ast}$ of
(\ref{eq decomposition for the restricted estimator}). Thus, $\hat{\theta
}^{\ast} $ has the same consistency rate as $\tilde{\theta}^{\ast}$ This
argument applies to any $\check{\theta}^{\ast}$which is measurable and
minimizes $q_{n}^{\ast}$ over $\partial\Theta^{\ast}\cap\Pi$ for outcomes in
$A_{n}$. This completes Step 3.\medskip

Finally, consider the first-order condition [foc] for minimization of
$q_{n}^{\ast}$ on $\partial\Theta^{\ast}$. As $\Vert\check{\theta}^{\ast
}-\theta_{0}\Vert(1-\mathbb{I}_{n}^{\ast})\leq\{\Vert\check{\theta}^{\ast
}-\tilde{\theta}^{\ast}\Vert+\Vert\check{\theta}^{\ast}-\theta_{0}%
\Vert\}(1-\mathbb{I}_{n}^{\ast})\overset{w^{\ast}}{\rightarrow}_{a.s.}0$, it
follows that $\mathbb{I}_{\{\check{\theta}^{\ast}\in\operatorname*{int}%
(\Pi)\}}(1-\mathbb{I}_{n}^{\ast})+\mathbb{I}_{n}^{\ast}\overset{w}%
{\rightarrow}_{a.s.}1$. As additionally $\dot{g}(\theta_{0})\neq0$, by
continuity of $\dot{g}(\theta):=(\partial g/\partial\theta^{\prime})(\theta)$,
the foc takes the form%
\[
P^{\ast}\left(  \{\dot{q}_{n}(\check{\theta}^{\ast})+\check{\delta}_{n}\dot
{g}(\check{\theta}^{\ast})\}(1-\mathbb{I}_{n}^{\ast})=0\right)  \overset
{a.s.}{\rightarrow}1,
\]
where $\check{\delta}_{n}\in\mathbb{R}$ are measurable and can be determined
from the constraint $h(\check{\theta}^{\ast})=0$.

\section{Simulation design}

\label{Appendix MC}We provide here a description of the Monte Carlo [MC]
simulation design used for the linear regression model of Section
\ref{sec example}. Data $D_{n}:=\{y_{t},x_{t}\}_{t=1}^{n}$ are generated
according to eq. (\ref{eq:lm})\ and the object of interest is inference on
$\beta$ based on $\tau_{n}:=n^{1/2}(\hat{\beta}-\beta)$, with $\hat{\beta}$
denoting the OLS estimator of $\beta$, see Section
\ref{sec example model and BS}. We consider the case where $x_{t}=\sum
_{s=1}^{t}\eta_{s}$ is a non-stationary ($I(1)$) process under the following
three different distributional structures for $(\varepsilon_{t},\eta_{t})$:

\begin{description}
\item (i) $(\varepsilon_{t},\eta_{t})$ is i.i.d. $N(0,I_{2})$ such that if the
true variance $\hat{\omega}_{\varepsilon}=1$ was used in eq. (\ref{eq:blm}),
the bootstrap would perform exact conditional inference (see Remark
\ref{Remark on exact conditional inference});

\item (ii) $\varepsilon_{t}=\zeta_{t}(1+0.3\varepsilon_{t-1}^{2}+0.3\eta
_{t-1}^{2})^{1/2}$ and $\eta_{t}=\xi_{t}(1+0.6\eta_{t-1}^{2})^{1/2}$, where
$(\zeta_{t},\xi_{t})$ is i.i.d. $N(0,I_{2})$; this corresponds to a stationary
and ergodic conditionally heteroskedastic process with non Gaussian
unconditional marginals;

\item (iii) $\eta_{t}=\xi_{t}(1+\delta\mathbb{I}_{\{\varepsilon_{t}\leq0\}})$,
where $(\varepsilon_{t},\xi_{t})$ is i.i.d. $N(0,I_{2})$ and $\delta=9$.
\end{description}

For model (ii) we initialize the process by setting the conditional variance
equal to its unconditional expectation (results do not change if a burn-in
method is employed instead). The bootstrap is implemented as in Section
\ref{sec example model and BS}, with $\hat{\omega}_{\varepsilon}$ chosen as
the OLS residual variance. For DGP (i) bootstrap inference is close to exact
(see Remark \ref{Remark on exact conditional inference}) and it holds that the
bootstrap \emph{p}-value $p_{n}^{\ast}$ satisfies $p_{n}^{\ast}\overset{d}%
{=}U\left(  0,1\right)  +O_{p}(n^{-1/2})$. DGP (ii) satisfies the conditions
discussed in Section \ref{sec intro to conditional bs validity}\textbf{\ }and
bootstrap inference is valid conditionally on $X_{n}:=\{x_{t}\}_{t=1}^{n}$;
i.e., $p_{n}^{\ast}|X_{n}\overset{w}{\rightarrow}_{p}U\left(  0,1\right)  $.
For DGP (iii), on the other hand, bootstrap inference is not valid
conditionally on this $X_{n}$, but it is valid unconditionally; see Section
\ref{sec uncond validity W/O conditional validity}.\textbf{\ }Hence,
$p_{n}^{\ast}\overset{w}{\rightarrow}U\left(  0,1\right)  $ while $p_{n}%
^{\ast}|X_{n}$ has a random limit distribution. Notice that since the
bootstrap statistic is conditionally Gaussian, \emph{p}-values can be obtained
without resorting to simulation.

Standard MC experiments generating $D_{n}=\left(  y_{1},...,y_{n}%
,X_{n}\right)  $ at each MC\ iteration allow estimation of the unconditional
distribution of the \emph{p}-value $p_{n}^{\ast}$, rather than its
distribution conditional on $X_{n}$ (see e.g. Hansen, 2000, footnote 11). To
simulate the distribution of $p_{n}^{\ast}$ conditional on $X_{n}$, we
implement a double MC design where, for each $m=1,....,M$, we generate the
regressors $X_{n}^{(m)}\sim X_{n}$ and then, for each $v=1,...,N $, we
generate data $(y_{1}^{(m,v)},...,y_{n}^{(m,v)})$ from their distribution
conditional on $X_{n}=X_{n}^{(m)}$. The respective statistics $\tau
_{n}^{(m,v)}$ and the associated bootstrap \emph{p}-values, $p_{n}^{\ast
(m,v)}$, are used to estimate the conditional distribution of $p_{n}^{\ast
}|\{X_{n}=X_{n}^{(m)}\}$, for each $m$, by the empirical cdf $N^{-1}%
\sum\nolimits_{v=1}^{N}\mathbb{I}_{\{p_{n}^{\ast(m,v)}\leq\cdot\}}$. We set
$M=1,000$ and $N=100,000$ throughout. Notice that for model (iii), once the
regressor $x_{t}$ (hence, $\eta_{t}$) is generated, simulation conditional on
$\{x_{t}\}$ requires drawing from the conditional distribution of
$\varepsilon_{t}$ given $\eta_{t}$. A simple application of the Bayes rule
yields that
\[
P\left(  \varepsilon_{t}\leq0|\eta_{t}\right)  =\frac{1}{1+(1+\delta
)e^{-\frac{\eta_{t}^{2}}{2}\frac{\delta(2+\delta)}{(1+\delta)^{2}}}}=:p_{t}%
\]
and hence that $\varepsilon_{t}|\{x_{t}\}\overset{d}{=}\varepsilon_{t}%
|\eta_{t}\sim|\xi_{t}|s_{t}$, where $s_{t}$ is a random sign equal to $-1$
with probability $p_{t}$.

All computations have been performed using \textsc{Matlab} R2019\emph{b}. Code
is available from the Authors upon request.

\section*{Additional references}

\begin{description}
\item \textsc{Berti, P., L. Pratelli and P. Rigo} (2006): Almost sure weak
convergence of random probability measures, $\emph{Stochastics}$ 78, 91--97.

\item \textsc{Cavaliere, G. and I. Georgiev} (2019): Inference under random
limit bootstrap measures. Under revision at \emph{Econometrica}.

\item \textsc{Dudley, R.M} (2004): \emph{Real Analysis and Probability},
Cambridge University Press, Cambridge.

\item \textsc{Ethier, S and T. Kurtz} (2005): \emph{Markov Processes.
Characterization and Convergence}, John Wiley\&Sons, Hoboken.

\item \textsc{Worm, D and s. Hille} (2011): Ergodic decompositions associated
with regular Markov operators on Polish spaces, \emph{Ergodic Theory and
Dynamical Systems} 31, 571-597.
\end{description}

\end{document}